\documentclass[tradiabstract]{aa} % for the research notes

\usepackage{natbib}
\bibpunct{(}{)}{;}{a}{}{,} % to follow the A&A style
\setcitestyle{citesep={,}}
\usepackage{graphicx}

\def\kms{\hbox{km$\;$s$^{-1}$}}

\newcommand\mancha{\textsc{Mancha3D~}}

\begin{document} 

\title{Two-fluid simulations of waves in the solar chromosphere II. Propagation and damping of fast magneto-acoustic waves and shocks}
\author{B. Popescu Braileanu
          \inst{1,2}
                \thanks{ \email{bpopescu@iac.es}}
          \and
          V. S. Lukin\inst{3}
          \and
          E. Khomenko\inst{1,2}
          \and
          \'A. de Vicente\inst{1,2}}
          
\titlerunning{Two-fluid simulations of waves in the solar chromosphere}
\authorrunning{Popescu Braileanu et al.}

\institute{Instituto de Astrof\'{\i}sica de Canarias, 38205 La Laguna, Tenerife, Spain
\and Departamento de Astrof\'{\i}sica, Universidad de La Laguna, 38205, La Laguna, Tenerife, Spain
\and National Science Foundation, Alexandria, VA, 22306, USA}

\date{Received 2019; Accepted XXXX}
 
\abstract{Waves and shocks traveling through  the solar chromospheric plasma are influenced by its partial ionization and weak collisional coupling, and may become susceptible to multi-fluid effects, similar to interstellar shock waves. In this study, we consider fast magneto-acoustic shock wave formation and propagation in a stratified medium, that is permeated by a horizontal magnetic field, with properties similar to that of the solar chromosphere. The evolution of plasma and neutrals is modeled using a two-fluid code  that evolves a set of coupled equations for two separate fluids. We observed that waves in neutrals and plasma, initially coupled at the upper photosphere, become uncoupled at higher heights in the chromosphere. This decoupling can be a consequence of either the characteristic spatial scale at the shock front, that becomes similar to the collisional scale, or the change in the relation between the wave frequency, ion cyclotron frequency, and the collisional frequency with height.  The decoupling height is a sensitive function of the wave frequency, wave amplitude, and the magnetic field strength.  We observed that decoupling causes damping of waves and an increase in the background temperature due to the frictional heating.  The comparison between analytical and numerical results allows us to separate the role of the nonlinear effects from the linear ones on the decoupling and damping of waves.}

\keywords{Sun: chromosphere -- Sun: waves -- Sun: magnetic field -- Sun: numerical simulations}

\maketitle
%
%________________________________________________________________

\section{Introduction}

The solar chromosphere is a very dynamic region that is constantly perturbed by propagating waves and shocks. Waves can be observed by measuring fluctuations in the radiation intensity due to fluctuations in the density and temperature of both compressible acoustic and magneto-acoustic waves, and by measuring the Doppler shift associated with fluctuations in velocity for incompressible waves.  Waves have been observed at different layers in the solar atmosphere, from the photosphere to the corona, but limited spatial and temporal  resolution restricts their detection to low-frequency waves.  Significant effort has been made to measuring waves in the solar corona \citep{2004Verwichte, Ofman2008, Gupta2014}. Photospheric and chromospheric oscillations in different magnetic structures have been measured for decades. Making interpretations about chromospheric measurements is especially difficult because of the fast evolving structures, supersonic motions, and high-frequency oscillatory phenomena.  Physical processes in the chromosphere happen at characteristic times dictated by the Alfv\'en speed that can reach several hundreds of \kms. A review of multi-wavelength studies of waves in the chromosphere is provided by \citet{2015Jess}. 

The following two main mechanisms are thought to be responsible for the heating of the solar atmosphere: magnetic reconnection and waves. These two mechanisms are not independent since the magnetic reconnection can be a source of waves \citep{2004Verwichte, 2008Jess, 2008Luna, 2012Li, Prov2018} and since waves can produce instabilities needed to trigger reconnection processes \citep{Isobe2006, 2014Lee, 2009McLaughlin}. Even though the chromosphere has a much lower temperature than the solar corona, it has much higher energy density and requires more energy to compensate for the radiative losses \citep{2015Jess, 1977Withbroe, 1989Anderson}. Therefore, mechanisms providing energy to the chromosphere have to be very efficient.

Contrary to previous studies, based on low resolution observations or on 1D simulations, high-frequency waves may play an important role in the energization of the solar chromosphere.  From an observational point of view, waves
with frequencies above 10-50 mHz are  extremely difficult to detect and are often referred to as high-frequency waves \citep[e.g., see][]{2006Fossum, 2009Bello}. \cite{1994Musielak} considered wave generation in the solar convection zone and computed the acoustic wave energy flux as a function of frequency.  Their theoretical calculations revealed that acoustic flux has a broad maximum in the high-frequency region around 100 mHz (periods $\approx$ 10 s).  However, firm detection of the high-frequency part of the solar oscillation spectrum is still uncertain. For example, \citet{2009Bello, 2010aBello} and  \citet{2010bBello}, using high resolution data   obtained by Gregor Fabry-P\'erot Interferometer (GFPI) at Vacuum Tower Telescope (VTT) and  Imaging Magnetograph eXperiment (IMaX) instrument aboard the Sunrise  mission, revealed a high acoustic flux in the chromosphere for high-frequency waves. This result is in contradiction to the earlier work by \citet{2006Fossum} that uses Transition Region and Coronal Explorer (TRACE)  data.  

The maxima of the observed solar wave spectra fall into 3-5 mHz range, depending on the height. This low-frequency part of the solar wave spectrum has been extensively studied both theoretically and observationally. From a theoretical point of view, a magnetohydrodynamic (MHD) approximation can be safely used for the studies of low-frequency wave phenomena. A review of theoretical studies of wave propagation through photospheric, chromospheric, and coronal structures can be found in \citet{Khomenko+Santamaria2013, Khomenko+Collados2015}. Waves are sensitive to the horizontal and vertical structure of the atmosphere. In particular, strong vertical gradients in the acoustic and Alfv\'en speed lead to phenomena of mode transformation \citep{Cally2006, Cally+Goossens2008} or fast magneto-acoustic mode refraction \citep{Bogdan+etal2003, Khomenko+Collados2006} with particularly interesting phenomena happening around singular magnetic field structures, such as null points \citep{2014Lee, Santamaria+etal2017, Prov2018}. 

When considering high-frequency waves and shocks, damping or dissipation effects become important. Small-scale disturbances can easily be damped through various non-ideal mechanisms such as viscosity, radiation, thermal conduction, electrical resistivity, or ion-neutral friction. Other mechanisms related to the atmospheric structure are also at play, such as resonant absorption or phase mixing. The importance of these effects depends on the height in the solar atmosphere. For example, damping due to thermal conduction has been studied under coronal conditions \citep{2004Klimchuk, Gupta2014, 2015Soler}, phase mixing, and resonant absorption and are thought to play a role in damping coronal loop oscillations \citep{vasquez2005, 2004Verwichte}. High-frequency compressible waves are heavily damped by radiative processes at the base of the photosphere where the energy that is dissipated by shocks is converted into radiation \citep{2002Carlsson}. Viscous damping of shock waves is reconsidered by \citet{Arber+etal2016}.\ Additionally,  resistive dissipation of Alfv\'en waves, enhanced through ion-neutral interaction, is considered by \citet{Song2011, 2016Shelyag,  2016Sykora}.

Under the conditions of the solar chromosphere, the reduction in the collisional coupling with height leads to partial decoupling of the neutral and charged components of the solar plasma, and produces a series of multi-fluid effects on waves \citep{2018Ballester}. In order to treat these effects, classical MHD modeling may be insufficient and extensions to this approach are needed, such as the use of single-fluid formalism together with the generalized Ohm's law, or a two-fluid (or multi-fluid) formalism. Ion-neutral effects on waves are extensively studied analytically under the assumption of homogeneous, unbounded plasma \citep{2011Zaq, 2013Soler, 2013Zaq, 2013Soler2, 2016Gomez1, 2017Gomez1, 2018Ballester2}, or by using simplified magnetic configurations, such as flux tubes \citep{2012Zaq, 2017Soler}. The main conclusions from these works is that ion-neutral effects produce important wave damping when the wave frequency is close to the ion-neutral collision frequency. In addition, new wave modes, that do not exist in a pure proton-electron plasma, can appear while cut-off wave numbers may appear for other modes. 

Complex nonlinear dynamic processes and the 3D structure of the chromosphere strongly limit the practicality of linear theory for interpretation of observations, and numerical modeling techniques must be applied. Such numerical models mainly use a single-fluid approach that introduce the partial ionization effects through a generalized Ohm's law \citep{Khomenko+Collados2012, Cheung+Cameron2012, Sykora2012, 2014bKh, 2016Sykora, 2016Shelyag, Khomenko+Vitas2017, Khomenko+etal2018}. One of the advantages of the numerical approach is the possibility to study wave dissipation and heating due to ion-neutral interactions. The heating is expressed via the ambipolar term in the single-fluid approach \citep{Song2011, 2016Khomenko2}, or via the frictional heating term in the two-fluid approach \citep{Leake2014}, and is dropped in the linear theory, being quadratic in the perturbed quantities. Nonlinear simulations allow us to fully take the effects of wave heating due to neutrals into
account. Most of such modeling has been done using the single-fluid approach, which is most appropriate for the dense layers of the solar atmosphere (photosphere and low chromosphere). Only recently has modeling been performed in the solar atmosphere for  waves in the two-fluid fully nonlinear assumption \citep{2017Maneva}. 

One particular aspect of the chromospheric physics that has been scarcely studied is the influence of multi-fluid effects on the formation and dissipation of chromospheric shock waves. \citet{2016Hillier} study the formation and evolution of slow-mode shocks driven by reconnection in a partially ionized plasma. A complex multi-fluid structure of the shock transition was modeled revealing structures similar to $C$-shocks or $J$-shocks in the classification by \citet{Draine1993}. The frictional heating associated with ion-neutral decoupling at the shock front was up to 2\% of the available magnetic energy. 

The aim of the present paper is to deepen our understanding of the multi-fluid nature of chromospheric shocks and their contribution to the chromospheric heating. To do so, we performed two-fluid modeling of propagation of fast magneto-acoustic waves generated at the base of the chromosphere for a particular case  with a horizontal magnetic field. Depending on the conditions, these waves steepen into shocks at the middle-to-upper chromosphere, leading to decoupling between ion and neutral fluid velocities. Unlike many studies mentioned above, we fully take nonlinear effects into account. The simulations are performed using a newly extended \mancha code \citep{Popescu+etal2018}. We study the effects of the wave frequency and amplitude, as well as the background magnetic field strength on the height where ions and neutrals become decoupled.  The results of the numerical simulations are compared to analytical solutions within the two-fluid and single-fluid descriptions, thus allowing us to separate the nonlinear, linear, and two-fluid effects from wave damping and dissipation.

%%%%%%%%%%%%%%%%%%%%%%%%%%%%%%%%%%%%%%%%%%%%%%%%%%%%%%%%%
\section{Numerical model} \label{awvalc}
%%%%%%%%%%%%%%%%%%%%%%%%%%%%%%%%%%%%%%%%%%%%%%%%%%%%%%%%%

In the present study we used a two-fluid model assuming a purely hydrogen plasma. This allowed us to greatly simplify the expressions for the interaction terms in the two-fluid equations \citep{Leake2014, Khomenko2014}. Nevertheless, it has to be noted that modeling based on hydrogen plasma cannot be consistently applied through atmospheric layers. The solar atmosphere is composed of multiple chemical species in different ionization states. Thermodynamic conditions in the photosphere are such that the main donors of electrons are metals, and therefore hydrogen plasma modeling cannot self-consistently take this feature into account. For this reason, modeling performed in this work only applies to the atmospheric layers starting from the upper photosphere and upward. Below, we describe the equations solved by the code, the equilibrium atmospheric model, and the wave driver.  

\subsection{Two-fluid equations}

The following set of equations is solved numerically by the two-fluid version of the code \mancha \citep{Popescu+etal2018}. Electrons and ions are evolved together by a single set of equations for charges (marked by subindex ``c'') while neutrals are evolved separately (marked by subindex ``n''). Here we neglected Hall, battery, and other smaller effects from the generalized Ohm's law written in the system of reference of charges. Also, we did not consider ionization and recombination processes, viscosity, thermal conduction, or radiation. The only non-ideal terms are neutral-charges, elastic collisional terms in the momentum, and energy equations,
\begin{eqnarray}  \label{eq:twoflnum}
\frac{\partial \rho_n}{\partial t} + \nabla \cdot (\rho_n\vec{u}_n) = 0,   \nonumber \\
\frac{\partial \rho_c}{\partial t} + \nabla \cdot (\rho_c\vec{u}_c) = 0, \nonumber \\
\frac{\partial (\rho_n\vec{u_n})}{\partial t} + \nabla \cdot (\rho_n\vec{u_n}\vec{u_n} +{p_n} )  = \rho_n\vec{g} +\vec{R}_n,  \nonumber \\
\frac{\partial (\rho_c\vec{u_c})}{\partial t} + \nabla \cdot (\rho_c\vec{u}_c\vec{u}_c + {p_c} ) =\vec{J}\times\vec{B} + \rho_c\vec{g}  -\vec{R}_n,  \nonumber \\
\frac{\partial}{\partial t}\left( e_n+\frac{1}{2}\rho_n u_n^2\right) +  \nabla \cdot \left( \vec{u}_n (e_n +  \frac{1}{2}\rho_n u_n^2) + {p_n} \vec{u}_n \right ) =  \nonumber \\ 
\rho_n\vec{u}_n\cdot\vec{g}  + M_n,  \nonumber \\
\frac{\partial}{\partial t} \left(e_c+\frac{1}{2}\rho_c u_c^2 \right) +  \nabla \cdot \left( \vec{u}_c(e_c + 
\frac{1}{2}\rho_c u_c^2) + {p_c} \vec{u}_c   \right) =    \nonumber \\ 
\vec{u}_c \cdot (\vec{J} \times \vec{B})  + \rho_c\vec{u}_c\cdot\vec{g} -M_n,  \nonumber \\
\frac{\partial\vec{B}}{\partial t} - \vec{\nabla} \times   (\vec{u}_c\times\vec{B}) =0, 
\end{eqnarray}
\noindent with
\begin{equation} \label{eq:m}
\vec{R}_n = \alpha\rho_n \rho_c (\vec{u}_c - \vec{u}_n),
\end{equation}
\begin{equation} \label{eq:e}
M_n = \frac{1}{2}\alpha\rho_n \rho_c ({u_c}^2 - {u_n}^2) +\frac{(T_c - T_n)}{\gamma-1} \frac{k_B}{m_n}\alpha\rho_n \rho_c.
\end{equation} 

\noindent In these equations, $\vec{u}_{c,n}$, $\rho_{c,n}$, $p_{c,n}$, $e_{c, n,}$ and $T_{c,n}$ are center of mass velocities, mass densities, pressures, internal energies, and temperatures of charges and neutrals,  $\vec{B}$ is the magnetic field, $\vec{J}$ is the current density, and $\vec{R}_n$ and $M_n$ are the neutral momentum  and energy collisional source terms, respectively. We neglected the contribution of electrons to the mass density of charges with $m_e/m_H < 10^{-3}$, $\rho_c=m_H n_e+m_e n_e\approx m_Hn_e$, with $n_e$ being the electron number density.  The gravitational acceleration $\vec{g}$ is oriented in the negative $z$ direction. The collisional parameter $\alpha$ is defined as follows,

\begin{equation} \label{eq:alpha}
\alpha=\frac{\rho_e \nu_{en} + \rho_i \nu_{in}}{\rho_n \rho_c},
\end{equation}
where $\nu_{en}$ and $\nu_{in}$ are the electron-neutral and ion-neutral collision frequencies, respectively,

\begin{equation} \label{eq:nu}
\nu_{in} = n_n\sqrt{\frac{16 k_BT_{cn}}{\pi m_H}}\sigma_{in}; \,\,\, \nu_{en} = n_e\sqrt{\frac{8 k_BT_{cn}}{\pi m_e}}\sigma_{en};
\end{equation}
with  $T_{cn}$ being the average temperature of neutrals and charges, $\sigma_{in}=5\times10^{-19}$ m$^2$ and $\sigma_{en}= 10^{-19}$ m$^2$ \citep{1965Braginskii}.

The relation between internal energy and pressure, and between the pressure, the number density, and the temperature are defined by the following ideal gas laws:
\begin{eqnarray}\label{eq:eos}
e_{c,n} &=& p_{c,n}/(\gamma -1),  \nonumber \\
p_{n} &=& n_{n} k_B  T_{n},   \nonumber \\
p_{c} &=& 2n_{e} k_B  T_{c}.
\end{eqnarray}
The factor of two in the last equation appears to be due to the contributions of both ions and electrons.\ Additionally, both temperatures are assumed to be equal to $T_c$. 

The equations were advanced in time using a semi-implicit scheme from \citet{2012Toth} with the collisional terms implemented implicitly, and the rest of the terms were implemented in a second order accurate Runge Kutta scheme. The details of the implementation are provided in \citet{Popescu+etal2018}. 
%The collisional timescales can be significantly smaller than the hydrodynamical timescale, making the numerical discretization of the equations stiff. The implicit implementation of the collisional terms overcomes a very small time step limitation imposed by the collisional time scale. In our semi-implicit implementation, the time step determined by the ideal MHD CFL condition can be used.  As an example, for one of the configurations used in the simulations below, the MHD CFL time step limit is $5.2 \times 10^{-5}$ s, while the time step that would be required to evaluate the collisional terms in a fully explicit scheme is $7 \times 10^{-7}$ s. Thus, our semi-implicit implementation allows to advance the solution in time about 75 times faster as compared to a fully explicit implementation.
The collisional timescales can be significantly smaller than the
hydrodynamical timescale, making the numerical discretization of the
equations stiff. The implicit implementation of the collisional terms
is not affected by  a very small time step limitation imposed by the collisional
time scale.\ This makes it possible to use the time step determined by the
ideal MHD CFL condition and thus advance in time about 75 times
faster as compared to a fully explicit implementation.

%%%%%%%%%%%%%%%%%%%%%%%%%%%%%%%%%%%%%%%%%%%%%%%%%%%%%%%%%
\begin{figure*}[t]
\centering
\includegraphics[width = 8.5cm]{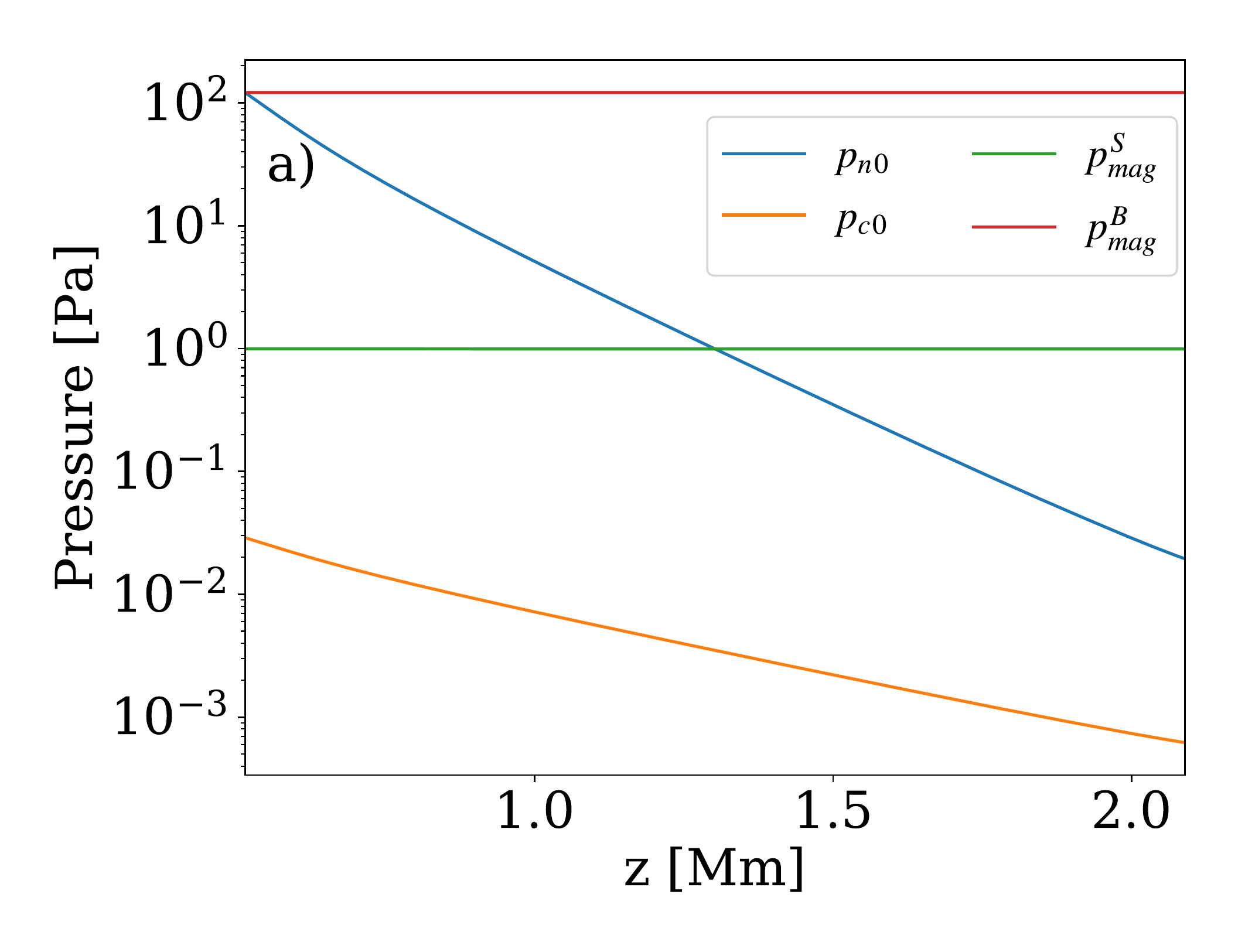}
\includegraphics[width = 8.5cm]{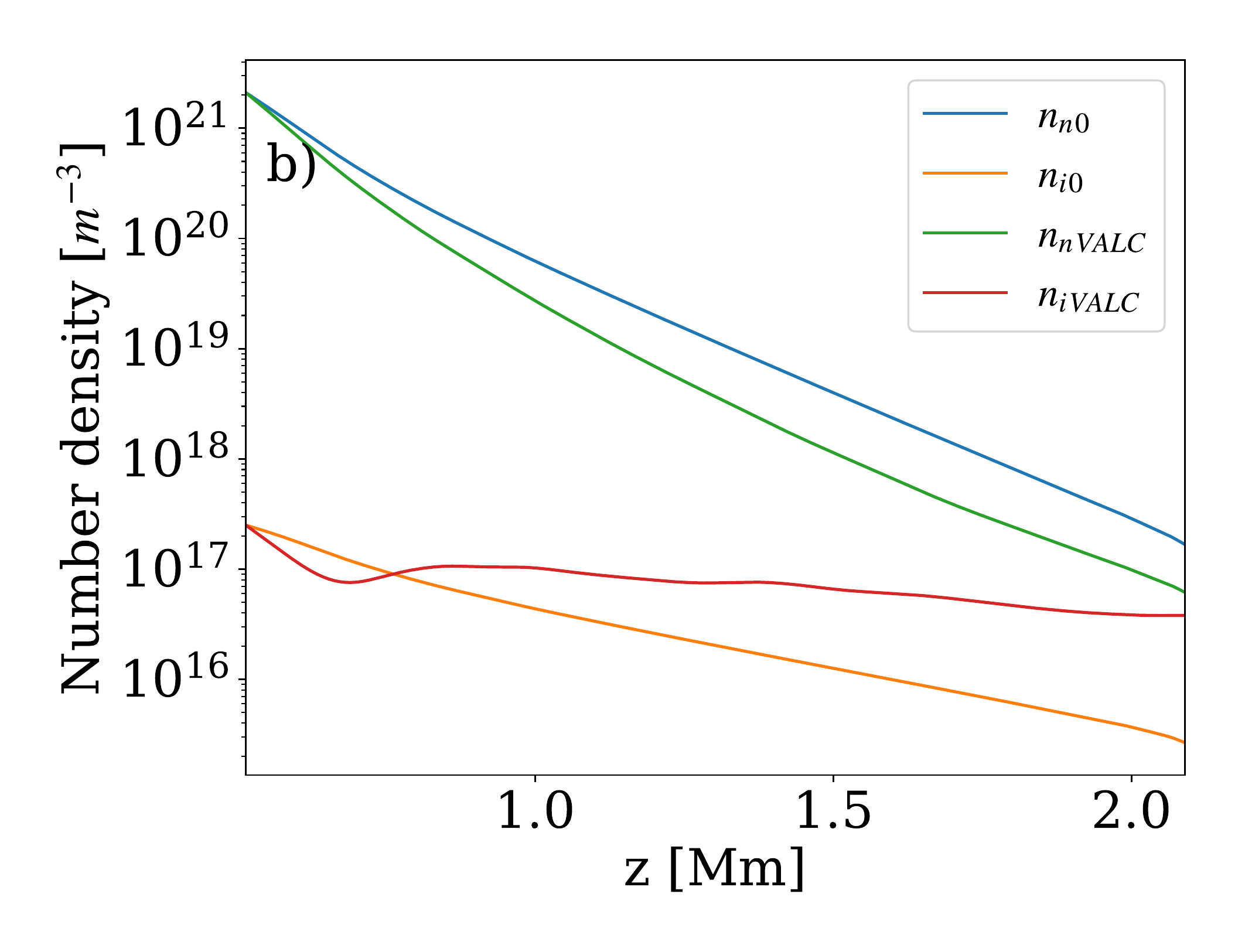}

\includegraphics[width = 8.5cm]{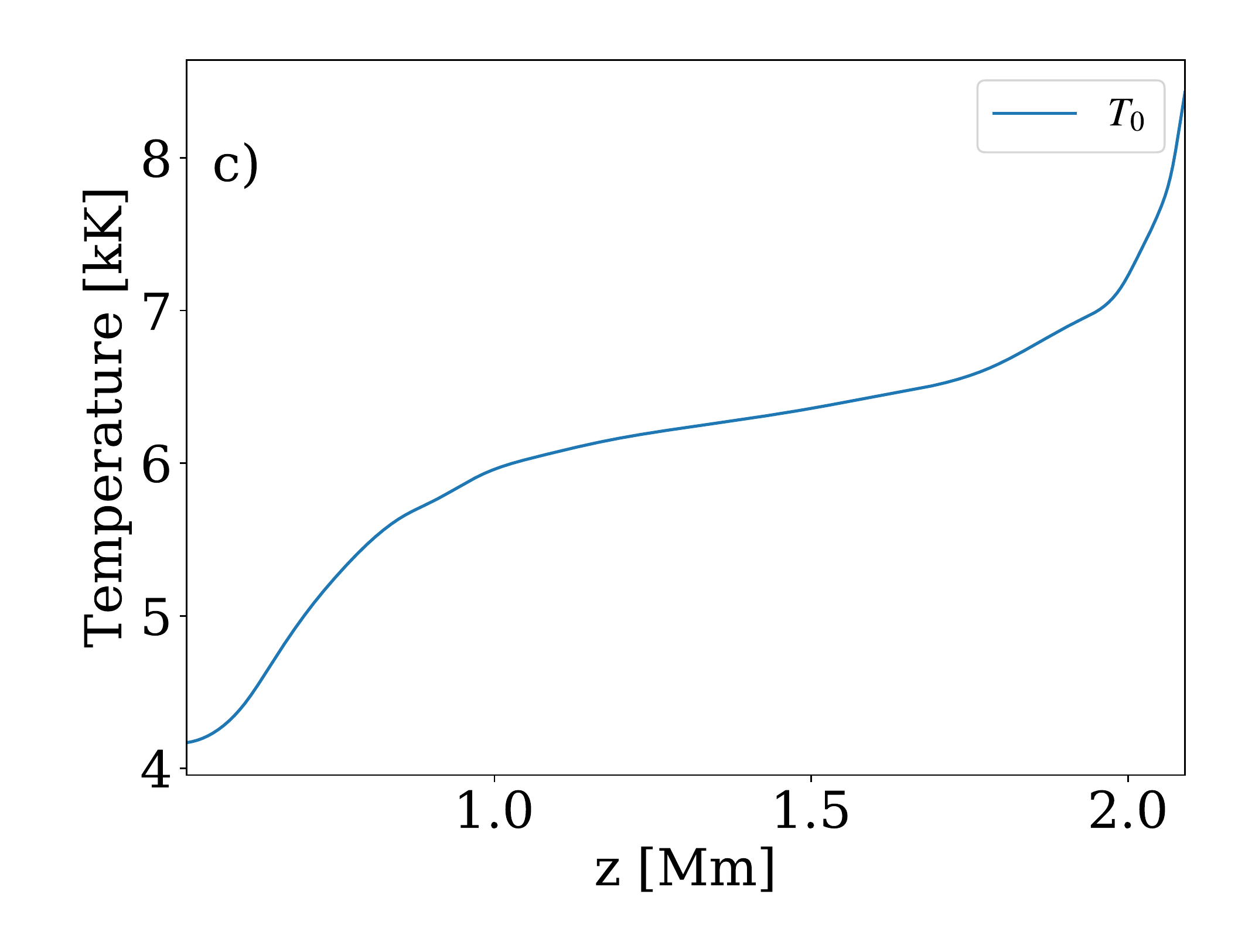}
\includegraphics[width = 8.5cm]{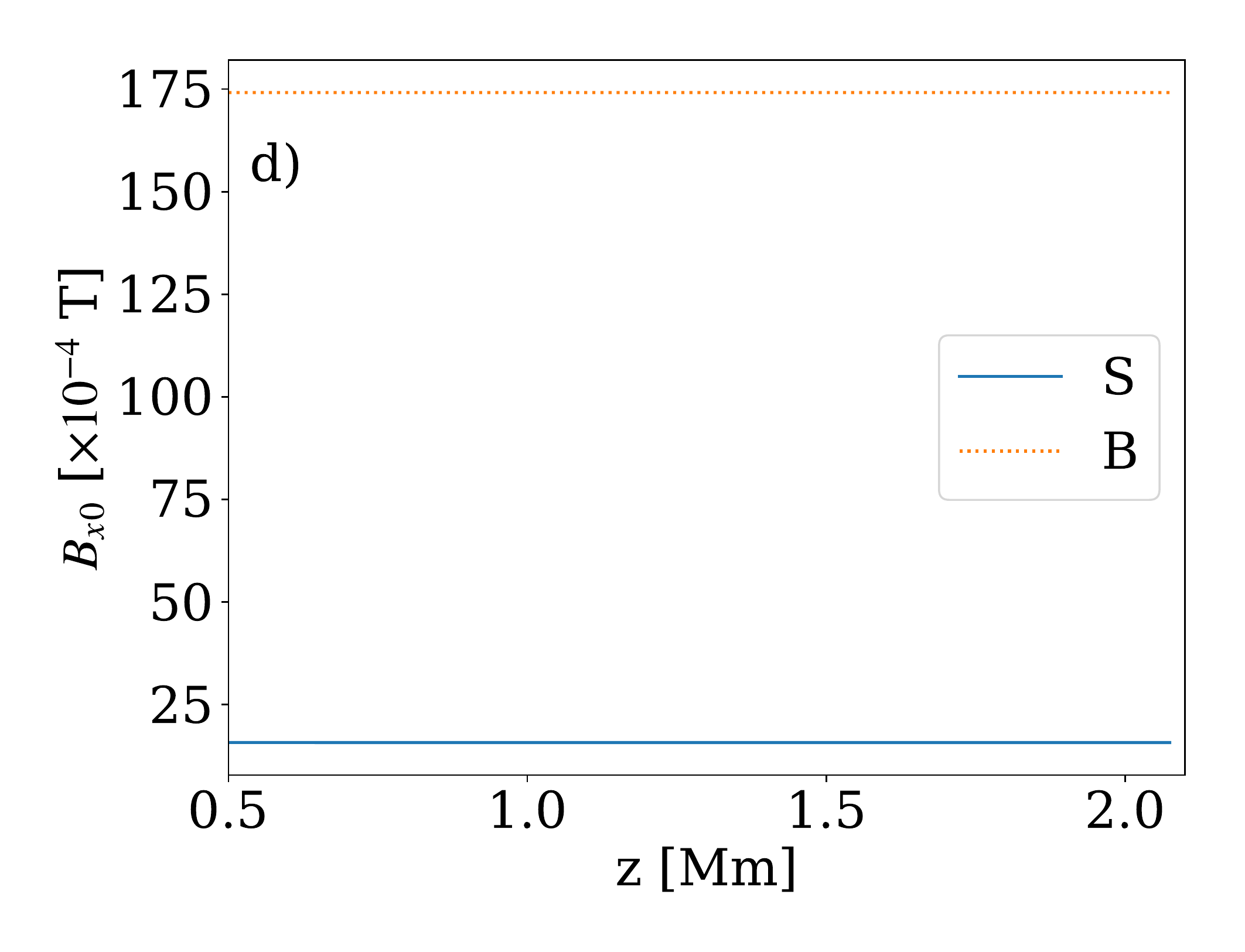}
\caption{Parameters of background model atmosphere as function of height. Panel (a): gas pressure of neutrals (blue), charges (orange), and magnetic pressures corresponding to the S magnetic field profile (green) and B profile (red). Panel (b): number density of neutrals (blue), and charges (orange). For comparison, corresponding number densities from VALC model atmosphere are plotted in green and red. Panel (c): temperature. Panel (d): horizontal magnetic field, $B_{x0}$, of the two background profiles, S and B. Here and below, zero of $z$ axis is considered to be the same as in VALC atmosphere. }
\label{fig:iniprofaw3}
\end{figure*}
%%%%%%%%%%%%%%%%%%%%%%%%%%%%%%%%%%%%%%%%%%%%%%%%%%%%%%%%%

%%%%%%%%%%%%%%%%%%%%%%%%%%%%%%%%%%%%%%%%%%%%%%%%%%%%%%%%%
\subsection{Equilibrium atmosphere}
%%%%%%%%%%%%%%%%%%%%%%%%%%%%%%%%%%%%%%%%%%%%%%%%%%%%%%%%%

We assumed a model atmosphere with all hydrodynamic variables and a purely horizontal magnetic field, $B_{x0}$, that is stratified in the vertical, $z$, direction, and subindex "0" refers to equilibrium variables.
 The \mancha code requires variables for the equilibrium atmosphere to be set separately for the neutral and charged components. In practice, any set of equilibrium variables can be used as long as they fulfill the conditions of purely hydrostatic (HS) equilibrium for neutrals, and magneto-hydrostatic (MHS) equilibrium for charges. 

 In equilibrium, we consider that the charges and neutrals have the same background temperature and we used the temperature profile from  the model proposed by
Vernazza, Avrett, and Loeser for the photosphere and Chromosphere 
(VALC) \citep{VALC}, starting from heights, $z$, above $z_0 \approx $ 500 km, and ending just below the transition region at a temperature of 9000 K, at $z_f \approx $ 2.1 Mm. Apart from the temperature structure, no other stratification from the VALC model can be used directly because the equations mentioned above are only valid for hydrogenate plasma. Therefore, we recalculated the stratification of pressure and density of both neutrals and charges, assuming pure hydrogen, in order to satisfy the requirements of the HS and MHS equilibrium as explained above. 

Apart from temperature, the only other parameter taken from VALC is the number density of charges and neutrals at the reference base level, $z_0\approx 500$ km, with $n_{c0}(z_0)=5\times10^{17}$ m$^{-3}$, and  $n_{n0}(z_0) = 2.1\times10^{21}$ m$^{-3}$. Then, the pressures of charges and neutrals at $z_0$, $p_{c0}(z_0)$ and $p_{n0}(z_0)$ are obtained from the ideal gas law from Eq. \ref{eq:eos}. As discussed above, we only considered the height range $z=\{z_0,z_f\}$ because the hydrogen plasma model does not provide correct electron number densities in the deeper layers as most available electrons come from metals.

%%%%%%%%%%%%%%%%%%%%%%%%%%%%%%%%%%%%%%%%%%%%%%%%%%%%%%%%%
\begin{figure}[ht]
\centering
\includegraphics[width=9cm]{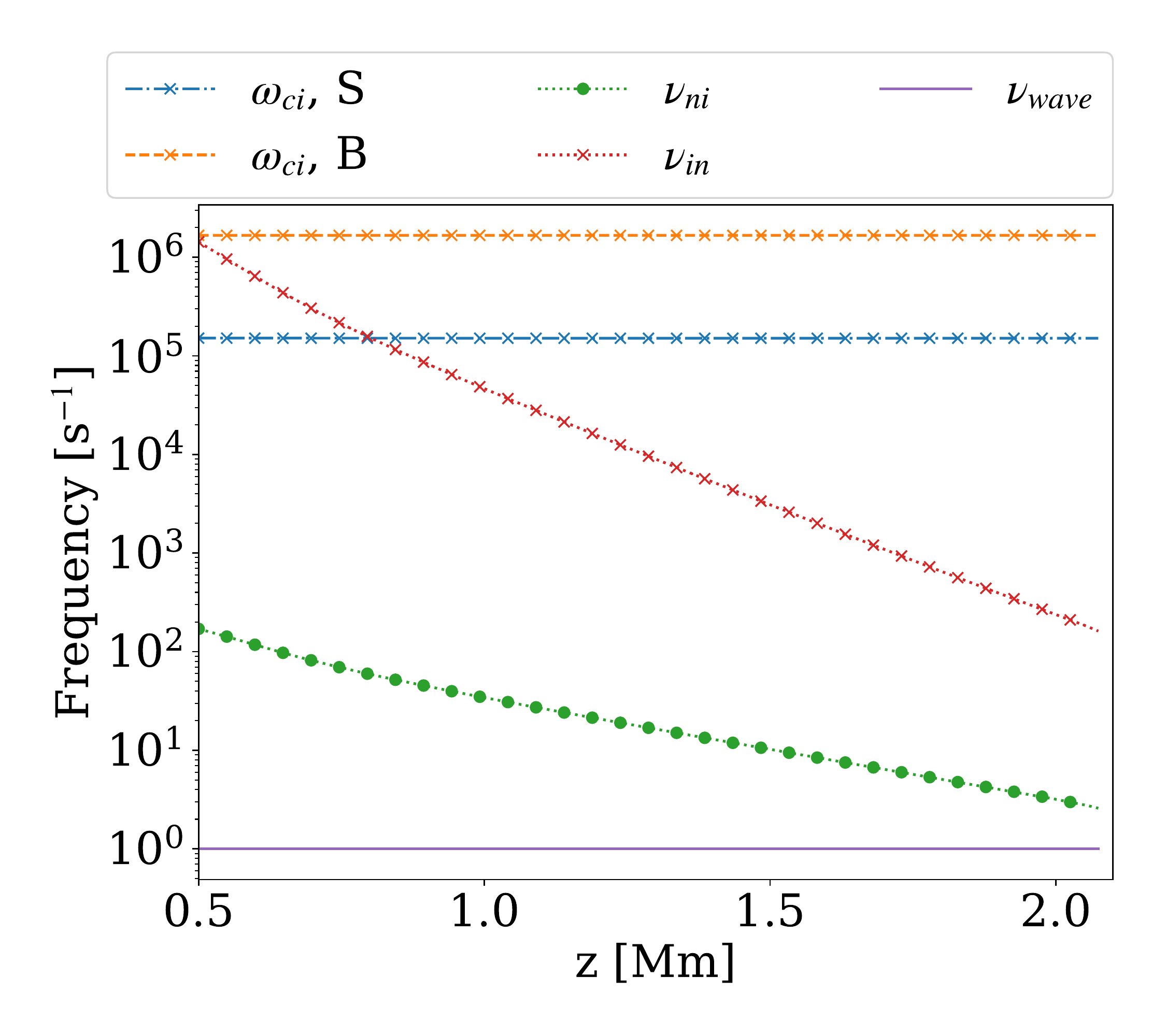}
\caption{Height dependence of characteristic frequencies calculated for background atmospheric model. The blue and orange lines are ion-cyclotron frequencies corresponding to the $S$ and $B$ magnetic field profiles, red and green lines are ion-neutral and neutral-ion collision frequencies, and violet is the highest wave frequency used is our work, corresponding to the wave period of 1 s.}
\label{fig:freq}
\end{figure}
%%%%%%%%%%%%%%%%%%%%%%%%%%%%%%%%%%%%%%%%%%%%%%%%%%%%%%%%%

Taking Eq.~\ref{eq:eos} into account, the equations of HS and MHS balance for neutrals and charges as follows,
\begin{equation} 
\frac{dp_{n0}}{dz} = -\frac{m_H g}{k_B T}p_{n0} ,
\end{equation}
\begin{equation}\label{eq:mhs} 
\frac{d(p_{c0} + p_{m0})}{dz} = -\frac{m_H g}{2 k_BT} p_{c0}, 
\end{equation}
where $p_{m0}$ is the magnetic pressure, which is the only contribution from the Lorentz force produced by the horizontal magnetic field.
The HS equation for neutrals is readily integrated, leading to:
\begin{equation}
p_{n0}(z)=p_{n0}(z_0) {\exp}\left( -\frac{m_H g }{k_B} \int_{z_0}^{z}{\frac{1}{T(z^\prime)} d z^\prime} \right).
\end{equation}

\noindent In order to be able to integrate the MHS equation for charges, we assumed that the charges pressure and the magnetic pressure have the same exponential dependence on height, 
\begin{equation}\label{eq:strati1}
p_{c0}(z)=p_{c0}(z_0){\exp}\left( - 2 F(z) \right) 
\end{equation}
\begin{equation} \label{eq:pmagEqui} 
p_{m0}(z) = \frac{B_{x0}(z)^2}{2 \mu_0}= \frac{B_{x0}(z_0)^2}{2 \mu_0}{\exp}\left( - 2 F(z) \right) + C
\end{equation}
where $F$ should satisfy  $F(z) \ge 0$ and $F(z_0)$ = 0, and $C$ is a constant. In introducing these expressions into Eq. \ref{eq:mhs}, the following expression for $F(z)$ was obtained,
\begin{equation}
F(z) =\frac{m_H g}{4 k_B}\frac{p_{c0}(z_0)}{p_{c0}(z_0) + B_{x0}(z_0)^2/2 \mu_0} \int_{z_0}^{z}{\frac{1}{T(z^\prime)} dz^\prime}.
\end{equation}
By setting the value of $B_{x0}(z_0) = 10^{-4}$ T, a pressure profile of the charged component similar to that of the VALC model was obtained. Given the temperature and pressure profiles with height, the densities were calculated afterward from the ideal gas laws for neutrals and charges, Eqs. \ref{eq:eos}, taking $\rho_{n0}=n_n m_H$, $\rho_{c0}=n_e m_H$.

The integration constant $C$ in Eq. \ref{eq:pmagEqui} allowed us flexibility in selecting the strength of the resulting magnetic field. We set,
\begin{equation}
C=p_{n0}\bigg\rvert_{z=z_0 + j L_z} - \frac{B_{x0}(z_0)^2}{2 \mu_0}{\exp}\left( - 2 F\bigg\rvert_{z=z_0 +j L_z} \right),
\end{equation} 
and $B_{x0}(z)$ was then recovered from Eq. \ref{eq:pmagEqui}. In this equation, $j \in [0,1]$ is a fraction of the total vertical domain length, $L_z=1.6$ Mm. By selecting $j$ inside the domain, we made sure that neutral pressure and magnetic pressure were equal to each other at $z=jL_z$, thus creating two zones in the atmosphere with either neutral gas pressure or magnetic pressure that is more prevalent. In the following, we used two magnetic field profiles, $S$-profile with $j=1/2$ and $B$-profile with $j=0$. 
In the case of the $S$ profile, the average field strength is smaller, about 1.5$\times 10^{-3}$ T, while for the $B$-profile, it reaches 1.74$\times 10^{-2}$ T. 
 We note that the integration constant dominates over the space varying term, having the value of 0.988 Pa for the $S$ profile and  120.73 Pa for the 
$B$ profile, thus making the magnetic field profiles almost flat. The difference between the maximum value of the magnetic field,
 obtained at the base of the atmosphere and the minimum value of the magnetic field obtained at the top of the atmosphere, 
is maximum for the minimum magnetic field profile that fulfills MHS. This happens when the integration constant is 
chosen  as 
\begin{equation}  \label{eq:nullC}
-\frac{B_{x0}(z_0)^2}{2 \mu_0}{\exp}\left( - 2 F\bigg\rvert_{z=L_z} \right).
\end{equation} 
In this case,  the magnetic field is zero at the top of the atmosphere and has a value of less than 1G at the base of the atmosphere.

The resulting stratification of pressures, number densities, temperature, and the magnetic field are displayed in Figure \ref{fig:iniprofaw3}.  It can be observed in panel $a$ that the neutral and magnetic pressures become equal in the middle of the domain for the $S$ profile, and at the bottom of the domain for the $B$ profile,  while the thermal pressure of charges always remains below the magnetic pressure. The number densities (panel $b$) are different as compared to those of VALC due to the above mentioned reasons. However, given all the simplifications of our model, these differences are considered to be acceptable. 
%Both magnetic field profiles, $S$ and $B$ are nearly flat. 

As is shown later, in order to understand the results of the simulations presented below, it is essential to know the characteristic frequencies associated with the background model atmosphere, such as the ion-cyclotron frequencies and the collision frequencies. The collision frequencies are calculated from Eq. \ref{eq:nu}, taking into account that $\nu_{ni} = \nu_{in}n_e/n_n$. These frequencies are given in Figure \ref{fig:freq} as a function of height. It is observed that for the $S$ magnetic field profile, the ion-neutral collision frequency and the cyclotron frequency become equal in the lower part of the domain at $z\approx 0.8$ Mm, while for the $B$ magnetic field profile, the ion-cyclotron frequency is higher than the collision frequency throughout the domain. The highest wave frequency considered in our models corresponds to the period of 1 s (see below). This frequency is lower than any of the characteristic frequencies at all heights. 

%%%%%%%%%%%%%%%%%%%%%%%%%%%%%%%%%%%%%%%%%%%%%%%%%%%%%%%%%
\begin{figure*}[t]
\centering
\includegraphics[width=8.5cm]{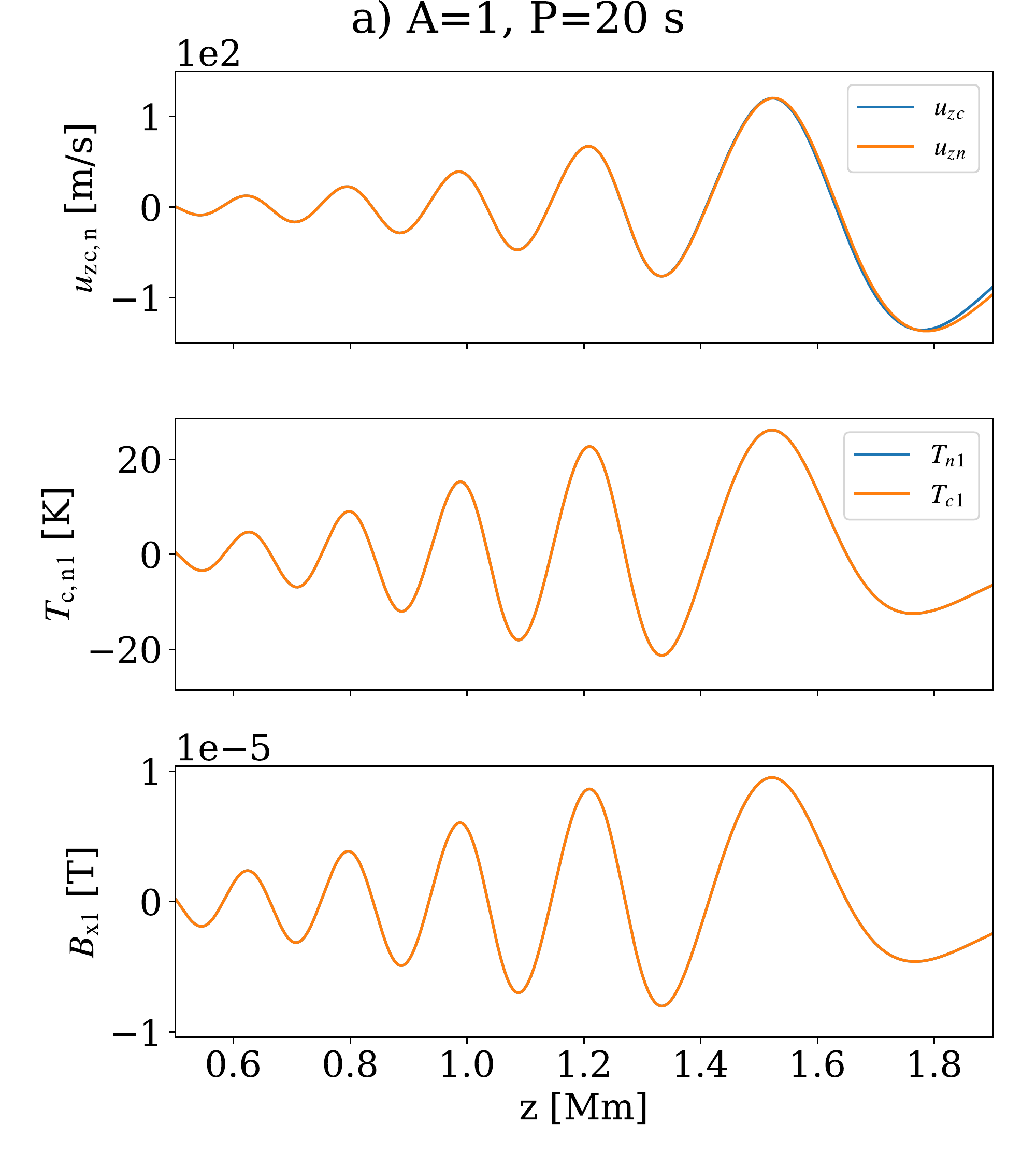}
\includegraphics[width=8.5cm]{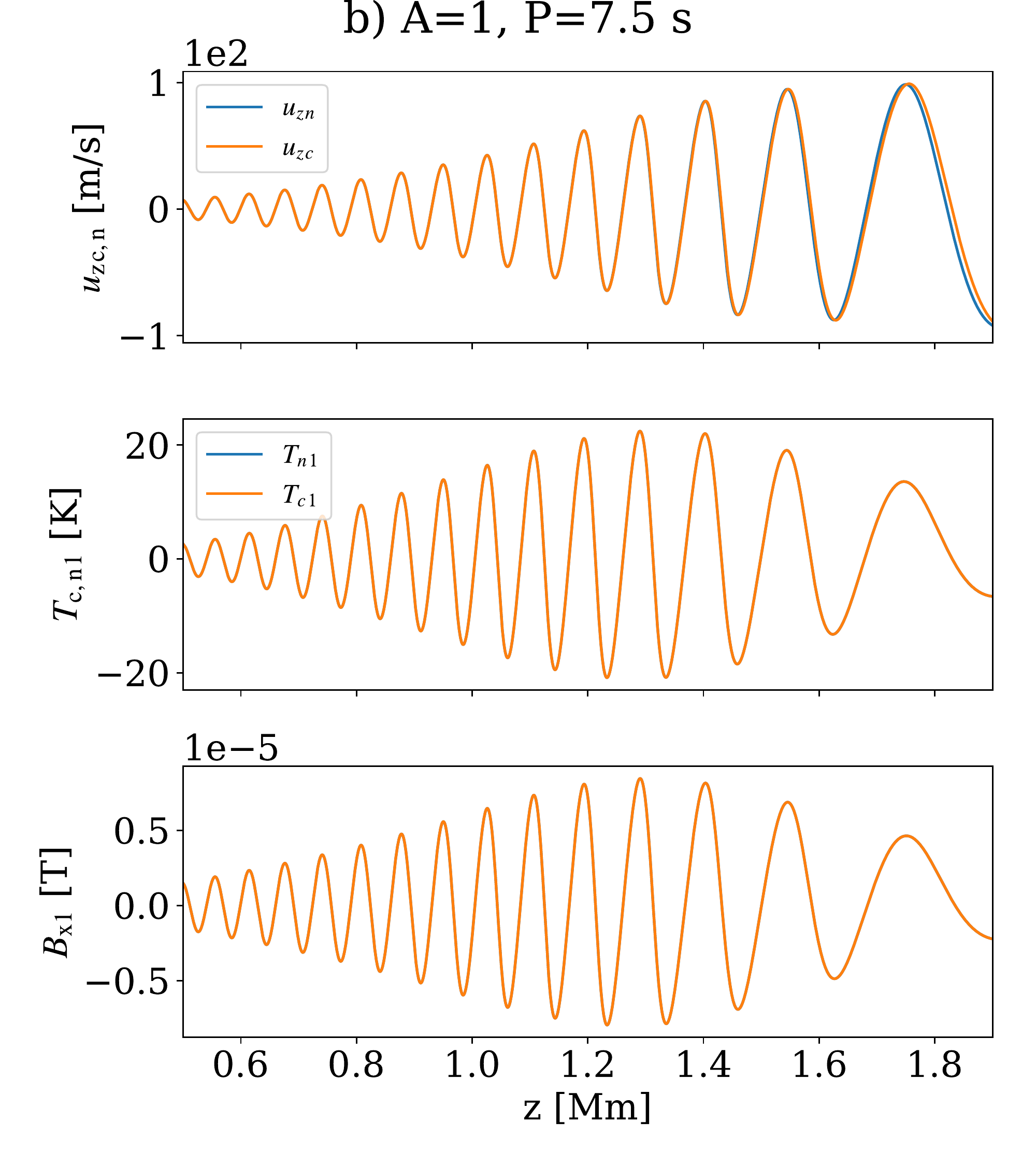}

\includegraphics[width=8.5cm]{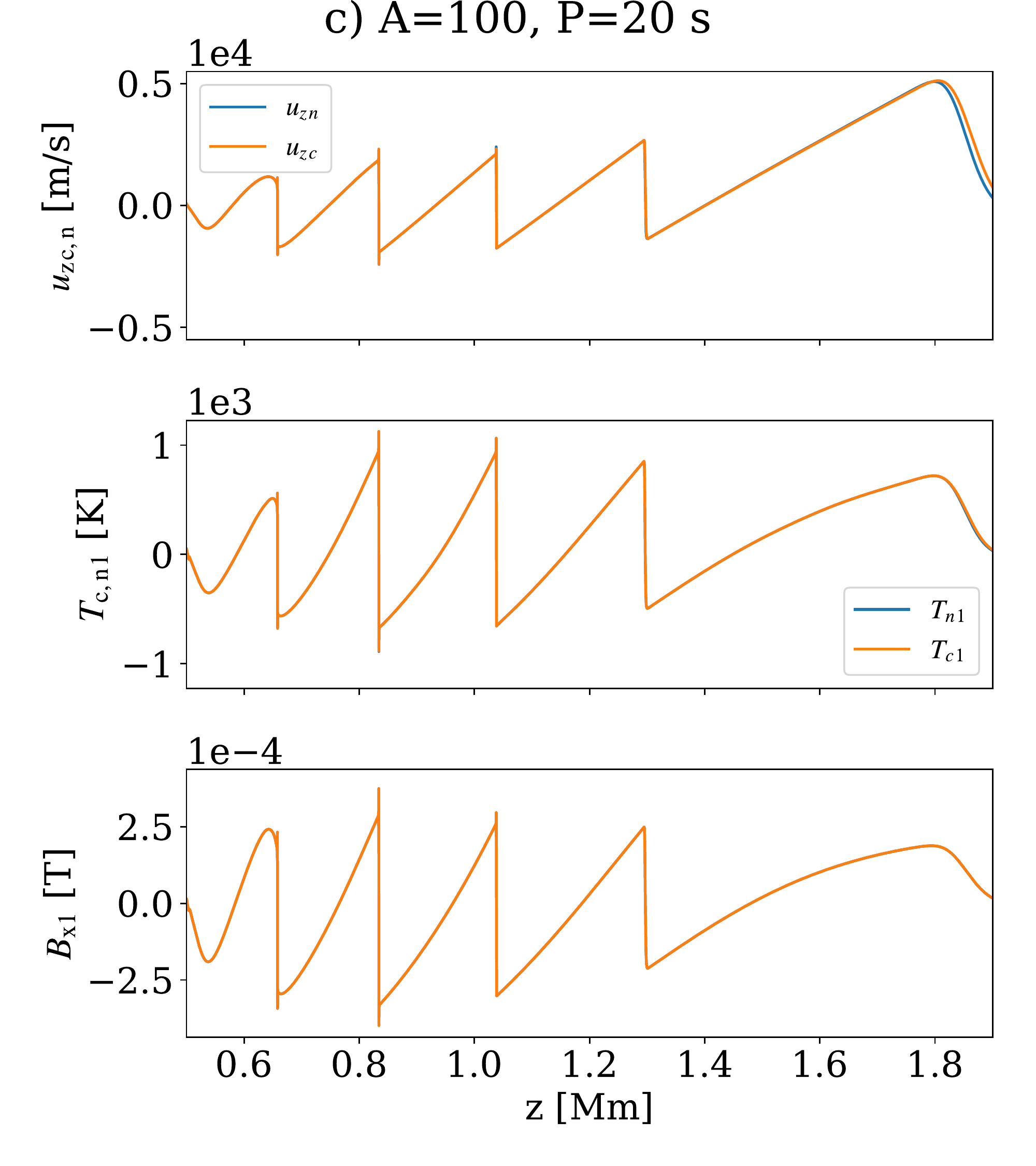}
\includegraphics[width=8.5cm]{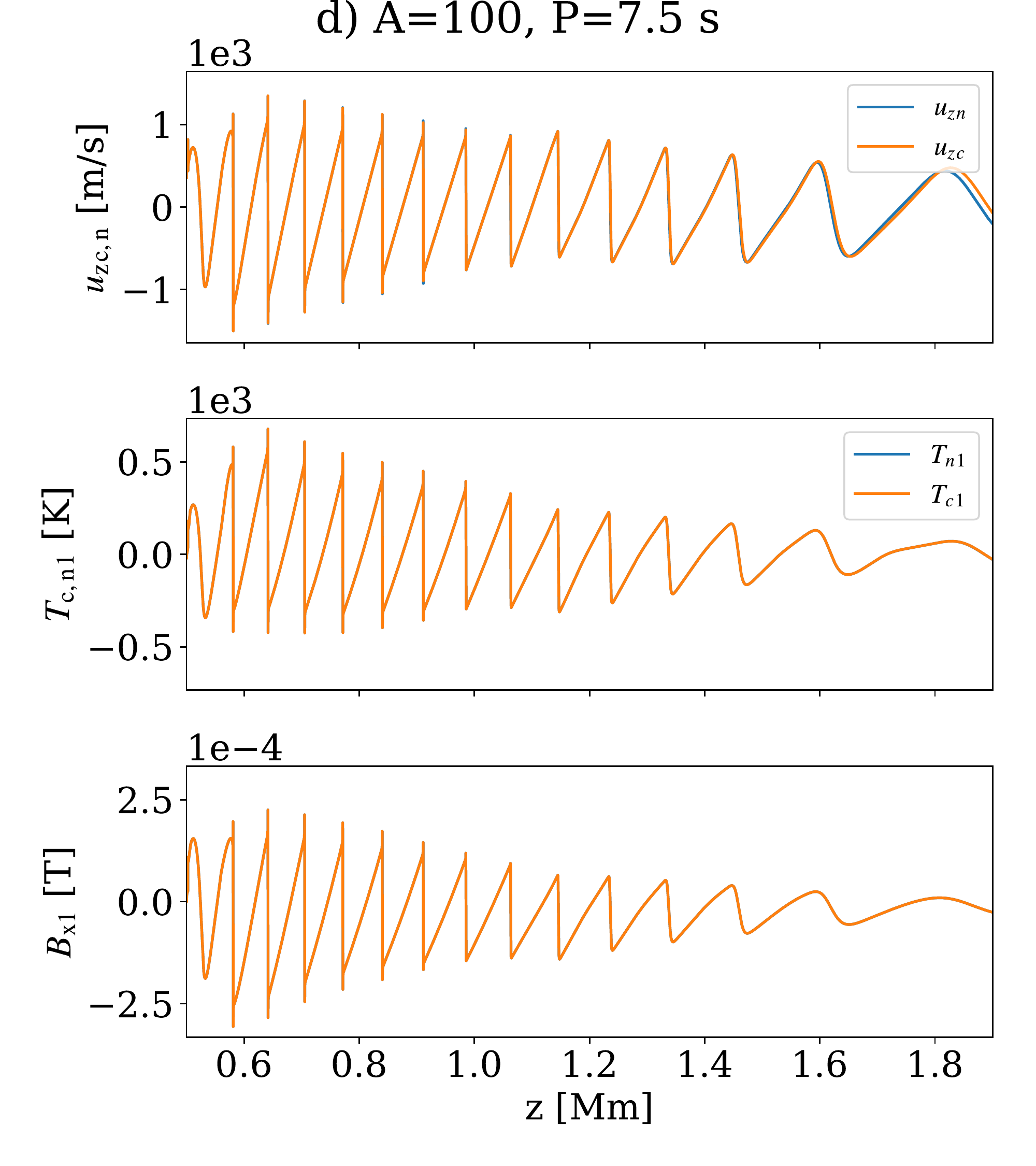}
\caption{Height dependence of oscillations in velocity and temperature of charges and neutrals, and in magnetic field for fixed moment of time, obtained from numerical solution of two fluid equations for magnetic field profile $S$.  Panel (a) and (b) are for the initial wave amplitude factor $A=1$; panels (c) and (d) are for $A=100$. The wave period is $P=20$ s for panels (a) and (c), and $P=7.5$ s for panels (b) and (d). Orange lines are for the parameters corresponding to charges, and blue lines are those for neutrals.  We note that shock overshooting observed in panels (c) and (d) are numerical artifacts since we ran the simulation without any shock capturing algorithms in this case. }
\label{fig:aw3per20Dec}
\end{figure*}
%%%%%%%%%%%%%%%%%%%%%%%%%%%%%%%%%%%%%%%%%%%%%%%%%%%%%%%%%

%%%%%%%%%%%%%%%%%%%%%%%%%%%%%%%%%%%%%%%%%%%%%%%%%%%%%%%%%
\begin{figure*}[t]
\centering
\includegraphics[width=8.5cm]{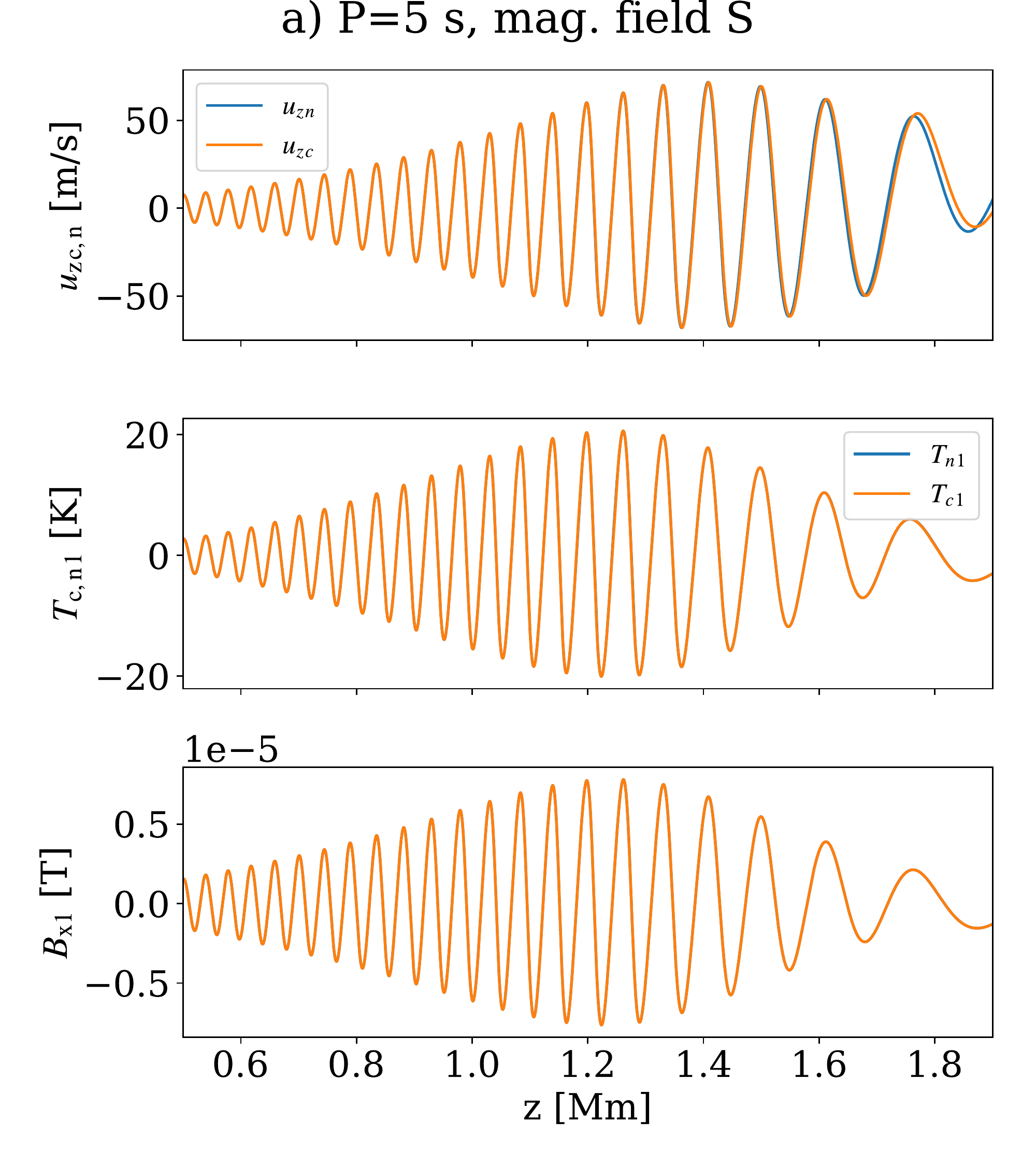}
\includegraphics[width=8.5cm]{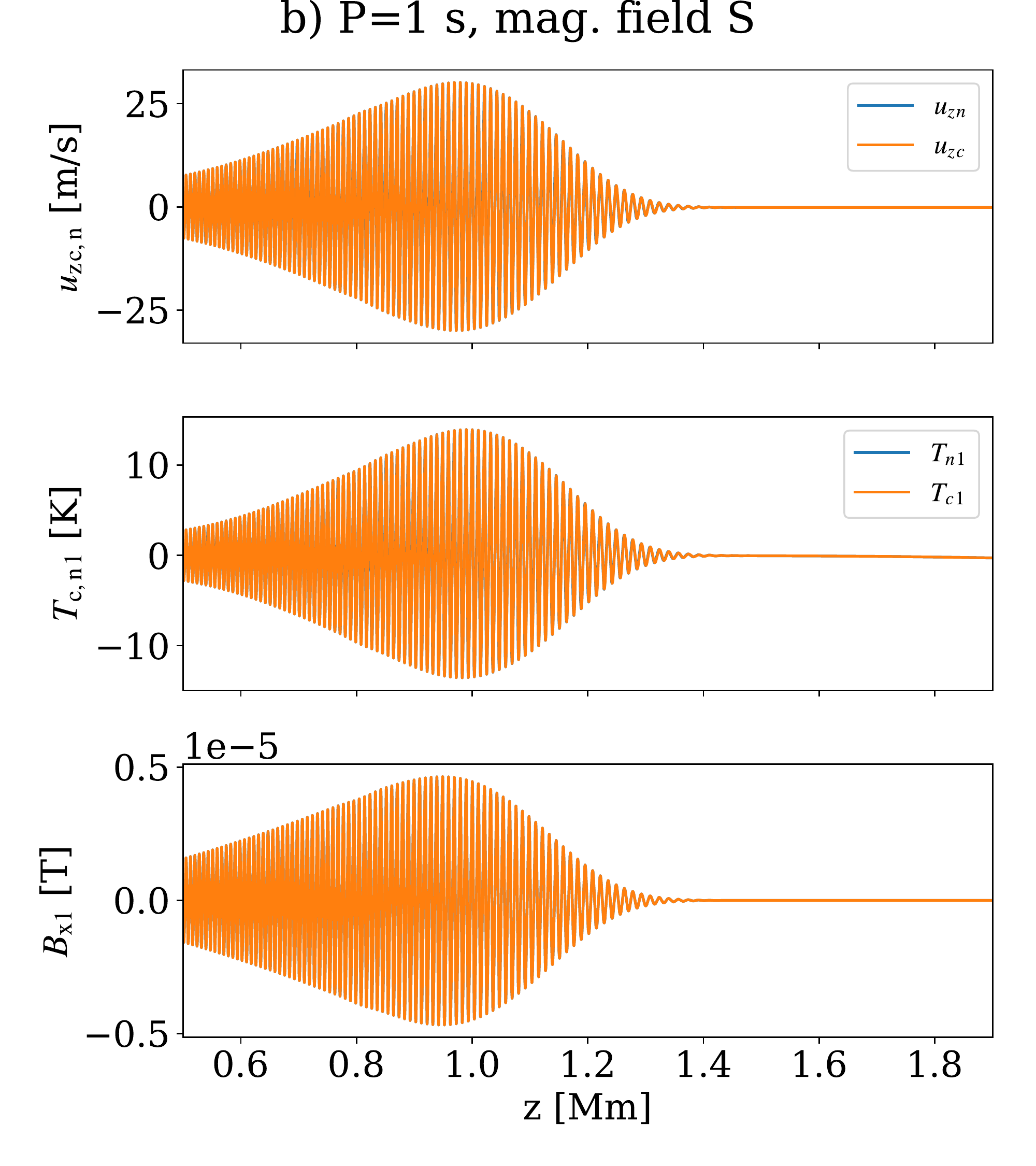}
 
\includegraphics[width=8.5cm]{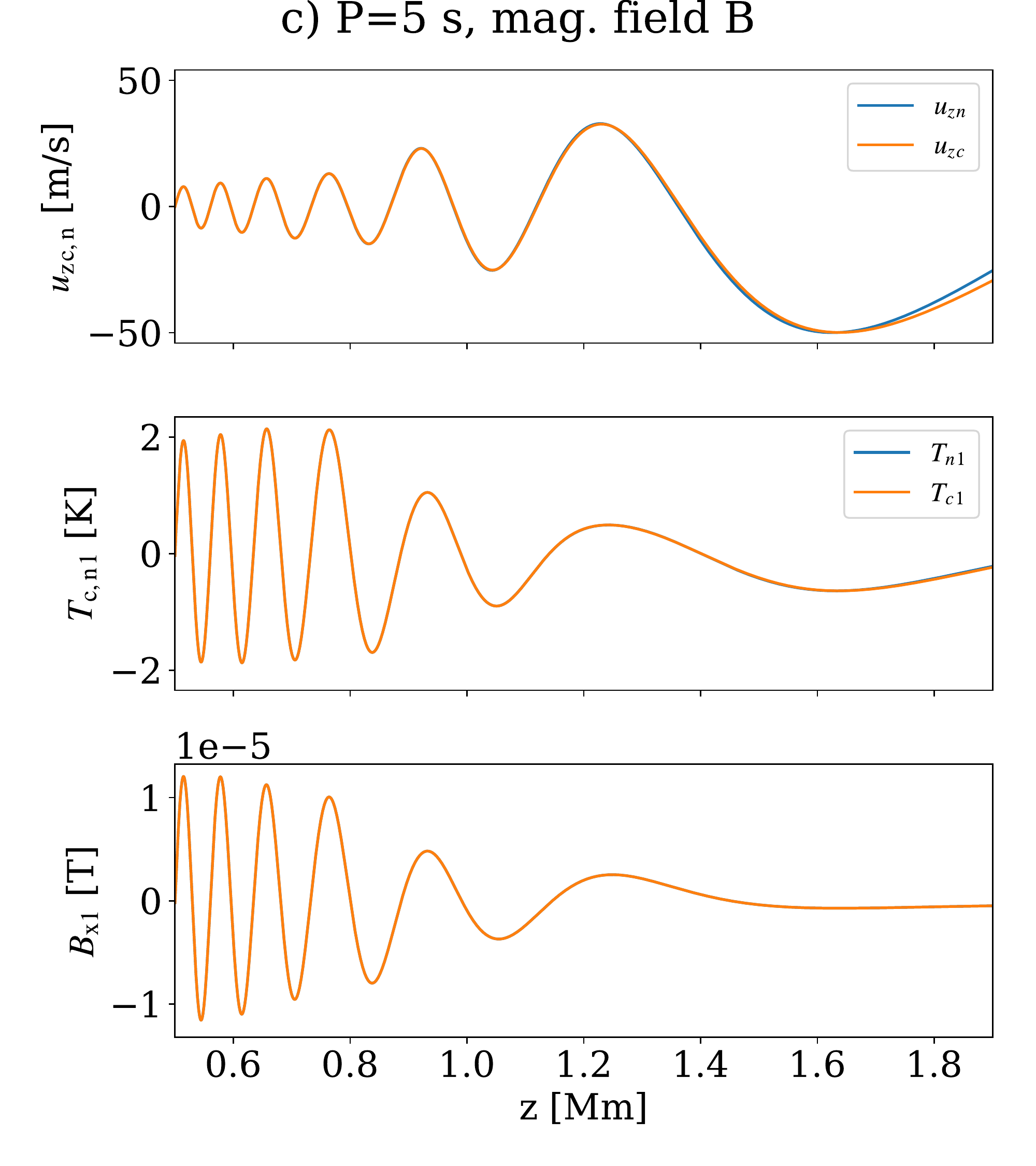}
\includegraphics[width=8.5cm]{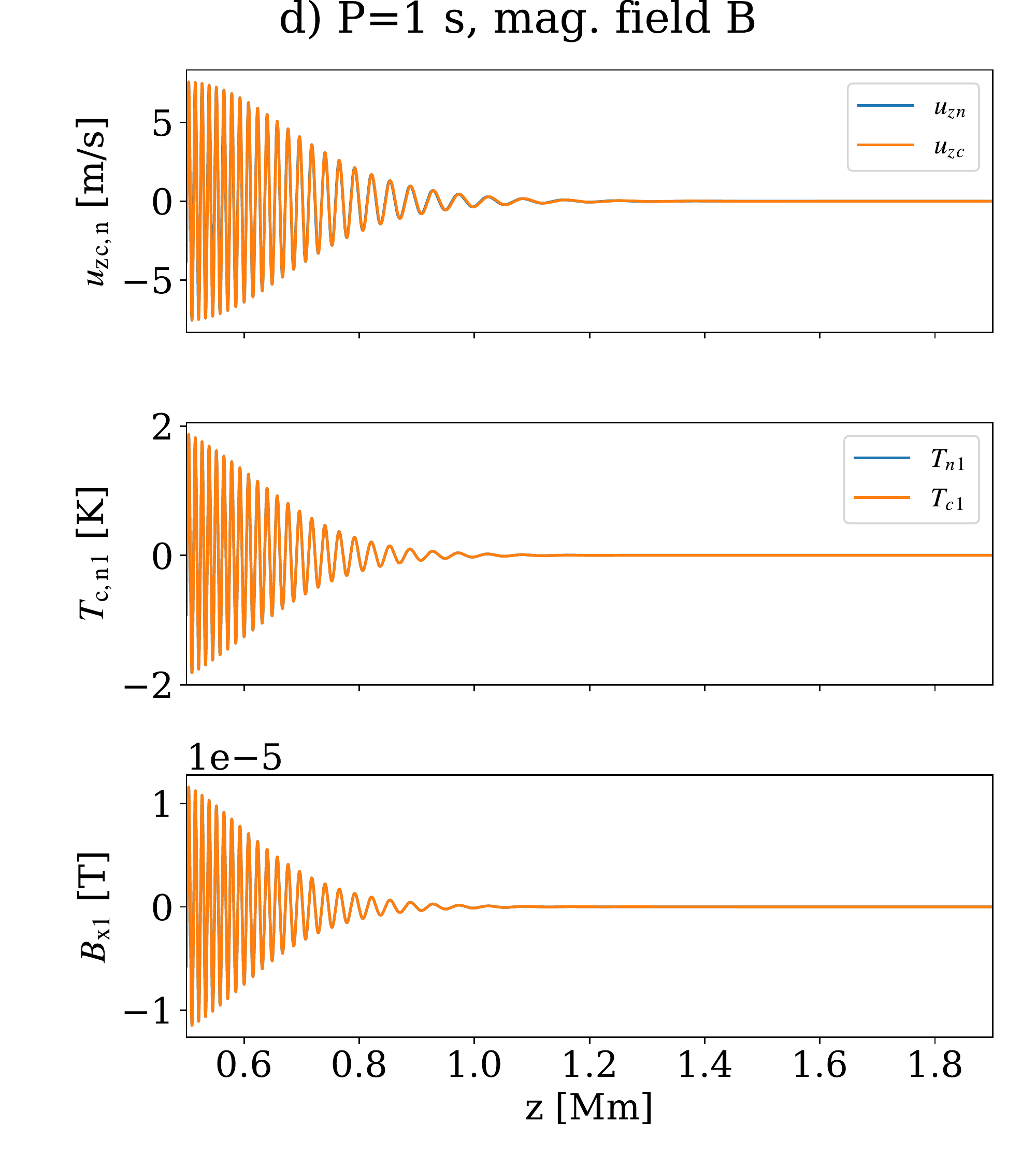}
\caption{Height dependence of oscillations in velocity and temperature of charges and neutrals, and in magnetic field for fixed moment in time, for initial wave amplitude factor $A=1$. The format of the figure is the same as for Fig. \ref{fig:aw3per20Dec}. Panels (a) and (b) are for the $S$ magnetic field profile; panels (c) and (d) are for the $B$ profile. The wave period is $P=5$ s for panels (a) and (c), and $P=1$ s for panels (b) and (d). }\label{fig:aw3per5Dec}
\end{figure*}
%%%%%%%%%%%%%%%%%%%%%%%%%%%%%%%%%%%%%%%%%%%%%%%%%%%%%%%%%

%%%%%%%%%%%%%%%%%%%%%%%%%%%%%%%%%%%%%%%%%%%%%%%%%%%%%%%%%
\subsection{Perturbation}
%%%%%%%%%%%%%%%%%%%%%%%%%%%%%%%%%%%%%%%%%%%%%%%%%%%%%%%%%

The perturbation is generated by a driver at the base of the atmosphere. It consists of an oscillatory perturbation in the vertical fluid velocity and the corresponding perturbations in thermodynamic variables and magnetic field that are obtained from an analytical solution. Since the driver is located in the deep and dense layers, we assumed strong collisional coupling.  The values we chose for the wave frequencies are much smaller  than the collisional frequencies (see Figure \ref{fig:freq}). Therefore, in order to generate the initial solution, we used single-fluid equations together with the generalized induction equation and kept only the leading effect due to ion-neutral interaction in the solar atmosphere, that is, the ambipolar diffusion,
\begin{eqnarray}
\frac{\partial \rho}{\partial t} + \mathbf{\nabla}\cdot \left(\rho\mathbf{u}\right) =  0, \nonumber  \\ 
\frac{\partial (\rho\mathbf{u})}{\partial t} + \mathbf{\nabla}\cdot (\rho\mathbf{u} \mathbf{u} +p)  = \mathbf{J}\times\mathbf{B} + \rho\mathbf{g},  \nonumber \\
\frac{\partial }{\partial t}\left(e + \frac{1}{2}\rho u^2 \right) + \mathbf{\nabla}\cdot  \left( \mathbf{u}\, ( e + \frac{1}{2}\rho u^2) +p\mathbf{u}  \right)   = \nonumber \\
\mathbf{J} \cdot \mathbf{E}   + \rho\mathbf{u} \cdot \mathbf{g},  \nonumber \\
\frac{\partial\mathbf{B}}{\partial t}  -  \mathbf{\nabla}\times (\mathbf{u}\times\mathbf{B}) -  \mathbf{\nabla}\times \left[\eta_A\frac{[(\mathbf{J} \times \mathbf{B}) \times \mathbf{B}]}{|B|^2} \right] = 0,
\end{eqnarray}
\noindent where the ambipolar diffusivity coefficient is defined as
\begin{equation} \label{eq:etaa-total}
\eta_A=\frac{\xi_n^2 |B|^2}{\alpha \rho_n \rho_c},
\end{equation}
being $\xi_n=\rho_n/\rho$ the neutral fraction, and the collisional parameter $\alpha$ as defined in Equation (\ref{eq:alpha}).

These equations are linearized when taking into account that the magnetic field is purely horizontal and its variation with height is very small compared to the rest of background variables, so that the gradients in $B_{x0}$ can be neglected. It is also important to take into account that both the stratification and the perturbations are only functions of one vertical coordinate, $z$, and that another independent horizontal direction is $x$,
\clearpage 
\begin{eqnarray} \label{eq:ambi1fl}
\frac{\partial \rho_1}{\partial t} &= &- u_z \frac{d \rho_0}{d z} - \rho_0  \frac{\partial u_z}{\partial z},  \nonumber  \\
\rho_0 \frac{\partial u_z}{\partial t} &=& -\rho_1 g - \frac{\partial p_1}{\partial z} - \frac{1}{\mu_0}\left ( \frac{\partial B_{x1}}{\partial z} B_{x0} \right ), \nonumber \\
\frac{\partial p_1}{\partial t} &=& c_s^2 \frac{\partial \rho_1}{\partial t} + c_s^2 u_z \frac{d \rho_0}{d z} - u_z \frac{d p_0}{d z} ,\nonumber \\
\frac{\partial B_{x1}}{\partial t} &=& - B_{x0} \frac{\partial u_z}{\partial z}  +\frac{\eta_A}{\mu_0}\frac{\partial^2 B_{x1}}{\partial z^2} + \frac{1}{\mu_0}\frac{d \eta_A}{d z}\frac{\partial B_{x1}}{\partial z} ,
\end{eqnarray}
where $c_s=\sqrt{\gamma p_0/\rho_0}$ is the sound speed.

An analytical solution for these equations is still complicated and several simplifications need to be made. Since we needed to generate the analytical solution in only a few bottom grid points (the ghost points), we assumed that the temperature is locally uniform, that is, $c_s$ is also locally uniform, and then the background pressure and density have the same vertical scale height $H$, as follows:
\begin{equation}
\{p_0,\rho_0\}=\{p_0(z_0),\rho_0(z_0)\} \times \exp{\left(-z/H\right)}.
\end{equation}

\noindent We further assumed that the ambipolar diffusion is negligible at the bottom layer. After these simplifications, Eq. \ref{eq:ambi1fl} can be combined into a single wave equation, similar to Eq. 10 provided by \cite{1976Nye} for the particular case of $k_x$ = $k_y$ = 0,
\begin{equation} \label{eq:waveeq}
 \frac{\partial^2 u_z}{\partial t^2} = \frac{\partial^2 u_z}{\partial z^2}\left( c_s^2 + {v_A}^2 \right) -  \frac{c_s^2}{H}  \frac{\partial u_z}{\partial z} . 
\end{equation}
When calculating the solution in the few grid points at the bottom boundary, we assumed that the quantity $v_A \approx B_{x0}/\sqrt{\mu_0\rho_0}$ is uniform. Then, the dispersion relation is obtained as usual, assuming the solution of the form, $u_z=V\exp\left(i (\omega t - k z)\right)$,
\begin{equation}\label{eq:disp-simple}
\omega^2=k^2(c_s^2 + {v_A}^2) - \frac{ik}{H}c_s^2.
\end{equation}
The polarization relations for the amplitudes of the rest of the variables can be derived from Eq. \ref{eq:ambi1fl} by substituting,
\begin{equation}
\left\{ \frac{\rho_1}{\rho_0}, \frac{p_1}{p_0},\frac{B_{x1}}{B_{x0}} \right\} =\{ \tilde R, \tilde P, \tilde B \}\exp{\left( i (\omega t - k z)  \right)  },
\end{equation}
where $\{ \tilde R, \tilde P, \tilde B \}$ are generally complex amplitudes. We obtained the following relations,
\begin{eqnarray}        \label{eq:relaw3}
\tilde R &=& \frac{V }{i \omega}\left( ik + \frac{1}{H} \right),  \nonumber \\
\tilde P &=& \frac{\gamma V }{i\omega} \left( i k + \frac{1}{\gamma H} \right), \nonumber  \\
\tilde B &=& \frac{Vk}{\omega} .        
\end{eqnarray}

The wave period $P=2\pi/\omega$ and the velocity amplitude $V$ at the base of the atmosphere are free parameters of the simulations. Given $\omega$, we obtained $k$ from the dispersion relation, Eq. \ref{eq:disp-simple}. We  chose the velocity amplitude at the base of the atmosphere as a fraction of the background sound speed:  $V(z_0) = A\cdot10^{-3}\cdot c_s(z_0)$, where the factor $A$ was varied from 1 to 100. Also, the wave periods $P=1, 5, 7.5$ and 20 s were used.

The initial condition for oscillations of neutrals and charges are set in the following way,
\begin{eqnarray} \label{eq:relaw32}
 V_n = V&;& \,\,\,  V_c = V  \nonumber\\
 \tilde R_n =\frac{{\rho_n}_0}{\rho_0}  \tilde R&;& \,\,\,   \tilde R_c =\frac{{\rho_c}_0}{\rho_0}  \tilde R; \nonumber  \\
   \tilde P_n =\frac{{\rho_n}_0}{\rho_0}  \tilde P&;& \,\,\,   \tilde P_c =\frac{{\rho_c}_0}{\rho_0}  \tilde P. 
\end{eqnarray}
For our 1D numerical calculation, we covered the domain of $L_z=1.6$ Mm with 32000 grid points. The analytical wave solution was generated  at each time step in the ghost points  as the lower boundary condition. The perfectly matched layer \citep[PML; see][]{Berenger1994, Felipe2010} was used as the upper boundary condition.

%%%%%%%%%%%%%%%%%%%%%%%%%%%%%%%%%%%%%%%%%%%%%%%%%%%%%%%%%
\begin{figure*}[ht]
\centering
\includegraphics[width=8.5cm]{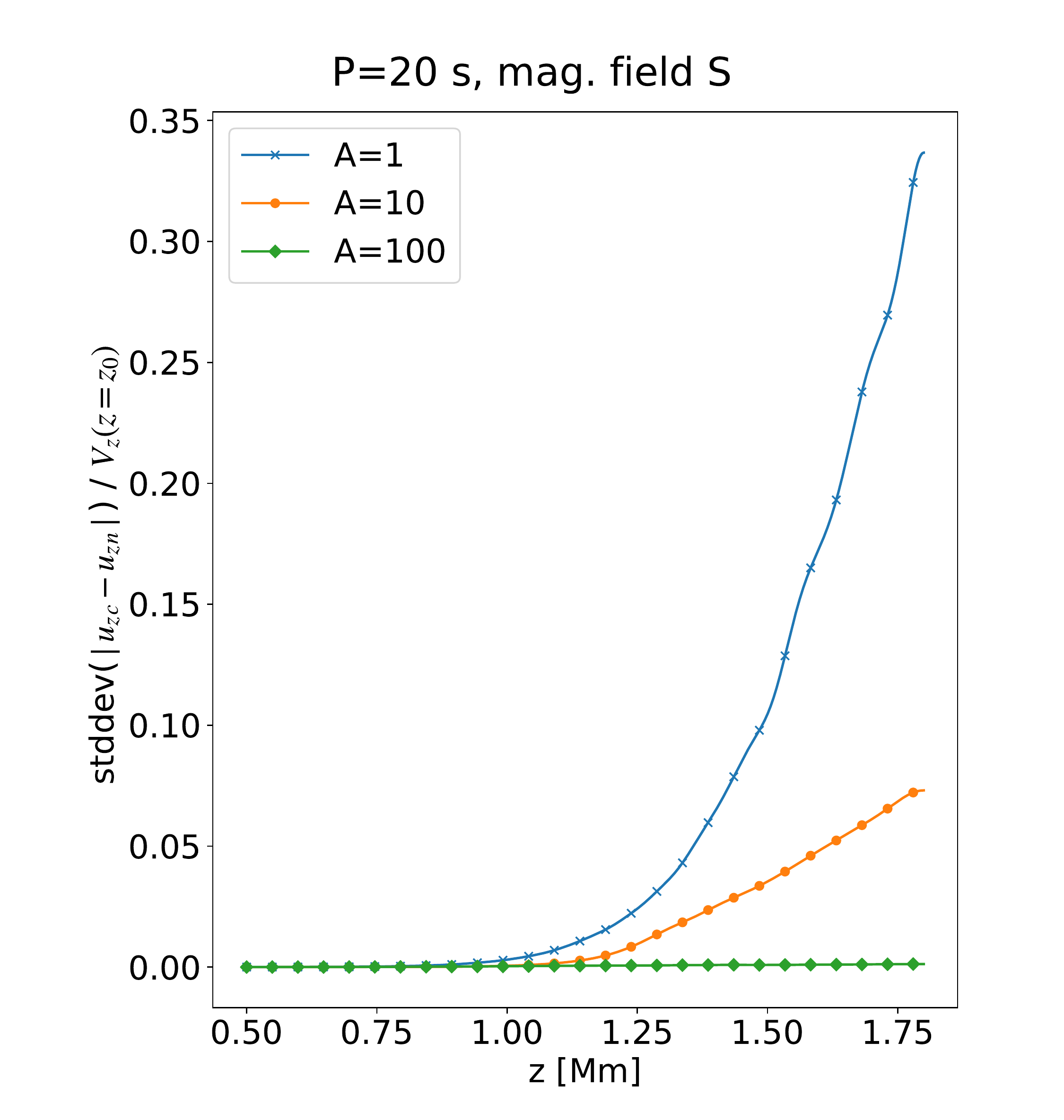}
\includegraphics[width=8.5cm]{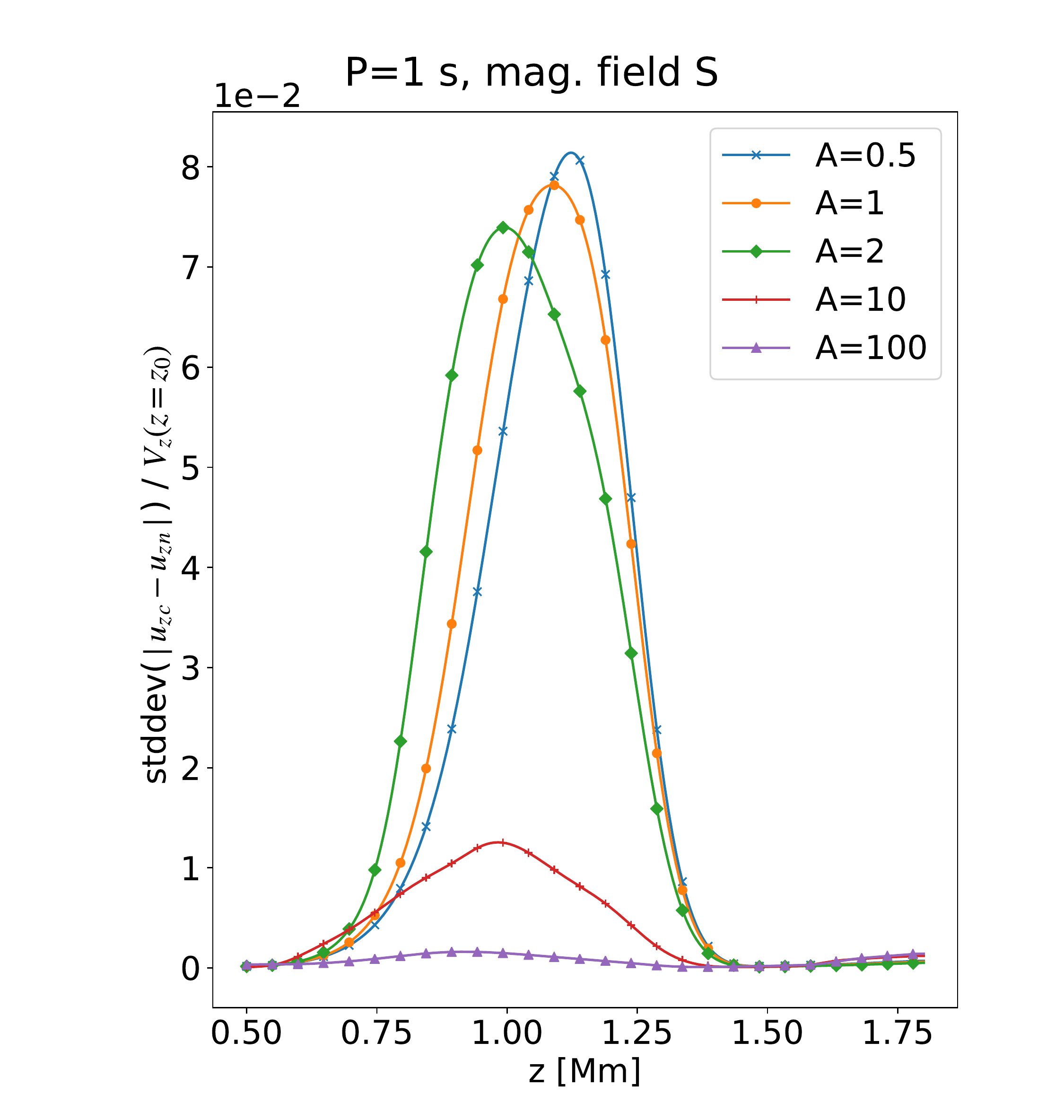}
\caption{Decoupling between charges and neutral velocity as function of height for magnetic field profile $S$. Left panel: wave period 20 s, amplitude factors $A=1$ (blue), $A=10$ (orange), $A=100$ (green). Right panel: wave period 1 s, amplitude factors $A=0.5$ (blue), $A=1$ (orange), $A=2$ (green).}
\label{fig:awDecQ1}
\end{figure*}
%%%%%%%%%%%%%%%%%%%%%%%%%%%%%%%%%%%%%%%%%%%%%%%%%%%%%%%%%

%%%%%%%%%%%%%%%%%%%%%%%%%%%%%%%%%%%%%%%%%%%%%%%%%%%%%%%%%
\section{Nonlinear wave propagation in stratified plasma}
%%%%%%%%%%%%%%%%%%%%%%%%%%%%%%%%%%%%%%%%%%%%%%%%%%%%%%%%%

The results of the numerical solution are given in Figures \ref{fig:aw3per20Dec} and \ref{fig:aw3per5Dec}. 
Throughout the simulations, we varied the following three parameters: the background magnetic field profile, the period, and the amplitude of the wave.
The change in the background magnetic field and in the period of the wave can be studied in the linear approximation,
while the change in the amplitude is related to  the  nonlinear effects.
These figures show variations with the height of the charges, neutral velocities, their corresponding temperatures, and the magnetic field for a fixed moment of time. Figure  \ref{fig:aw3per20Dec} gives the results for two different initial wave amplitudes (factor $A=1$ and 100) and two different periods ($P=20$ and 7.5 s) for the background model with $S$ magnetic field profile. Figure  \ref{fig:aw3per5Dec} shows oscillations for two magnetic field profiles ($B$ and $S$) and two wave periods ($P=1$ and 5 s) for the same wave amplitude factor $A=1$.  

The first impression from Fig. \ref{fig:aw3per20Dec} is that the oscillation curves for neutrals and charges are nearly the same. The temperatures of both species are completely collisionally coupled via the thermal exchange process and, therefore, oscillations in both temperatures are the same. Nevertheless, one can observe some decoupling in the charges and neutral velocities at the upper part of the atmosphere, at heights above approximately 1.6 Mm. At these heights the collision frequency decreases because of the very low background densities. This strong drop of densities compensates for the increase in the background temperature and, consequently, in the value of the collisional parameter $\alpha$. This decoupling is present, independently of the wave amplitude factor, and for both wave periods shown in the figure. It can be observed that velocity oscillations in neutrals slightly lag behind the oscillations in charges. The decoupling is visible not only at the shock wave fronts, but over the entire oscillation curve. 

It is important to note that the vertical wavelength of the fast waves considered in this numerical experiment are dependent on the Alfv\'en speed. The increase of the Alfv\'en speed with height produces oscillations with significantly larger wavelength in the upper part of the atmosphere. This longer wavelength affects the formation of nonlinearities (panels $c$ and $d$). Oscillations are strongly nonlinear in nature already at the bottom part of the domain for the case $A=100$. Nevertheless, the shock fronts become smoother at bigger heights. 

In general, velocity oscillations in Fig. \ref{fig:aw3per20Dec} increase their amplitude with height. However,  by comparing the panels ($a$) and ($c$) it can also be seen that the amplitude increase   due to larger  height is significantly lower, or even absent, in the case of strong perturbation. Perturbations in the temperature and magnetic field behave differently, and their amplitude starts to decrease roughly after the same height as where the charges-neutral velocity decoupling becomes visible. 

When the wave period for the $S$ magnetic field profile decreases further, a new effect becomes visible, namely oscillation damping (two upper panels of Figure \ref{fig:aw3per5Dec}). As a result of this damping, the velocity amplitude decreases after about 1.4 Mm for $P=5$ s wave, and above 1 Mm for $P=1$ s wave.  The damping is also apparently sensitive to the magnetic field profile. For the stronger field $B$ (two bottom panels), the $P=1$ s oscillation is damped from the very beginning, and the wave completely disappears after 0.8 Mm. For the $P=5$ s, oscillations in the temperature and magnetic field also become visibly damped, while the velocity amplitude increase is still present. The difference between $P=1$ and $P=5$ is seen in significantly larger wavelength of oscillations in the case of the latter. This means that they are less affected by collisions.\ Furthermore, the effect on the damping is discussed in section \ref{sect:comparison}.
%The reason of the damping can be traced back to Figure \ref{fig:freq} where the cyclotron and the collisional frequencies are compared. For the $S$ magnetic field, the collisional frequency is still larger than the cyclotron frequency at the lower part of the atmosphere, while for the $B$ magnetic field the cyclotron frequency dominates at all heights. 

In summary, the following two main effects on waves are apparent from our simulations: decoupling in charges-neutral velocity and wave damping. These two effects are not independent from each other. The charges-neutral velocity decoupling can produce wave damping. However, the damping can also be produced by more mechanisms, which is demonstrated later in the paper. As for the case considered here, wave damping is a consequence of the charges-neutral decoupling and nonlinear effects, to a lesser extent.

%%%%%%%%%%%%%%%%%%%%%%%%%%%%%%%%%%%%%%%%%%%%%%%%%%%%%%%%%
\begin{figure}[t]
\centering
\includegraphics[width=8.5cm]{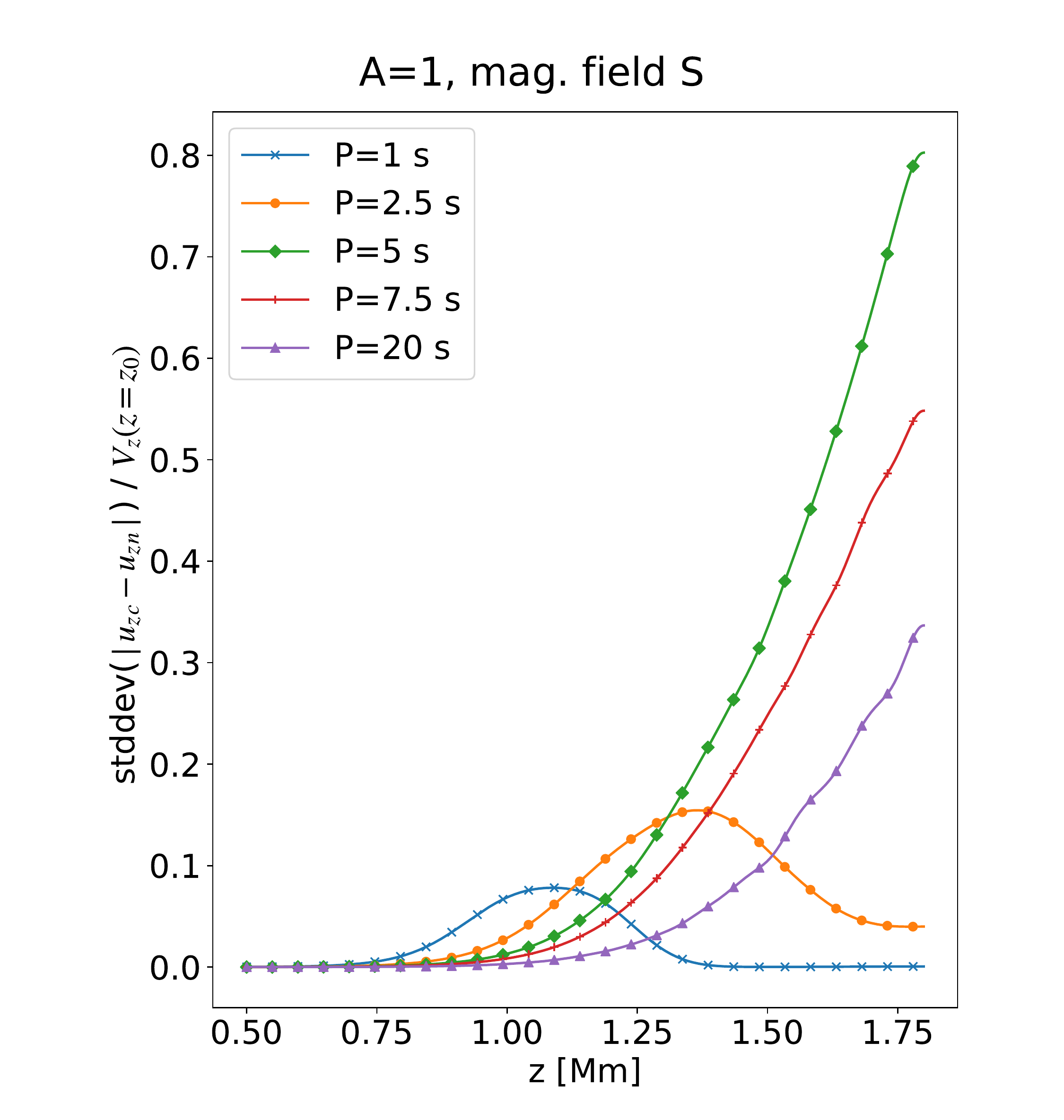}
\caption{Decoupling between charges and neutral velocity as function of height for magnetic field profile $S$ and wave amplitude factor $A=1$. Wave periods $P=20$ s (violet), $P=7.5$ s (red), $P=5$ s (green), $P=2.5$ s (orange), and $P=1$ s (blue).}
\label{fig:awDecQ2}
\end{figure}
%%%%%%%%%%%%%%%%%%%%%%%%%%%%%%%%%%%%%%%%%%%%%%%%%%%%%%%%%

%%%%%%%%%%%%%%%%%%%%%%%%%%%%%%%%%%%%%%%%%%%%%%%%%%%%%%%%%
\begin{figure}[t]
\centering
\includegraphics[width=8.5cm]{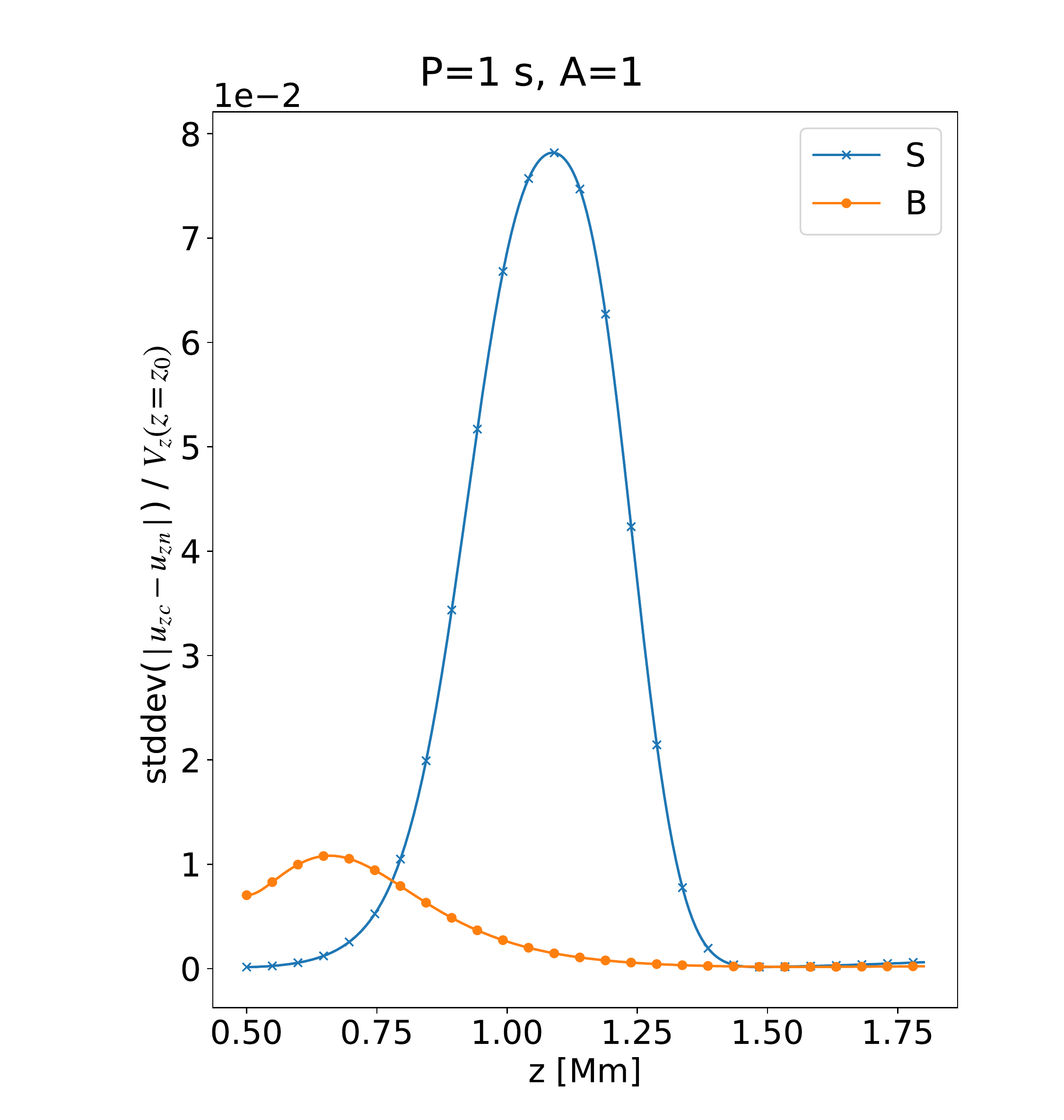}
\caption{Decoupling between charges and neutral velocity as function of height for  amplitude factor $A=1$ and  wave period $P=1$ s. The dependence on the magnetic field profiles is shown with different curves: $S$ (blue) and $B$ (orange). 
}
\label{fig:awDecQ22}
\end{figure}
%%%%%%%%%%%%%%%%%%%%%%%%%%%%%%%%%%%%%%%%%%%%%%%%%%%%%%%%%

%
%%%%%%%%%%%%%%%%%%%%%%%%%%%%%%%%%%%%%%%%%%%%%%%%%%%%%%%%%
\subsection{Decoupling}
%%%%%%%%%%%%%%%%%%%%%%%%%%%%%%%%%%%%%%%%%%%%%%%%%%%%%%%%%

The dependence of the degree of decoupling on the wave period and nonlinearities is difficult to evaluate just with the naked eye for Figs. \ref{fig:aw3per20Dec} and \ref{fig:aw3per5Dec}. In order to quantify the effects, we adopted a measure of decoupling by computing the standard deviation of $\mid u_{\rm cz} - u_{\rm nz} \mid$ in a stationary regime of the simulations (when the waves reached the upper boundary and several periods have passed through it), which is relative to the initial amplitude of the waves at the bottom level $z_0$. Such normalization gives us the possibility to better visualize the influence of the nonlinear effects. Figures \ref{fig:awDecQ1},  \ref{fig:awDecQ2}, and \ref{fig:awDecQ22} compare the amount of decoupling as a function of several parameters. In all of the cases discussed below, the amount of decoupling, normalized to $u_{\rm cz}(z_0),$ varies from a few percent of the initial wave amplitude at the lower layers to a significant fraction at the upper layers, which is due to the growth of the wave amplitude with height.

Figure \ref{fig:awDecQ1} reveals that the decoupling is a sensitive function of the wave amplitude. For the wave period of 20 s, when oscillation velocities are not visibly damped, the decoupling smoothly grows with height for all the wave amplitudes considered. For progressively more nonlinear cases (increasing $A$) the decoupling starts at lower heights and increases up to about
1.2 Mm. Above this height, nonlinear effects play an important role. 
Comparing the $A=1$, $A=10,$ and $A=100$ cases, the decoupling measured relative to the initial wave amplitude is smaller in the latter cases.

 The action of nonlinear effects increases the relative difference between the velocity of charges and neutrals due to the formation of discontinuities associated with shocks.  Nevertheless, a further increase of $A$ reverses this tendency. The shock waves in the $A=100$ case become strongly collisionally damped. Because of that, the amplitude of both $u_{\rm cz}$ and $u_{\rm nz}$, and of their difference, decreases compared to the $A=10$ case. The values of the decoupling measured with respect to the initial wave amplitude become lower at heights where strong shock damping takes place. However, the absolute value of the decoupling, without normalization, are still larger in the strongly nonlinear case $A=100$.

The right panel of Figure \ref{fig:awDecQ1}  shows the case of $P=1$ s where strong wave damping was observed. In this case, the decoupling has a broad maximum between 0.8 and 1.2 Mm, and is zero approximately above 1.4 Mm since no waves exist at higher heights. Similar to the previous case, the amount of decoupling depends on the wave amplitude. We observed that the peak becomes slightly lower and it is progressively shifted to lower heights with an increasing $A$ due to nonlinear effects.

The amount of decoupling also depends on the wave period. As is seen in Figure \ref{fig:awDecQ2}, the values of the decoupling smoothly decrease with the period for $P=5 - 20$ s, showing the same height dependence. The decoupling is maximum for the strongly damped case, $P=1$ s at intermediate heights; however, it disappears completely above a certain height for the reasons explained above. The increase of the magnetic field causes lower decoupling, as shown in Fig. \ref{fig:awDecQ22}, and shifts the peak to  lower heights. 

\begin{figure}[t]
\centering
\includegraphics[width=8.5cm]{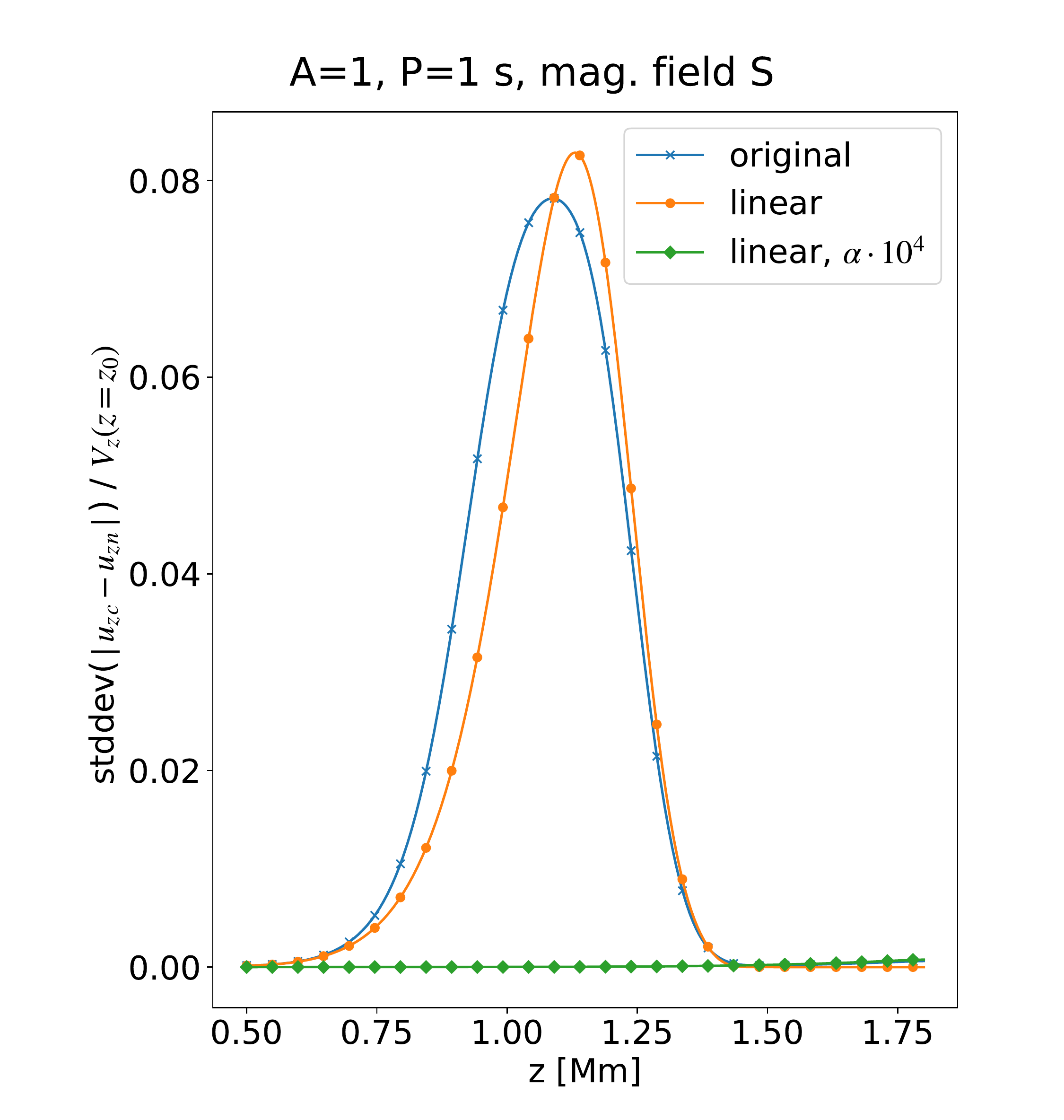}
\caption{Decoupling between charges and neutral velocity as function of height for $S$ magnetic field profile and amplitude factor $A=1$. Blue line: nonlinear simulation; orange line: linear simulation; green line: linear simulation with artificially increased collisional parameter $\alpha$. }
\label{fig:awDecQ3}
\end{figure}

To better understand the role of the nonlinear effects and collisions for the decoupling, we ran the simulation with $P=1$ s again. This time we only evolved linear equations with either the original collisional parameter $\alpha$ (Eq. \ref{eq:alpha}), corresponding to the atmospheric model, or artificially increased the collisions by a factor  of 10$^4$. The results of this experiment are given in Figure \ref{fig:awDecQ3}, together with the original nonlinear case. It can be seen that the decoupling is larger and its maximum is shifted to higher heights in the linear case (orange line), showing the same tendency as in Fig. \ref{fig:awDecQ1} (right panel). In the case when the collisional parameter $\alpha$ is  artificially increased, the decoupling completely disappears (green line). 

Therefore, the overall conclusion from the analysis above is that the decoupling increases with decreasing the wave period due to the relative increase of the importance of the collisional time scales compared to the wave scales. Similarly, the absolute value of the decoupling increases with the wave amplitude since nonlinearities bring shorter scales to the problem due to the formation of shock waves. However, the value of the decoupling relative to the initial wave amplitude is lower in the strongly nonlinear cases due to the collisional damping of the shock waves and the subsequent decrease of the oscillation amplitude of neutrals and charges.

%%%%%%%%%%%%%%%%%%%%%%%%%%%%%%%%%%%%%%%%%%%%%%%%%%%%%%%%%
%%%%%%%%%%%%%%%%%%%%%%%%%%%%%%%%%%%%%%%%%%%%%%%%%%%%%%%%%

%%%%%%%%%%%%%%%%%%%%%%%%%%%%%%%%%%%%%%%%%%%%%%%%%%%%%%%%%%
\begin{figure}[!ht]
\includegraphics[width=9cm]{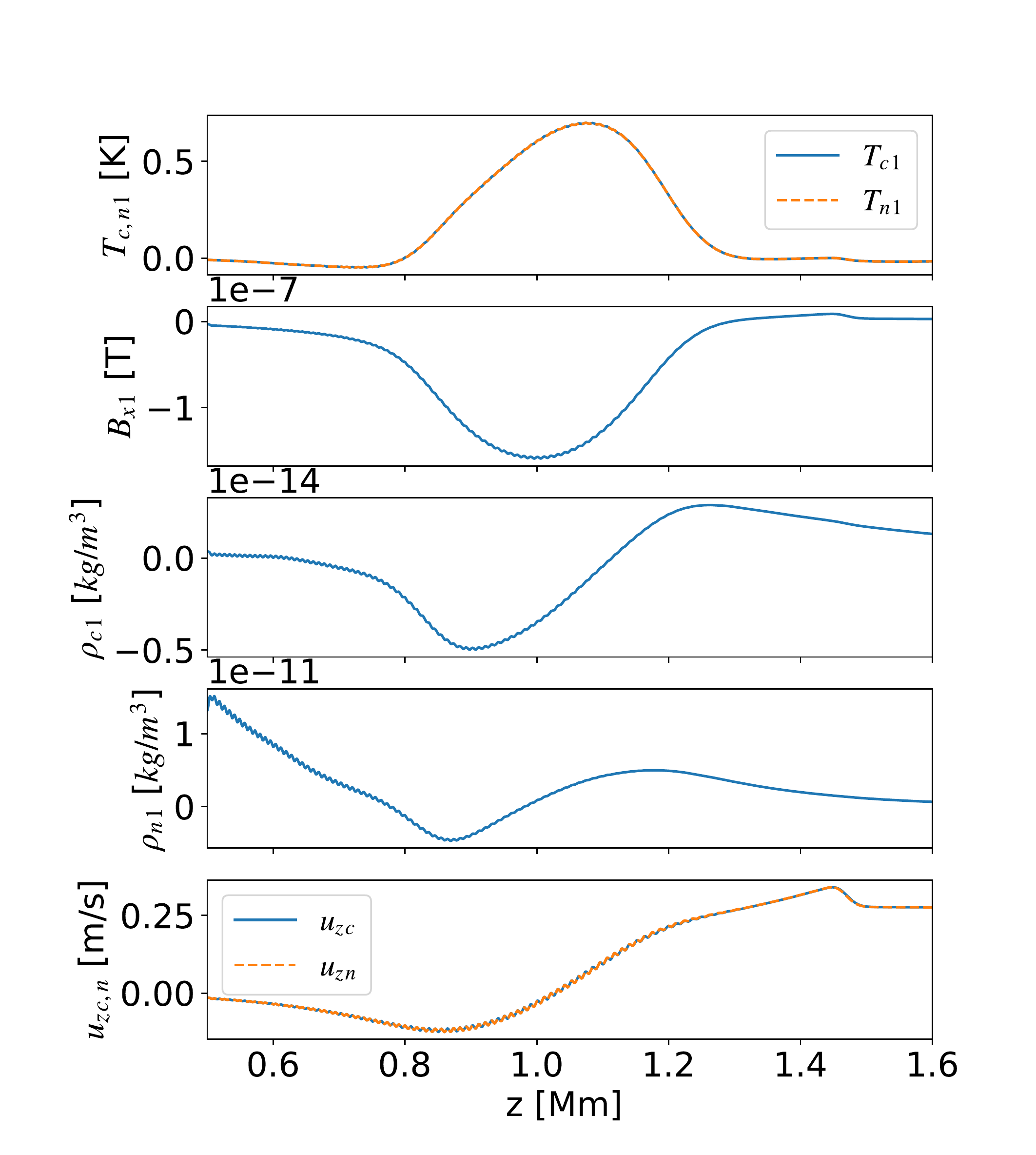}
\caption{Time average of perturbed variables as function of height for simulation with $S$ magnetic field profile, wave period $P=1$ s and amplitude factor $A=2$. From top to bottom:  temperatures of neutrals (orange) and charges (blue), magnetic field, density of charges, density of neutrals, velocity of neutrals (orange) and charges (blue).}
\label{fig:varsTimeAvg}
\end{figure}

%%%%%%%%%%%%%%%%%%%%%%%%%%%%%%%%%%%%%%%%%%%%%%%%%%%%%%%%%%

%%%%%%%%%%%%%%%%%%%%%%%%%%%%%%%%%%%%%%%%%%%%%%%%%%%%%%%%%%
\begin{figure}[!ht]
\includegraphics[width=9cm]{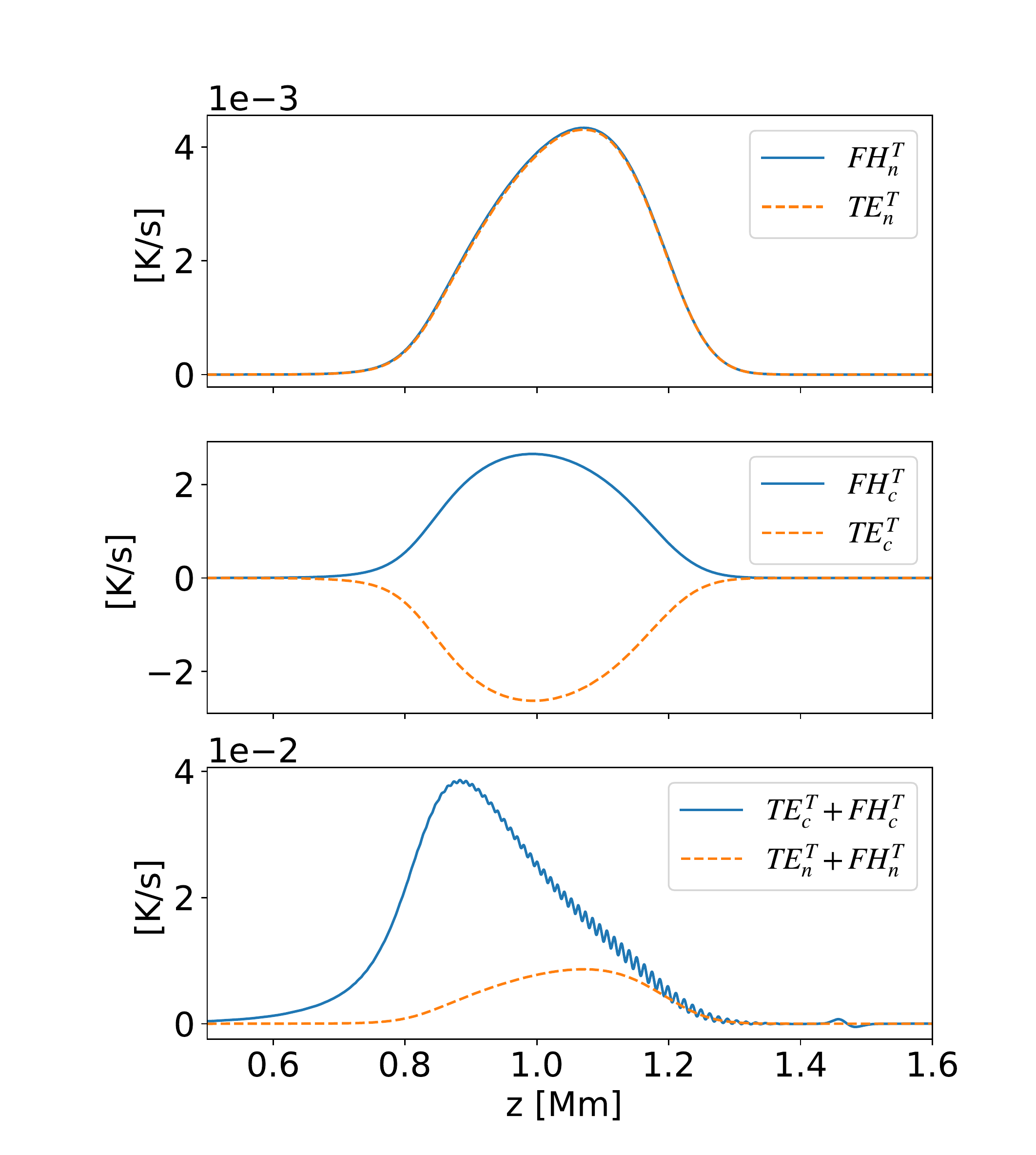}
\caption{Dissipative terms in temperature equations (Eq. \ref{eq:t}) as function of height for simulation with $S$ magnetic field profile, wave period $P=1$ s, and amplitude factor $A=2$. Top panel: thermal exchange term (TE, orange line) and frictional heating term (FH, blue line) for neutrals. Middle panel: same for charges. Bottom panel: total energy exchange terms for charges (blue), neutrals (orange).  We note that apparent oscillations in the bottom panel are numerical artifacts due to averaging the signal with relatively low temporal resolution.}
\label{fig:heatTimeAvg}
\end{figure}
%%%%%%%%%%%%%%%%%%%%%%%%%%%%%%%%%%%%%%%%%%%%%%%%%%%%%%%%%%

%%%%%%%%%%%%%%%%%%%%%%%%%%%%%%%%%%%%%%%%%%%%%%%%%%%%%%%%%
\subsection{Evolution of background variables} \label{subsec:bgEv}
%%%%%%%%%%%%%%%%%%%%%%%%%%%%%%%%%%%%%%%%%%%%%%%%%%%%%%%%%

The difference in the velocity of charges and neutrals leads to frictional dissipation of the kinetic energy of both fluids and, therefore, can produce net variations in the atmospheric parameters, such as the temperature. To check the amount of these net variations, we computed the time averages of the perturbed variables as a function of height. The results are given in Figure \ref{fig:varsTimeAvg} for the case of $P=1$ s, where the maximum damping was observed for the $S$ magnetic field profile.

Figure \ref{fig:varsTimeAvg} shows that the temperature has, on average, increased at heights between 0.8 and 1.2 Mm, which coincides with the height range where the decoupling was maximum for this case (Fig. \ref{fig:awDecQ1}). The amount of the temperature increase is about 0.5 K in this case; however, it must be compared to the amplitude of the temperature oscillations (about 10 K). Therefore, the temperature increase is not negligible. The average temperature increase is accompanied by the decrease of the magnetic field strength and variations of density and net velocity of the charges and neutrals. As the temperature increases with time, the pressure increases as well producing gas expansion and a corresponding decrease in the densities. The plasma tied to the magnetic field  expands and compresses the field lines. Therefore, the increase in temperature is  accompanied by the decrease in the magnetic field. The existence of the non-zero net flow in neutrals and charges (bottom panel) confirms this picture. 

A better understanding of the above result is gained through considering the equations of evolution of the internal energy, which are  obtained after removing the kinetic energy part from the corresponding equations of the two-fluid system (Eq. \ref{eq:twoflnum}),
\begin{eqnarray}
\frac{\partial e_n}{\partial t} &+&  \mathbf{\nabla}\cdot\left( \mathbf{u}_n e_n \right) + p_n\cdot\nabla\mathbf{u}_n   =  Q_n,  \nonumber \\
\frac{\partial e_c}{\partial t} &+&  \mathbf{\nabla}\cdot\left( \mathbf{u}_c e_c \right)+ p_c\cdot\nabla\mathbf{u}_c  =   \mathbf{J}\cdot\mathbf{E}^* + Q_c,
\end{eqnarray}

\noindent where $\mathbf{E}^*=[\mathbf{E}+\mathbf{u}_c\times \mathbf{B}]$. The expressions for collisional terms, $Q_n$ and $Q_c$  have the following form,

\begin{eqnarray} 
Q_n=  \frac{1}{2}\alpha\rho_n \rho_c(\mathbf{u}_c - \mathbf{u}_n)^2 + \frac{1}{\gamma-1} \frac{k_B}{m_n}\alpha\rho_n \rho_c(T_c - T_n),  \nonumber \\
Q_c =   \frac{1}{2}\alpha\rho_n \rho_c(\mathbf{u}_c - \mathbf{u}_n)^2 - \frac{1}{\gamma-1} \frac{k_B}{m_n}\alpha\rho_n \rho_c(T_c - T_n).
\end{eqnarray} 

These collisional terms can be compared with $M_n$ in the total energy equation. The $Q_n$ and $Q_c$ terms do not strictly compensate each other as is the case of $M_n$. There are contributions proportional to the square of the velocity difference, $(\mathbf{u}_c - \mathbf{u}_n)^2$, that are positive and are added to both charges and neutral energy equations, called frictional heating (FH). The second term, that is, the thermal exchange (TE),  has the same absolute value, but different sign for the neutrals and charges. 

%%%%%%%%%%%%%%%%%%%%%%%%%%%%%%%%%%%%%%%%%%%%%%%%%%%%%%%%%
\begin{figure*}[ht]
\centering
\includegraphics[width=8.5cm]{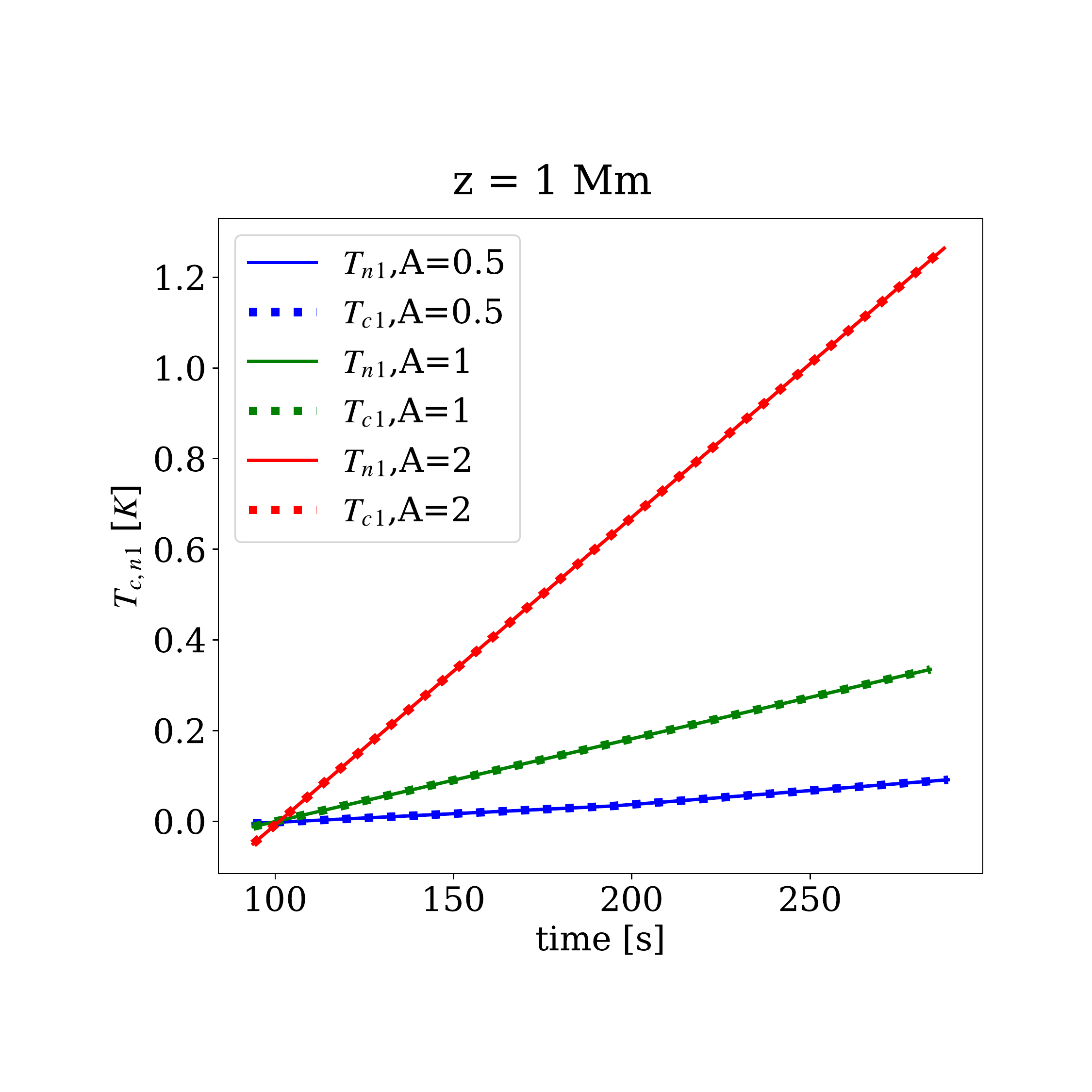}
\includegraphics[width=8.5cm]{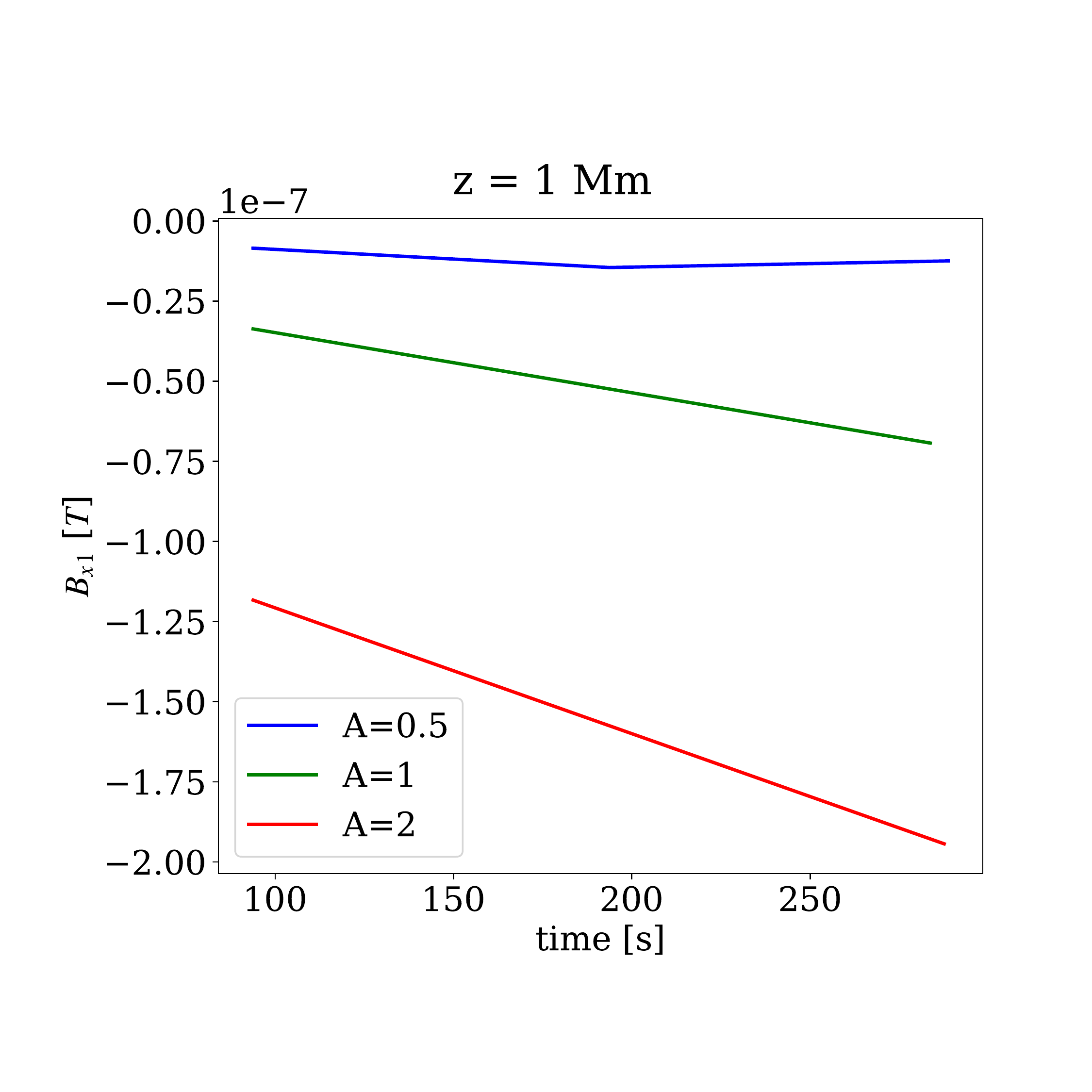}
\caption{Running average of perturbed temperature (left) and magnetic field (right) as function of time, for fixed height $z=1$ Mm for simulation with $S$ magnetic field profile, wave period $P=1$ s. The averaging is taken over three consecutive wave periods. The amplitude factor is $A=0.5$ (blue), $A=1$ (green), and $A=2$ (red). Temperature of neutrals and charges closely match together.}
\label{fig:tempMagTime}
\end{figure*}
%%%%%%%%%%%%%%%%%%%%%%%%%%%%%%%%%%%%%%%%%%%%%%%%%%%%%%%

In order to translate the contribution of $Q$-terms into the temperature increase, one can rewrite the internal energy equation in terms of temperature by substituting $e_{n}=n_nk_b T/(\gamma-1)$ and $e_{c}=2n_ek_b T/(\gamma-1)=n_ck_b T/(\gamma-1)$. By operating the energy equation using the continuity equations, the temperature evolution equations are obtained,
\begin{eqnarray} \label{eq:t}
\frac{\partial T_n}{\partial t} +  \mathbf{\nabla}\cdot\left( \mathbf{u}_n T_n \right) &+& (\gamma-1)T_n\nabla\cdot\mathbf{u}_n  =  Q_n\frac{\gamma-1}{k_bn_n}, \nonumber \\
\frac{\partial T_c}{\partial t} +  \mathbf{\nabla}\cdot\left( \mathbf{u}_c T_c \right)&+& (\gamma-1)T_c\nabla\cdot\mathbf{u}_c  =   \nonumber \\
&& \left(\mathbf{J}\cdot\mathbf{E}^* + Q_c\right)\frac{\gamma-1}{k_bn_c}. 
\end{eqnarray}

The contributions to the terms $Q_{n,c}(\gamma-1)/k_bn_{n,c}$ are computed from the same simulation as above. These terms  appear to be constant in time. Figure \ref{fig:heatTimeAvg} shows the height dependence of the frictional heating and thermal exchange terms from Eq. \ref{eq:t}, for neutrals (upper panel) and charges (middle panel), together with their joint contribution (bottom panel). As expected, the  thermal exchange is a positive quantity for the neutrals and a negative quantity for the charges. The negative sign means that the energy would be transferred from charges to neutrals. Due to significantly lower number density of charges, the absolute values of both FH and TE terms in the temperature equation for charges are three orders of magnitude larger than for neutrals. Therefore, the collisions lead to efficient frictional heating of charges, and then this energy is transferred to neutrals via the thermal exchange. 
%The charges and neutrals become efficiently thermally coupled. The evolution of the temperature of the fluid as a whole is dominated by the evolution of the neutral temperature because of their larger density. 
With the charges and neutrals thermally coupled, the evolution of the temperature of the fluid as a whole is dominated by the evolution of the neutral temperature.

The constant action of the frictional heating terms leads to an uninterrupted energy supply and an increase of temperature in time. Figure  \ref{fig:tempMagTime} shows the time evolution of temperature (left panel) and the magnetic field (right panel) at a fixed point $z=1$ Mm for the case $P=1$ s, $S$ magnetic field profile, and varying the amplitude of the perturbation, $A$. The curves are obtained through calculating the running average over three consecutive wave periods. It can be observed that the increase in temperature is a time cumulative effect, proportional to the square velocity amplitude difference, and that the temperature grows linearly in time. The growth is  linear while the modification of the background is of the same order or smaller than the perturbation amplitude.  
This linear growth is consistent with the value of $Q_n(\gamma-1)/k_bn_n \approx 8 \times 10^{-3}$ K/s, see the bottom panel of Figure \ref{fig:heatTimeAvg} at the location $z=1$ Mm. The corresponding drop of the magnetic field strength is also linear with time. 
 The gas is heated and it expands, expelling the field lines, which is consistent with the picture from Fig. \ref{fig:varsTimeAvg}. 
%Even if the mechanisms are different, the overall picture resembles the action of the ambipolar diffusion term in the single-fluid treatment \citep{Khomenko+Collados2012}, and the comparison of the two models  
%will be the object of the further study.
%While the reasons of this drop are different in the two-fluid model, the overall picture resembles the action of the ambipolar diffusion term in the single-fluid treatment \citep{Khomenko+Collados2012}, that allows to convert the magnetic energy into heat through dissipation of perpendicular currents. 

%%%%%%%%%%%%%%%%%%%%%%%%%%%%%%%%%%%%%%%%%%%%%%%%%%%%%%%%
\section{Reasons for wave damping}
%%%%%%%%%%%%%%%%%%%%%%%%%%%%%%%%%%%%%%%%%%%%%%%%%%%%%%%%

One of the interesting results obtained above is the wave damping, that is, the decrease of the wave amplitude (or even a complete disappearance of the wave) after a certain height. This effect is especially pronounced for the high-frequency perturbations. 

There can be several reasons for this damping. One of them can be ion-neutral collisions, which are well investigated in linear theory \citep{2018Ballester}. Another reason can be wave reflection due to the gradient of the Alfv\'en speed with height. It is well known that fast magneto-acoustic waves refract and reflect in the solar atmosphere due to this reason \citep{Cally2006, Khomenko+Collados2006}.\ This is a phenomenon that is actually thought to be responsible for the formation of high-frequency acoustic halos in observations \citep{Khomenko+Collados2009, Rijs+etal2016}. 

In order to check if the waves are reflected in our experiment, we computed the acoustic wave energy fluxes, as $p_{c1}u_{zc}$ for charges, and as $p_{n1}u_{zn}$ for neutrals. It turns out that both fluxes are positive at all heights, meaning vertical wave propagation and no apparent reflection. The energy fluxes vanish after a certain height above 1.2 Mm, which means that the waves became non-propagating. The reasons of this behavior are better highlighted with the help of the analytical solution, developed below. 

%%%%%%%%%%%%%%%%%%%%%%%%%%%%%%%%%%%%%%%%%%%%%%%%%%%%%%%%
\subsection{Comparison between analytical and numerical results}
\label{sect:comparison}
%%%%%%%%%%%%%%%%%%%%%%%%%%%%%%%%%%%%%%%%%%%%%%%%%%%%%%%%

In order to distinguish  between the  damping produced by the decoupling of the neutral and charged fluids in the linear regime, and the damping caused by the decoupling at the shock fronts or the steepening in the wave profile in the nonlinear regime, we compared the numerical solution of the linear and nonlinear two-fluid equations with the analytical solution.  
The analytical solution is obtained by solving the linearized two fluid  equations with a nonuniform background for each species. We introduced the coupling term in the momentum equations and assumed the collisional parameter $\alpha$ to be  constant in time, which is equal to the value corresponding to the equilibrium atmosphere and consistent with the linear approximation. The linearized equation takes the following form,
\begin{align} 
\frac{\partial \rho_{\rm c1}}{\partial t} + u_{\rm cz} \frac{d \rho_{\rm c0}}{d z} + \rho_{\rm c0}  \frac{\partial u_{\rm cz}}{\partial z} = 0, \nonumber  \\
\frac{\partial \rho_{\rm n1}}{\partial t} + u_{\rm nz} \frac{d \rho_{\rm n0}}{d z} + \rho_{\rm n0}  \frac{\partial u_{\rm nz}}{\partial z} = 0, \nonumber  \\
\rho_{\rm c0} \frac{\partial u_{\rm cz}}{\partial t} = -\rho_{\rm c1} g - \frac{\partial p_{\rm c1}}{\partial z} - \frac{1}{\mu_0}\left ( \frac{\partial B_{\rm x1}}{\partial z} B_{\rm x0} + \frac{d B_{\rm x0}}{d z} B_{\rm x1}\right ) \nonumber \\ + \alpha \rho_{\rm n0} \rho_{\rm c0} (u_{\rm nz} - u_{\rm cz} ), \nonumber \\
\rho_{\rm n0} \frac{\partial u_{\rm nz}}{\partial t} = -\rho_{\rm n1} g - \frac{\partial p_{\rm n1}}{\partial z} +  \alpha \rho_{\rm n0} \rho_{\rm c0}  (u_{\rm cz} - u_{\rm nz} ), \nonumber \\
\frac{\partial p_{\rm c1}}{\partial t} = c_{\rm c0}^2 \frac{\partial \rho_{\rm c1}}{\partial t} + c_{\rm c0}^2 u_{\rm cz} \frac{d \rho_{\rm c0}}{\partial z} - u_{\rm cz} \frac{d p_{\rm c0}}{d z}, \nonumber \\
\frac{\partial p_{\rm n1}}{\partial t} = c_{\rm n0}^2 \frac{\partial \rho_{\rm n1}}{\partial t} + c_{\rm n0}^2 u_{\rm nz} \frac{d \rho_{\rm n0}}{\partial z} - u_{\rm nz} \frac{d p_{\rm n0}}{d z}, \nonumber \\
\frac{\partial B_{\rm x1}}{\partial t} = - B_{\rm x0} \frac{\partial u_{\rm cz}}{\partial z} - u_{\rm cz} \frac{d B_{\rm x0} }{d z} . 
\end{align}

These equations are combined into two coupled partial differential equations for the vertical velocity of each species and coupled by the collisional terms as follows:

\begin{align} 
 \frac{\partial^2 u_{\rm cz}}{\partial t^2} & = a_c(z) \frac{\partial^2 u_{\rm cz}}{\partial z^2} + b_c(z) \frac{\partial u_{\rm cz}}{\partial z} + \alpha \rho_{\rm n0} \left (\frac{\partial u_{\rm nz}}{\partial t} - \frac{\partial u_{\rm cz}}{\partial t} \right), \nonumber \\
 \frac{\partial^2 u_{\rm nz}}{\partial t^2} & = a_n(z) \frac{\partial^2 u_{\rm nz}}{\partial z^2} + b_n(z) \frac{\partial u_{\rm nz}}{\partial z} + \alpha \rho_{\rm c0} \left (\frac{\partial u_{\rm cz}}{\partial t} - \frac{\partial u_{\rm nz}}{\partial t} \right), \label{eq:waveeq22fl}
\end{align}
where
\begin{align}
a_c(z) &=  c_{\rm c0}^2 + {v_A}_0^2, \quad a_n(z) =  c_{\rm n0}^2, \nonumber \\
b_{c,n}(z)  & = \frac{1}{{\rho_{c,n}}_0} \frac{d \left( {{\rho_{c,n}}_0} a_{c,n}   \right)}{d z},\nonumber \\
{c_{c,n}}_0^2 &= \gamma \frac{{p_{c,n}}_{0}}{{\rho_{c,n}}_{0}}, \quad {v_A}_0^2 = \frac{{B_x}_{0}^2}{\mu_0{\rho_c}_{0}}. \label{eq:Hz2fl}
\end{align}

Assuming the functional dependence on space and time of the vertical velocity of the form $\{u_{\rm cz}(z,t), u_{\rm nz}(z,t) \} = \{ \tilde u_{\rm cz}(z), \tilde u_{\rm nz}(z) \} \times \text{exp}\left( i \omega t \right )$, where the frequency $\omega$ is fixed, the above coupled system can be rewritten as, 

\begin{align} 
 - \omega^2 \tilde u_{\rm cz} & = a_c(z) \frac{d^2 \tilde u_{\rm cz}}{d z^2} + b_c(z) \frac{d\tilde u_{\rm cz}}{d z} + i \omega \alpha \rho_{\rm n0} \left (\tilde u_{\rm nz}  - \tilde u_{\rm cz} \right), \nonumber \\
- \omega^2 \tilde u_{\rm nz} & = a_n(z) \frac{d^2 \tilde u_{\rm nz}}{d z^2} + b_n(z) \frac{d \tilde u_{\rm nz}}{d z} + i \omega \alpha \rho_{\rm c0} \left (\tilde u_{\rm cz} - \tilde u_{\rm nz} \right). \label{eq:waveeq22fl3}
\end{align}
In combining these two equations, we obtained the folllowing:
\begin{align}\label{eq:waveeq22fl4}
\frac{d^4  \tilde u_{\rm cz}}{d z^4} a_c a_n   + \frac{d^3  \tilde u_{\rm cz}}{d z^3} \left (a_n b_c + a_c b_n \right ) + \nonumber \\
 \frac{d^2  \tilde u_{\rm cz}}{d z^2} \left( b_c b_n + \omega^2 (a_c + a_n) - i \alpha \omega (a_c \rho_{c0} + a_n \rho_{n0}) \right )  +\nonumber\\
\frac{d  \tilde u_{\rm cz}}{d z} \omega \left( \omega (b_c + b_n) - i \alpha (b_c \rho_{c0} + b_n \rho_{n0})  \right) +\nonumber \\
 \tilde u_{\rm cz} \omega^3 \left(\omega - i \alpha (\rho_{c0} + \rho_{n0}) \right ) =0.
\end{align}

\noindent The coefficients of the spatial derivatives that appear in  Eq. (\ref{eq:waveeq22fl4}) are not uniform and there is no apparent exact analytical solution to Eq. (\ref{eq:waveeq22fl4}).
%If the temperatures were uniform and the pressures and densities had exponential profiles, then $a$, $b$ and $\alpha$ would be uniform. However, there are coefficients proportional to the multiple of $\alpha$ and densities (either $\rho_{n0}$ or $\rho_{c0}$) that will not be uniform even in this simplified case. Therefore, it prevents the above equation from having simple solutions in terms of $\text{exp}\left( i k z \right )$. 
The waves we consider in this study are short-period waves with wavelengths that are shorter than equilibrium gradient scales. 
Thus, we used the WKB approximation to find solutions for Eq. (\ref{eq:waveeq22fl4}), as follows: 
\begin{equation}
\left\{ \tilde u_{\rm nz}, \tilde u_{\rm cz}\right \} = \left\{ V_n(z), V_c(z) \right \} \cdot \text{exp} \left[i \phi(z)\right],
\end{equation}
where $V_n$ and $V_c$ are space dependent complex amplitudes, and 
\begin{equation}
\phi(z)=-\int_0^z{k(z^\prime) dz^\prime}.
\end{equation}
Similar to the WKB approximation, we assumed that  the velocity amplitude and  the wavenumber gradients are small and of the same order,
\begin{eqnarray}
\frac{d \phi}{d z} = -k = -k_0 - \epsilon k_1(z), \nonumber \\
V_{c,n}(z) = V_{c,n0}+\epsilon V_{c,n1}(z),
\end{eqnarray}
with $\epsilon$ being a small parameter and the second derivatives of $k_1(z)$ and $ V_{c,n1}$ being zero. It is then straightforward when making the following calculations:
\begin{eqnarray} \label{eq:coso}
\frac{d \tilde u_{\rm cz}}{d z} &=&  \left (\epsilon \frac{d V_{c1}}{d z} -i k V_{c1}  \right)\cdot \text{exp} (i \phi), \nonumber \\
\frac{d^2 \tilde u_{\rm cz}}{d z^2} &=&  \left (-2 i \epsilon k \frac{d V_{c1}}{d z} + i \epsilon V_{c1}   \frac{d k_1}{dz} - k^2 V_{c1} \right) \cdot \text{exp} (i \phi), \nonumber \\
\frac{d^3 \tilde u_{\rm cz}}{d z^3} &=&  \left (-3 \epsilon k \frac{d V_{c1}}{d z} - 3 \epsilon V_{c1} \frac{d k_1}{dz} +i k^2 V_{c1} \right)\cdot k \cdot \text{exp} (i \phi), \nonumber \\
\frac{d^4 \tilde u_{\rm cz}}{d z^4} &=&  \left (4i \epsilon k \frac{d V_{c1}}{d z} + 6i \epsilon V_{c1} \frac{d k_1}{dz} +k^2 V_{c1} \right)\cdot k^2 \cdot \text{exp} (i \phi). \nonumber \\
\end{eqnarray}

By substituting these expressions into Eq. \ref{eq:waveeq22fl4} and in only keeping 0th-order terms, the dispersion relation for $k$ is obtained,

\begin{eqnarray}  \label{eq:dispRel2Fl}
(\omega^2   - k^2 a_c  - i k b_c )(\omega^2  - k^2 a_n - i k b_n ) +\\ \nonumber
i\omega \alpha \rho_{\rm 0}(-\omega^2 + k^2 a +ikb) = 0
\end{eqnarray}
 \noindent where the remaining coefficients are defined as,
 \begin{eqnarray}
 b  = (\rho_{\rm n0} b_n + \rho_{\rm c0} b_c)/ {\rho}_0,\quad 
 a  = (\rho_{\rm n0} a_n + \rho_{\rm c0} a_c)/\rho_0
\end{eqnarray}
\noindent with $\rho_0=\rho_{\rm c0} + \rho_{\rm n0}$.

Next, the relations (\ref{eq:coso}) are replaced into equations (\ref{eq:waveeq22fl3}). Separating the 0th-order terms from the 1st-order terms, the expressions for the velocity amplitudes are obtained as follows:
\begin{equation} \label{eq:intV2}
V_{c,n}(z) = V_{c,n}|_{z=0}\cdot \text{exp} \left( \int_0^z{\frac{d k}{d z}  \frac{i a_{c,n}}{b_{c,n} - 2 i k a_{c,n}} dz^\prime}\right).
\end{equation}

In the limit of  $ \frac{\alpha {\rho}_0}{\omega} \gg 1 $, Eq. (\ref{eq:dispRel2Fl}) reduces to the single-fluid dispersion relation, 
\begin{equation}  \label{eq:d1}
-\omega^2 + k^2 a +ikb = 0.
\end{equation}
In the opposite limit, when $\frac{\alpha {\rho}_0}{\omega} \ll 1,$ we recovered the dispersion relations of either neutrals or charges. Otherwise, if  $ \frac{\alpha {\rho}_0}{\omega} \approx 1$, both of the terms in the equation are equally important.

%%%%%%%%%%%%%%%%%%%%%%%%%%%%%%%%%%%%%%%%%%%%%%%%%%%%%%%%%
\begin{figure}[t]
\centering
\includegraphics[width=9cm]{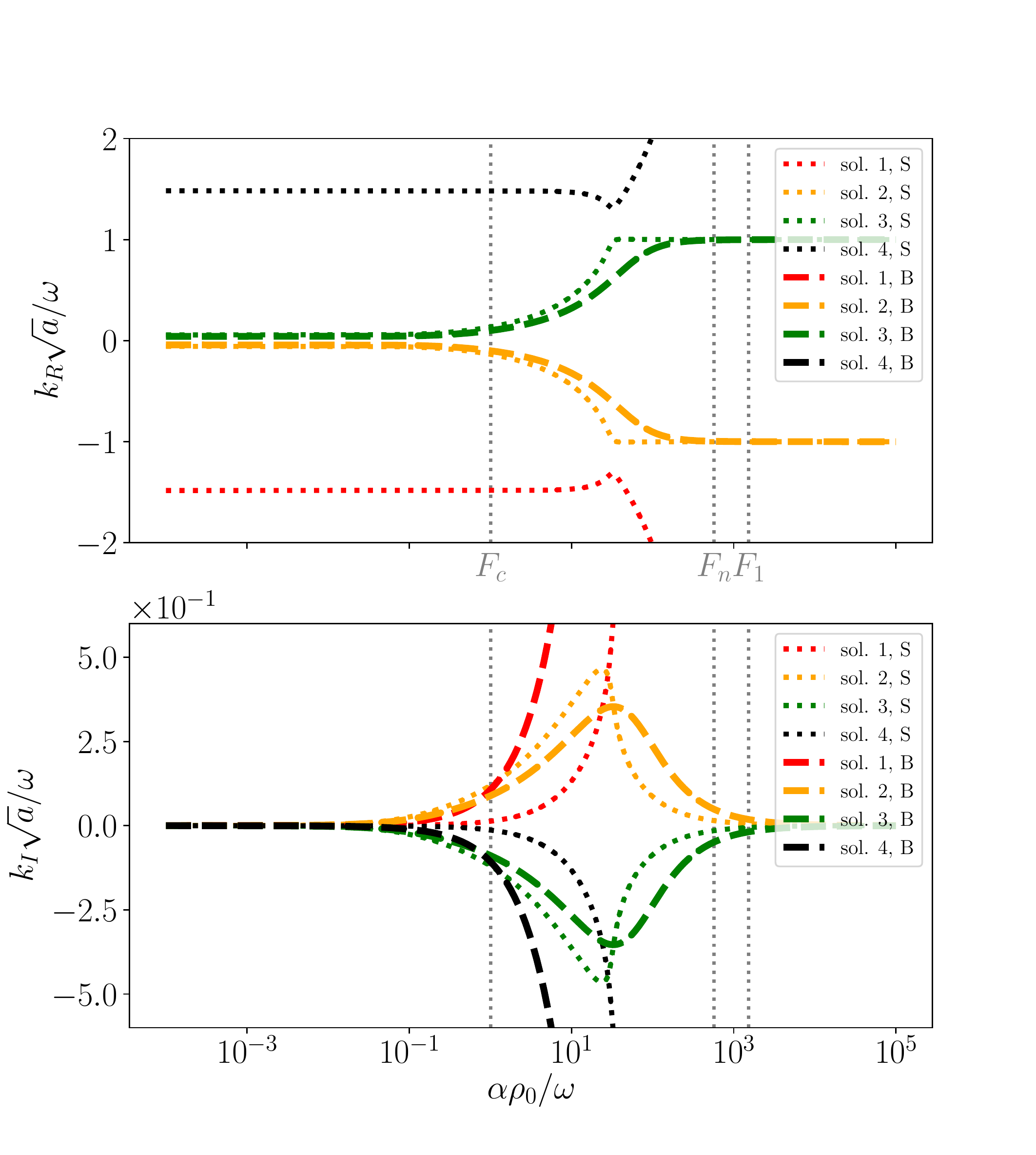}
\caption{Solutions of non-dimensional dispersion relation without stratification, Eq. (\ref{eq:d2_adim}).  The upper and lower panels show the real and imaginary part of $k$, scaled with $\sqrt{a}/\omega$ as a function of the ratio between the collisional and wave frequencies, scaled with $\rho_0$. Solutions labeled as ``S'' (dashed lines) are for $S$ magnetic field; solutions labeled ``B'' (solid lines) are for $B$ magnetic field. The vertical dotted lines mark values of $\omega=\nu_{in}$ ($F_c$), $\omega=\nu_{ni}$ ($F_n$), and $\omega=2\pi/1$ ($F_1$) for the 1 s period wave. }
\label{fig:adim_k_om}
\end{figure}
%%%%%%%%%%%%%%%%%%%%%%%%%%%%%%%%%%%%%%%%%%%%%%%%%%%%%%%%%

%%%%%%%%%%%%%%%%%%%%%%%%%%%%%%%%%%%%%%%%%%%%%%%%%%%%%%%%%
\begin{figure}[t]
\centering
\includegraphics[width=9cm]{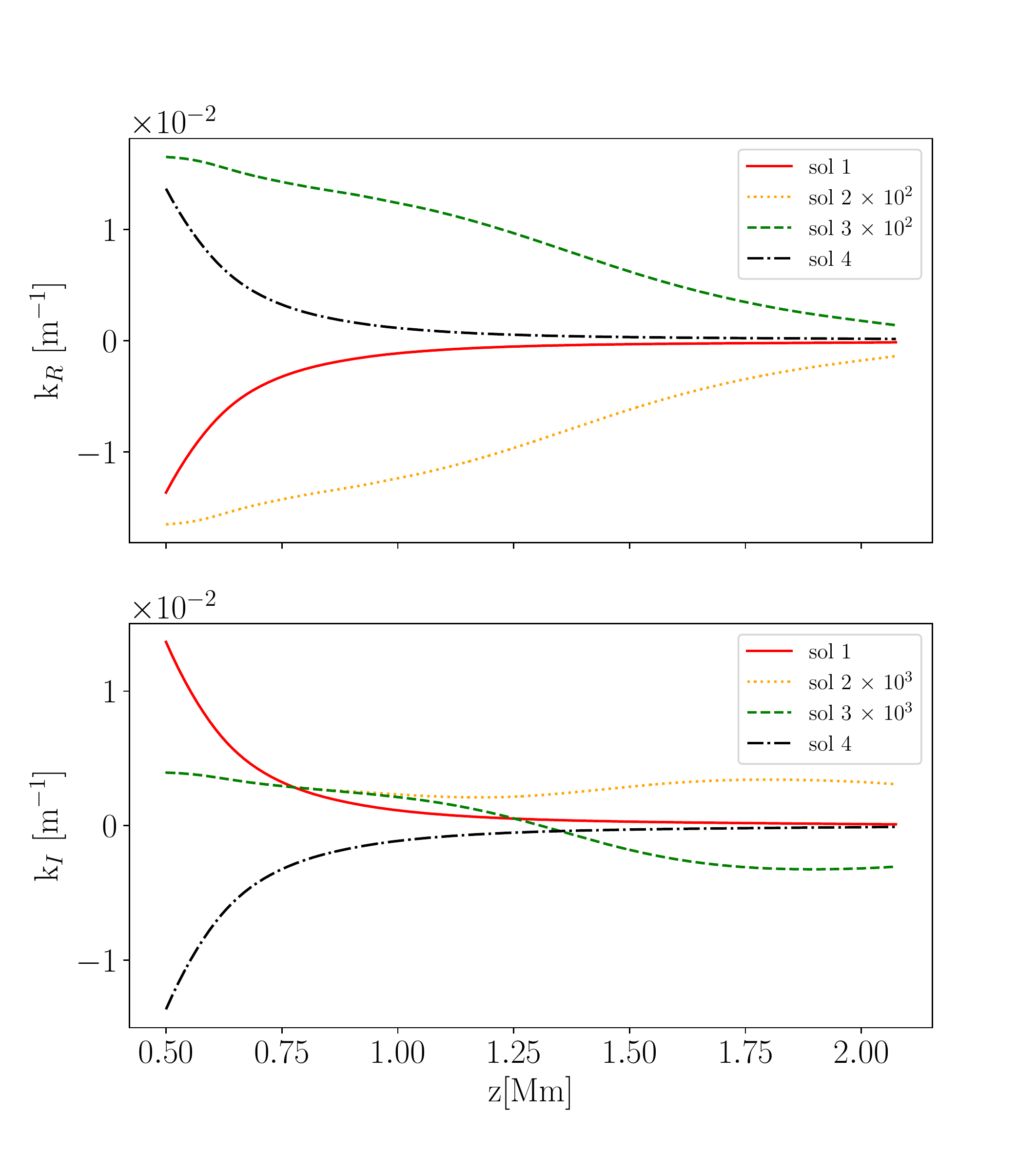}
\caption{Solutions of dispersion relation (\ref{eq:dispRel2Fl}) for two-fluid linearized equations. The upper and lower panels show the real and imaginary part of $k$ as a function of height, calculated for the magnetic field profile S and 5 s period.}
\label{fig:dispRel2Fl4sol}
\end{figure}
%%%%%%%%%%%%%%%%%%%%%%%%%%%%%%%%%%%%%%%%%%%%%%%%%%%%%%%%%

 It is interesting to analyze the effect of collisional damping separately from the effect of stratification. The effect of stratification is independent of the wave frequency for high-frequency waves considered here. By neglecting the stratification, the  dispersion relation, Eq. (\ref{eq:dispRel2Fl}) becomes the dispersion relation in a uniform atmosphere,
\begin{equation}  \label{eq:d2}
(\omega^2   - k^2 a_c  )(\omega^2  - k^2 a_n  ) + i\omega \alpha \rho_{\rm 0}(-\omega^2 + k^2 a ) = 0.
\end{equation}

\noindent We considered the solutions of this equation as a function of the wave period. Using the non-dimensional variables,
\begin{equation} \label{eq:d2_adim_vars}
F=\alpha \rho_0/\omega;  \qquad E=k\sqrt{a}/\omega, 
\end{equation}
the dispersion relation, Eq. (\ref{eq:d2}), becomes
\begin{equation}  \label{eq:d2_adim}
(1  - E^2 a_c/a  )(1  - E^2 a_n/a  ) - i F (1-E^2)= 0.
\end{equation}

Figure \ref{fig:adim_k_om} shows the real and imaginary part of the  four solutions, $E$, of the fourth order dispersion relation, 
Eq. (\ref{eq:d2_adim}), as a function of the non-dimensional variable $F$,  which is  the ratio between the collisional and the wave frequencies, scaled with $\rho_0$. This dispersion relation was solved using  the values for $a_n$, $a_c$, and $a$  corresponding to a point located in the middle of the atmosphere for $S$ and $B$ magnetic field profiles. The values on the  horizontal axis cover a large range,  from $F = 10^{-4}$, for the uncoupled case,  and up to  $F = 10^5,$ for the strongly coupled case.  We marked the values of $F$  corresponding to the the cases of $\omega=\nu_{in}$ (ion-neutral collision frequency), $\omega=\nu_{ni}$ (neutral-ion collision frequency), and $\omega=2\pi/1$ (frequency corresponding to the period of a 1 s wave) with $F_c$, $F_n$, and $F_1$, respectively. The values of the frequencies were  calculated for the height corresponding to the middle of the domain. 

We can observe that for small $F$, corresponding to high frequencies $\omega$ (uncoupled case), the waves  propagate with the characteristic speed of either charges ($\sqrt{a_c}$, downward solution \#1, and upward solution \#4) or with the characteristic speed of neutrals ($\sqrt{a_n}$, downward solution \#2, and  upward solution \#3). The imaginary part of $k$ in the uncoupled case is zero, meaning that there is no damping. This result is similar to \cite{Popescu+etal2018}.

For the large values of $F$, corresponding to low frequencies $\omega$ (strongly coupled case), there are two solutions, which propagate with the characteristic speed in the fluid as a whole, $\sqrt{a}$, both upward ( \#2) and downward (\#3). For the other two solutions,  \#1 and \#4,  the real and the imaginary parts of $k$  are very large. These are waves with very large phase speed, but also extremely large damping, and we consider them as unphysical. 

Solutions \#2 and \#3 (yellow and green lines) have maximum damping, $k_I$, for waves in the frequency range between $\nu_{ni}$ and $\nu_{in}$. This is true for both magnetic field profiles $B$ and $S$. For the larger magnetic field $B$, the maximum damping shifts to higher frequencies. We checked that for even higher magnetic fields the dispersion curves remain practically equal to the case $B$. For the 1 s period wave, the value of damping is decreased from its maximum, but it is still non-zero, which explains the damping we observe in our simulations. 

%%%%%%%%%%%%%%%%%%%%%%%%%%%%%%%%%%%%%%%%%%%%%%%%%%%%%%%%%
\begin{figure*}[t]
\centering
\includegraphics[width=8.5cm]{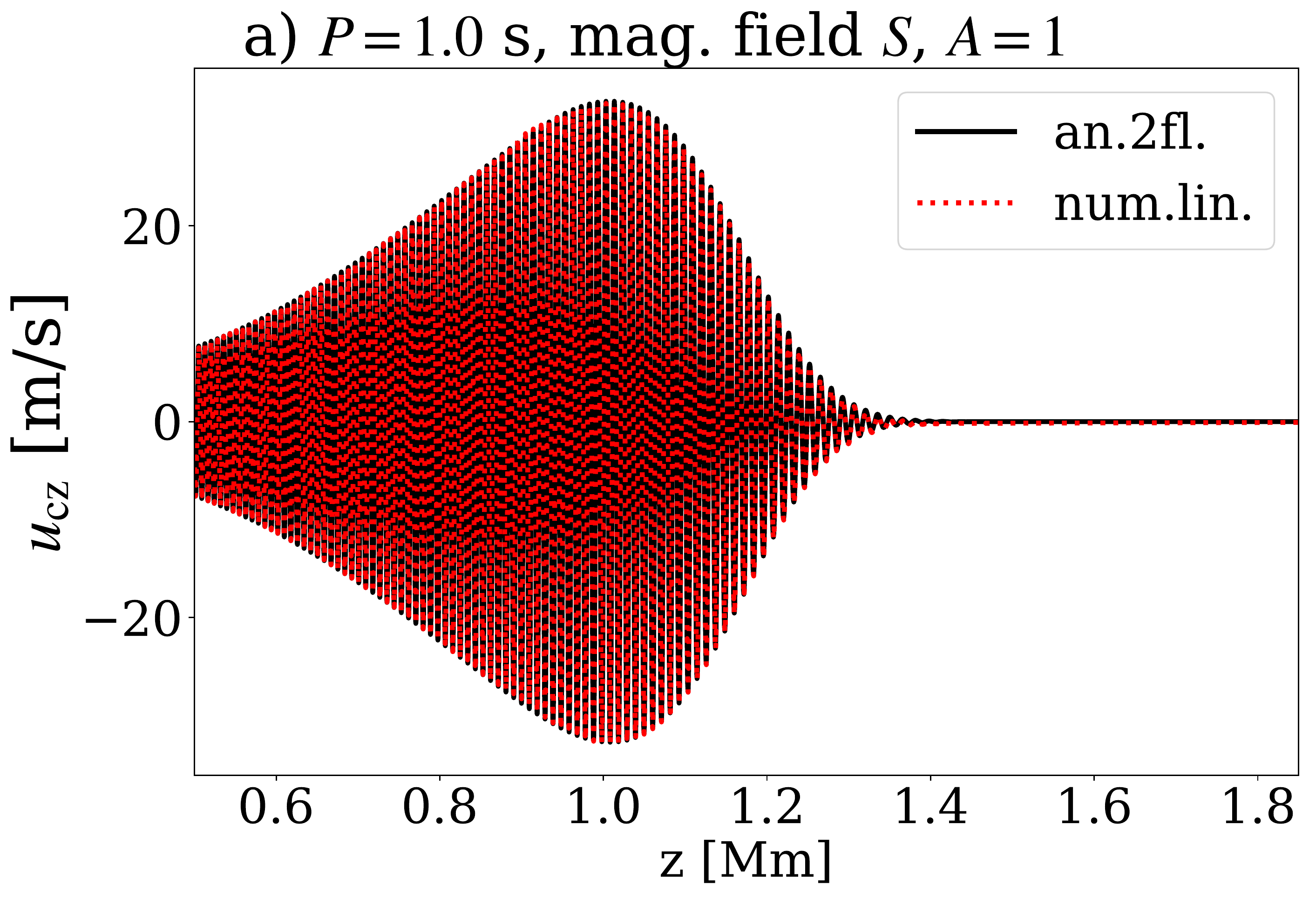}
\includegraphics[width=8.5cm]{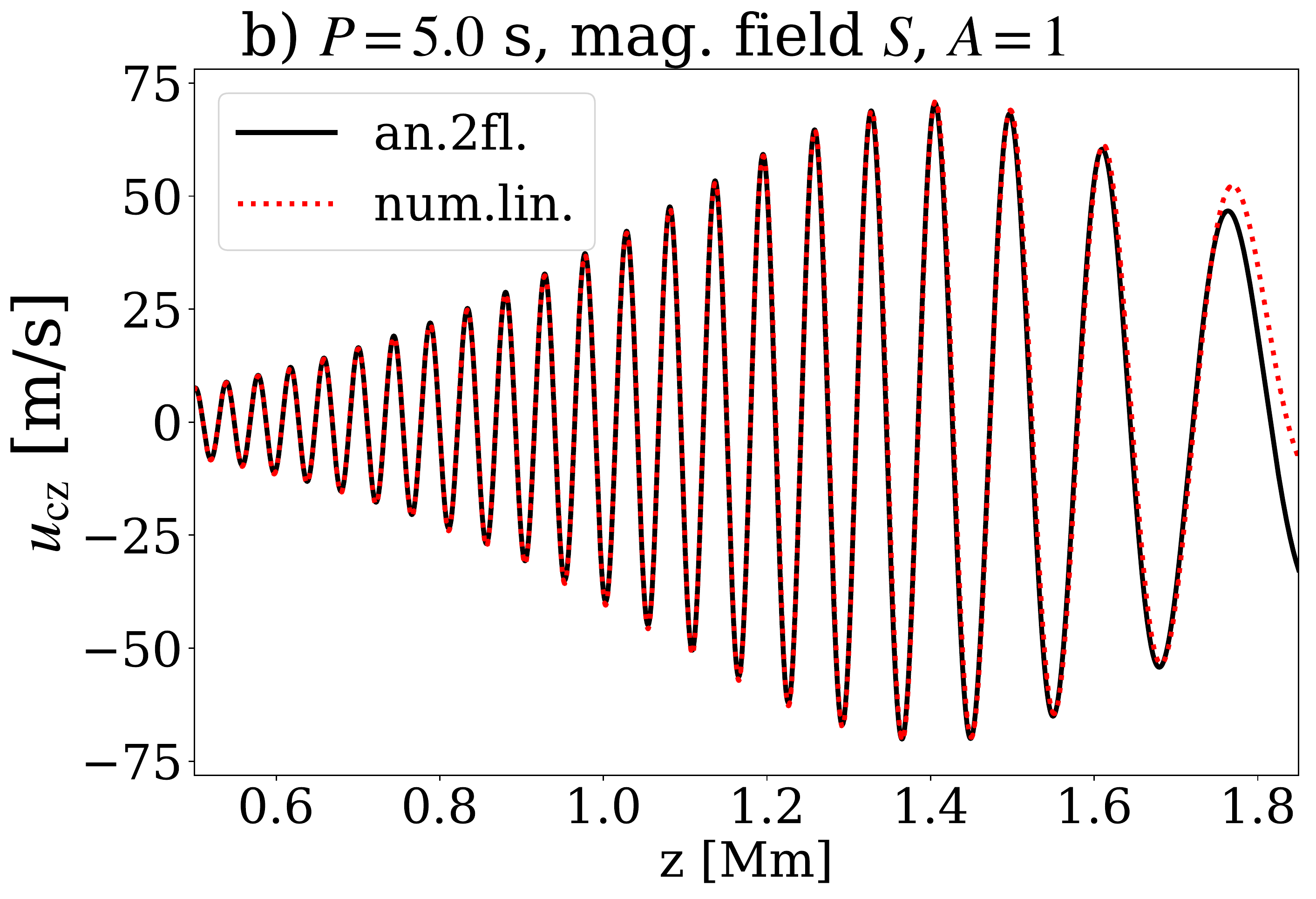}
\includegraphics[width=8.5cm]{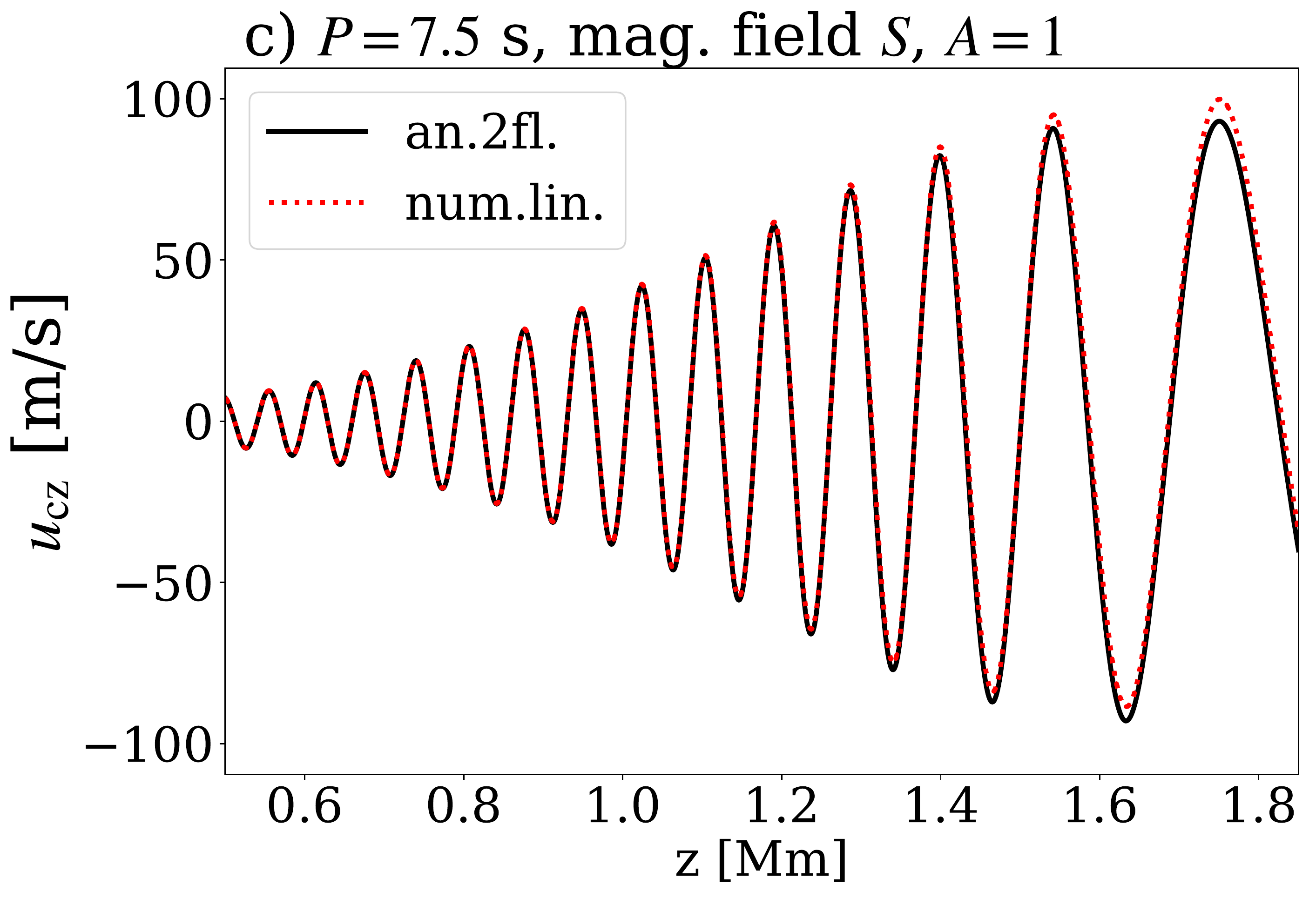}
\includegraphics[width=8.5cm]{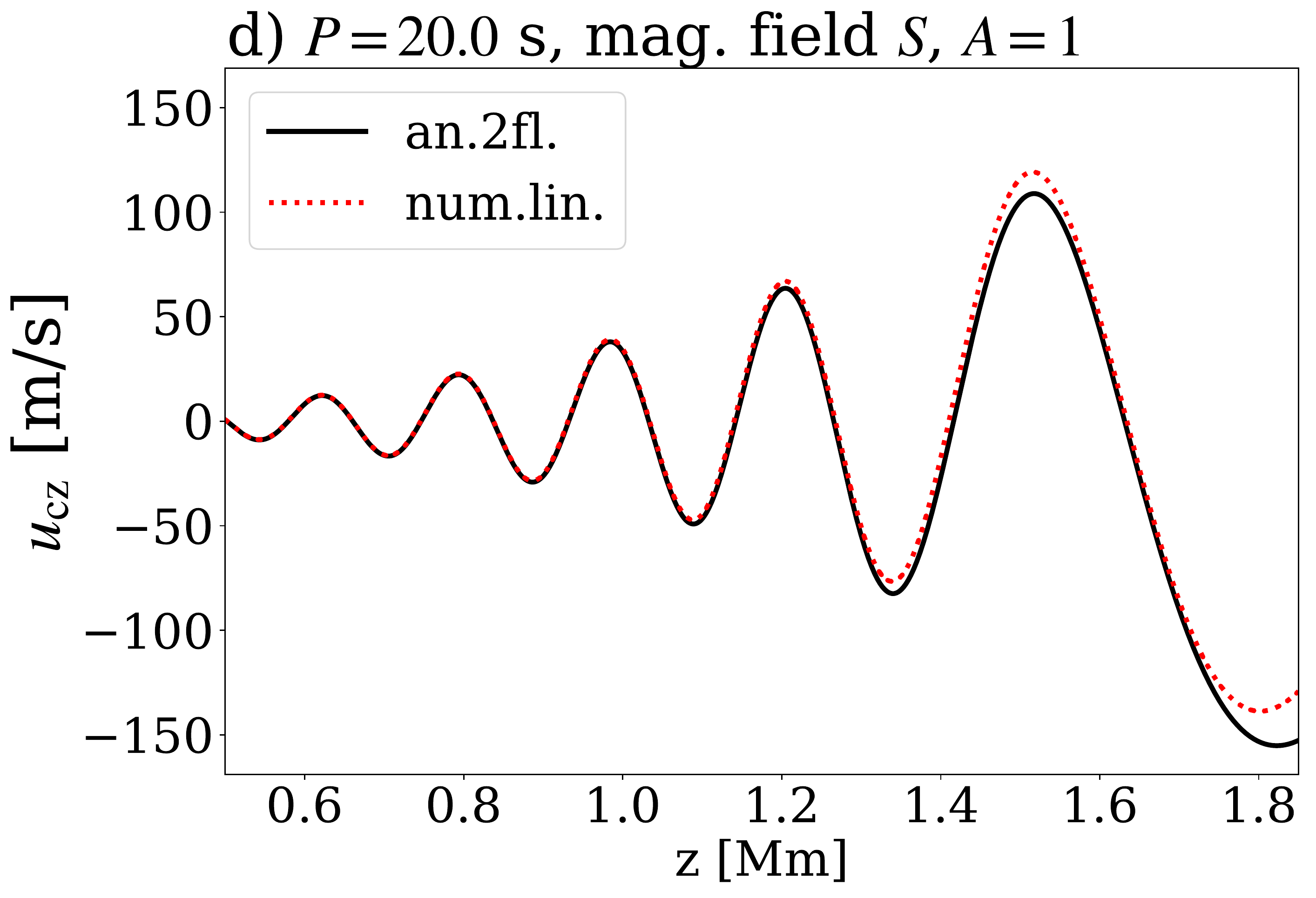} 
\caption{Comparison between numerical solution in linear regime (black lines) and analytical solution of two-fluid equations (red dotted line). Individual panels show snapshots of the velocity of charges as a function of height at fixed time moments in the stationary regime of the simulations. Panels from left to right, from top to bottom are for wave periods of 1, 5, 7.5, and 20 s, and S magnetic field profile. It is important to note that the analytical solution slightly overshoots the numerical solution, the effect being more pronounced toward the
upper part of the domain. This is due to the fact that the analytical solution is an approximate solution.}
\label{fig:l-an-S}
\end{figure*}
%%%%%%%%%%%%%%%%%%%%%%%%%%%%%%%%%%%%%%%%%%%%%%%%%%%%%%%%%

The  full dispersion relation, which  includes stratification,  (\ref{eq:dispRel2Fl}) is also fourth order in $k$ and has four different roots. Examples of these four different solutions   as a function of the height in the atmosphere are illustrated in Fig. \ref{fig:dispRel2Fl4sol} and were calculated for the S magnetic field profile and the wave period of 5 s.   Accordingly to Figure \ref{fig:adim_k_om}, from the point of view of the collisions, this is the coupled case. It can be seen that the solutions  \#1 and \#4 both have very large imaginary and real parts. These modes therefore have very low propagation speeds and very high damping, which is similar to the case without stratification. We did not consider these solutions further. The other two solutions marked \#2 and \#3 represent waves that propagate either up (solution \#3) or down (solution \#2) and their damping is moderate. It can be seen that $k_I$ of both solutions matches for the first 0.5 Mm from the bottom, and has positive values, corresponding to the wave amplification. At these heights, the charges-neutral decoupling is not strong enough to damp the waves and the growth of the wave amplitude with height is caused simply by the gravitational stratification of the atmosphere. Higher up, $k_I$ turns negative for the solution \#3, meaning wave damping. The same amount of damping is also present in the solution \#2, however $k_I$ is positive because the wave propagates downward (negative $k_R$). For the comparison below, we use the solution \#3. 

Figure \ref{fig:l-an-S} compares the numerical solutions of the linearized equations with the analytical solution for different wave periods and the  magnetic field profile S. One can observe that both solutions match rather well and small differences are only observed in the upper layers. The analytical solution in the WKB approximation is able to describe rather precisely the damping of waves due to collisions in the linear regime. Therefore, we can conclude that the wave damping observed in these simulations is determined mainly by the linear decoupling.

\begin{figure*}[t]
\centering
\includegraphics[width=8.5cm]{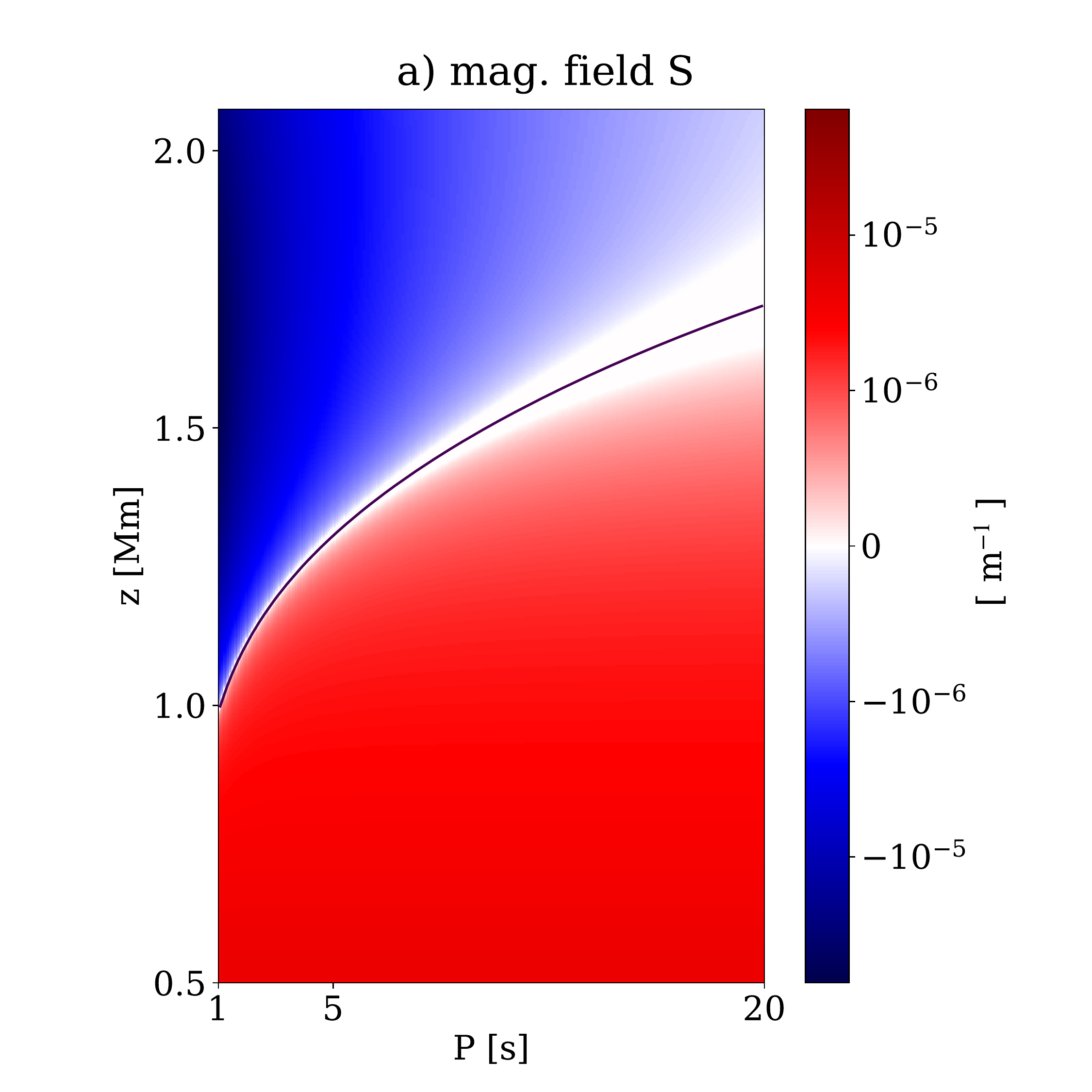}
\includegraphics[width=8.5cm]{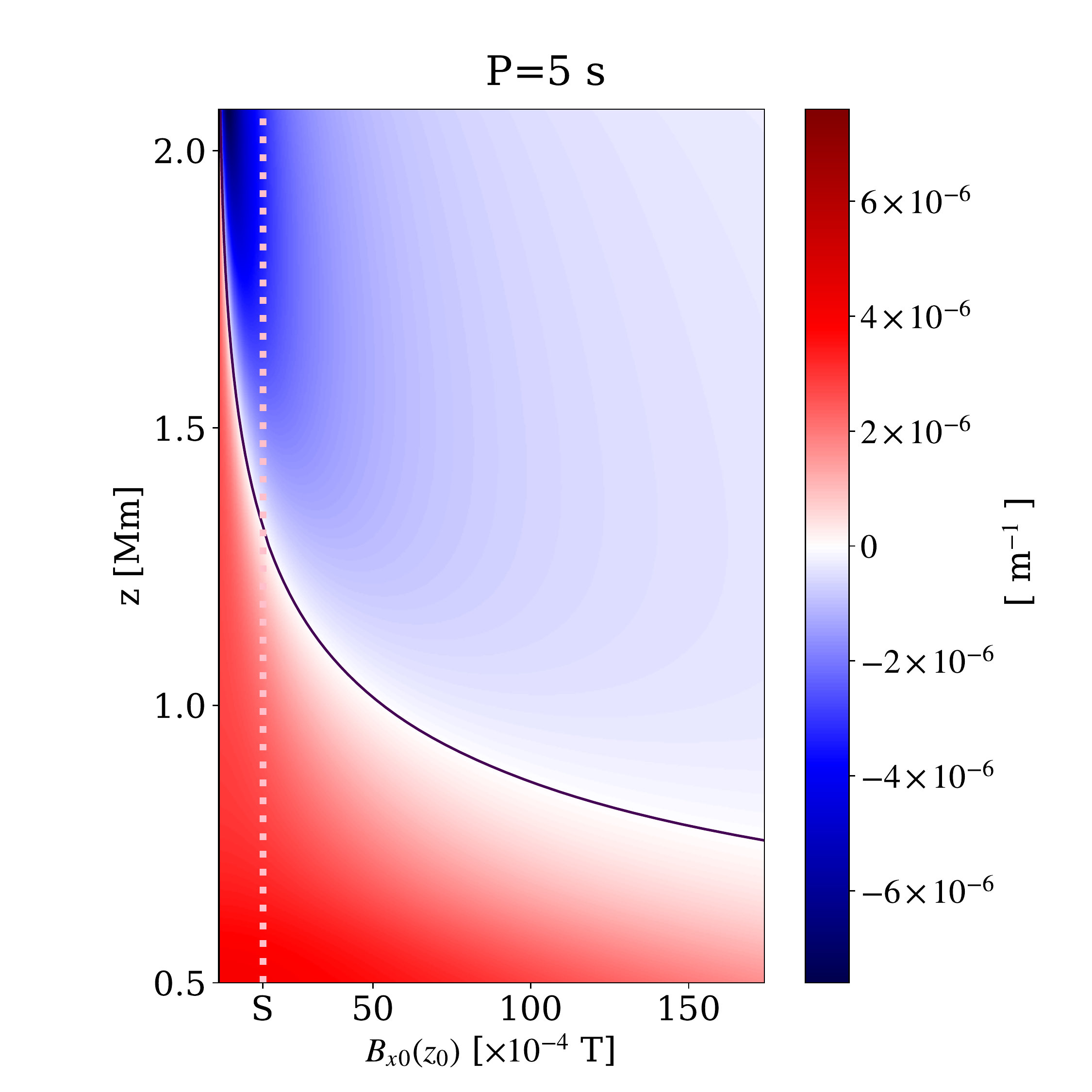} 
\caption{Left panel: imaginary part of k as function of wave period (horizontal axis) and height (vertical axis) obtained after solving dispersion relation for two-fluid equations (Eq. \ref{eq:dispRel2Fl}) for S magnetic field profile. Solution \#3 from Figure \ref{fig:dispRel2Fl4sol} is used. The period varies between 1 s and 20 s (the minimum and maximum values used in the simulations). The solid, black line represents the contour of $k_I$ = 0.  Right panel: imaginary part of k as function of background magnetic field at base of atmosphere (horizontal axis) and height (vertical axis). The wave period is fixed to 5 s. }
\label{fig:kI_P}
\end{figure*}

Figure \ref{fig:kI_P} further illustrates the effects of linear decoupling on the wave amplitudes. It shows the imaginary part of the wave vector, $k_I,$ for the solution \#3 as a function of the wave period (left) and of the magnetic field strength (right). The value and sign of $k_I$ is directly related to the modification of the wave amplitude with height.  It can be seen that in all cases, $k_I$ changes sign at some height in the atmosphere. At the bottom part of the atmosphere, $k_I$ is positive, meaning that the growth of the wave amplitude with height is due to gravitational stratification, as explained above. Once the collision frequency decreases with height, the effects of the linear wave damping become progressively more important and $k_I$ changes sign. 
The damping starts at lower heights for shorter period waves. This happens because the wave frequency  approaches the neutral-ion collisional frequency ($\nu_{ni}$) and the collisions alter the wave propagation,  consistently with Figure \ref{fig:adim_k_om}, which is where we did not take the stratification into account.  The damping also starts at lower height for stronger magnetic fields, which is consistent with the results of the simulations shown in the previous sections. However, the absolute value of damping is higher for lower B (so the damping length is shorter). Decreasing the value of the magnetic field for a fixed wave frequency results in a decrease  of the wavelength, which facilitates the effects of collisions on the wave damping. 

 Considering the propagation for different magnetic field strengths, we observed the following two effects: one due to the stratification, which increases the amplitude of the wave, and the other due to the collisions which damps the wave. The amplification effects are stronger in the lower part of the atmosphere because of the lower temperatures and, consequently, the pressure scale height. The damping effects are stronger in the upper part of the atmosphere. The interplay between two effects can be observed in Fig. \ref{fig:dispRel2Fl4sol} by analyzing the  imaginary part of solutions \#2 and \#3. In the lower part of the atmosphere the values are positive (meaning an increase of the amplitude) because of the stratification, and are identical for \#3 and \#2 because the effects of collisional damping are negligible. When the collisional damping become important in the upper part of the atmosphere, 
the two curves split and become symmetric with respect to the $k_I=0$ line, meaning that the stratification does not play a major role in amplifying the wave amplitude anymore.  The absolute values of both, amplification and  damping are smaller for larger magnetic fields. The height when the damping effects overcome those of the stratification is lower for smaller magnetic fields.

In approaching the limit of the absence of the magnetic field, $B\approx0$ since the temperature of charges and neutrals are the same.\ The propagation speeds of both species also become very close to each other. In this case, there is very little damping for the periods considered here, as we can see in panel b of Figure \ref{fig:kI_P}.

\begin{figure*}[t]
\centering
\includegraphics[width=8.5cm]{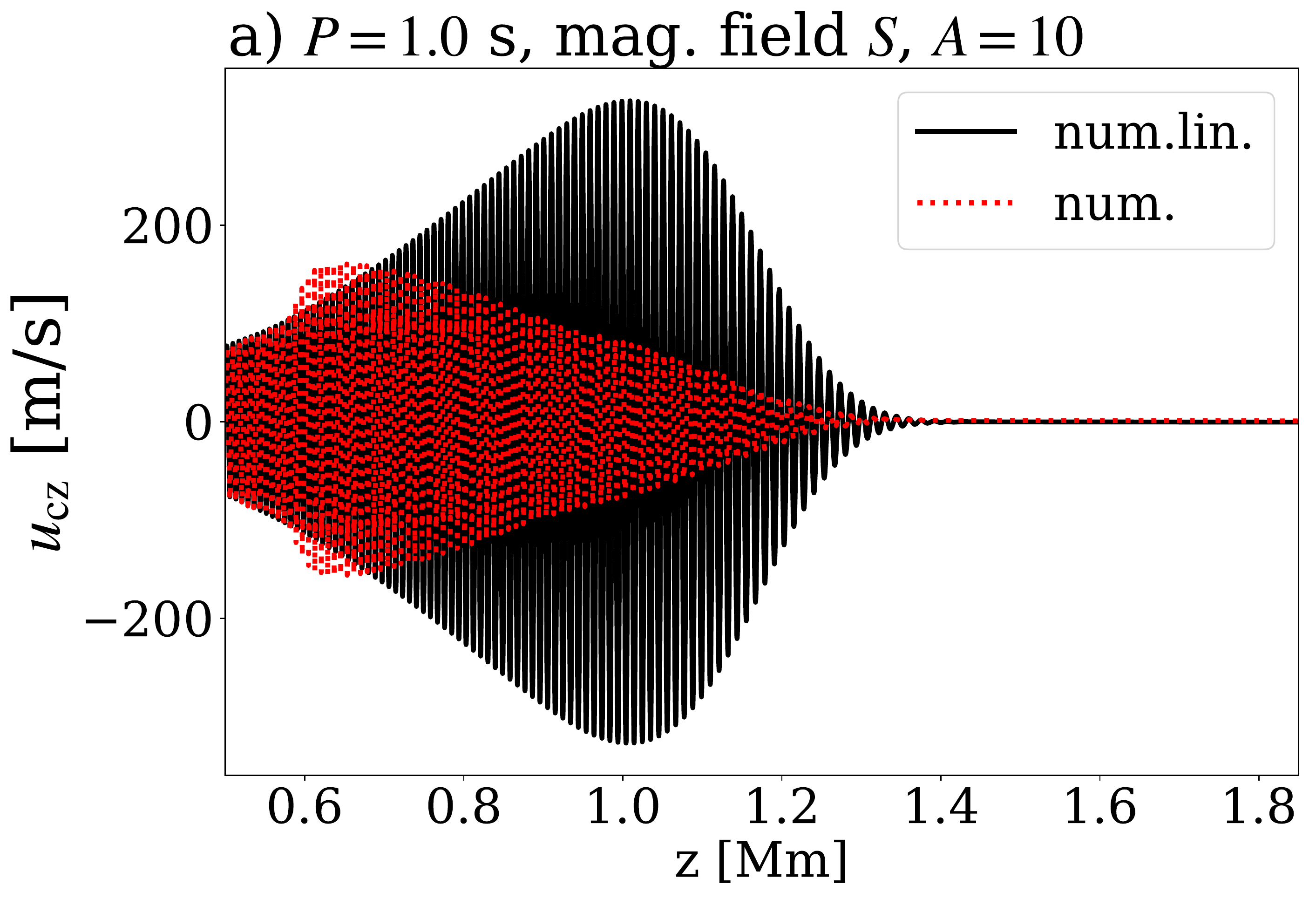}
\includegraphics[width=8.5cm]{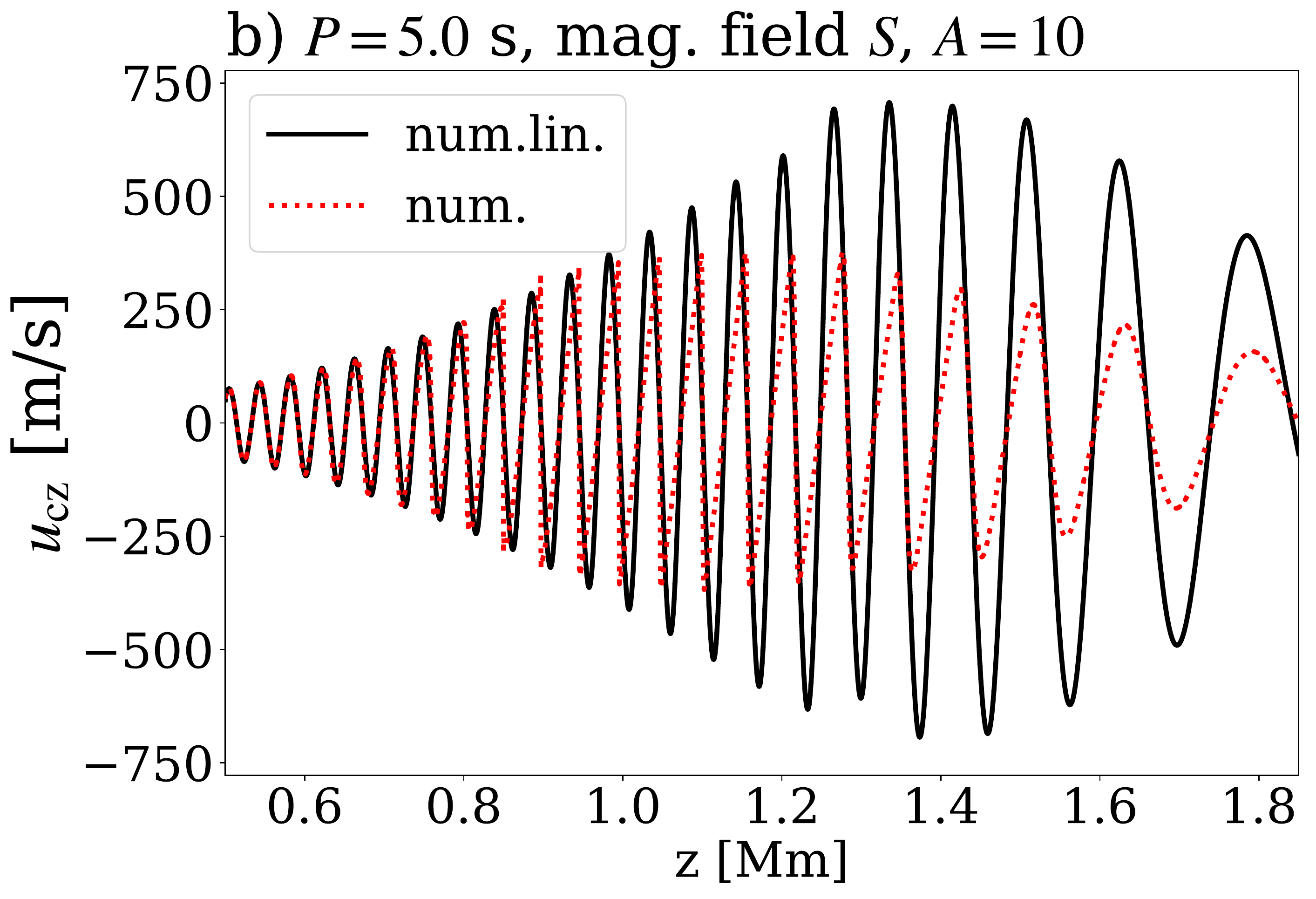}
\includegraphics[width=8.5cm]{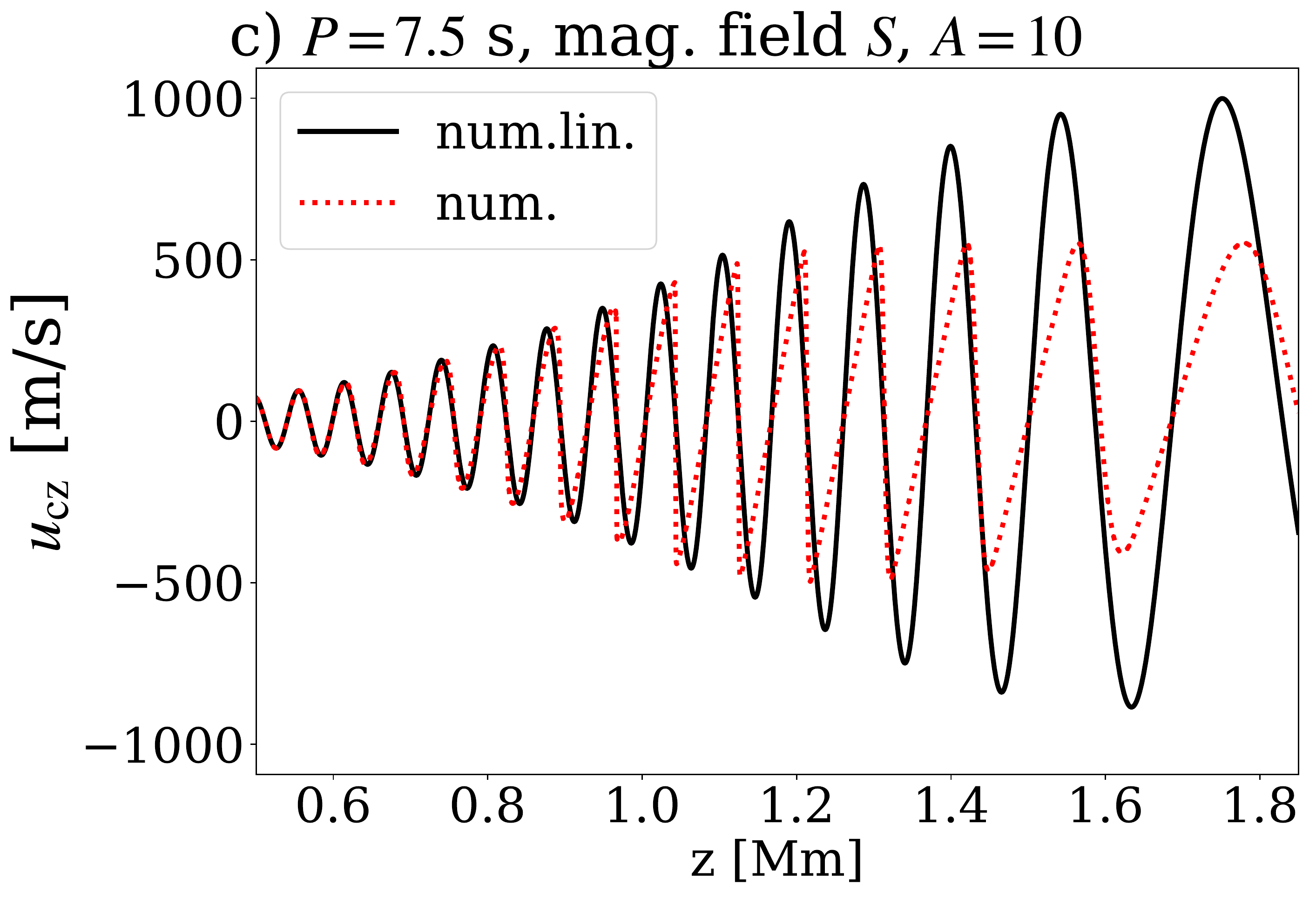}
\includegraphics[width=8.5cm]{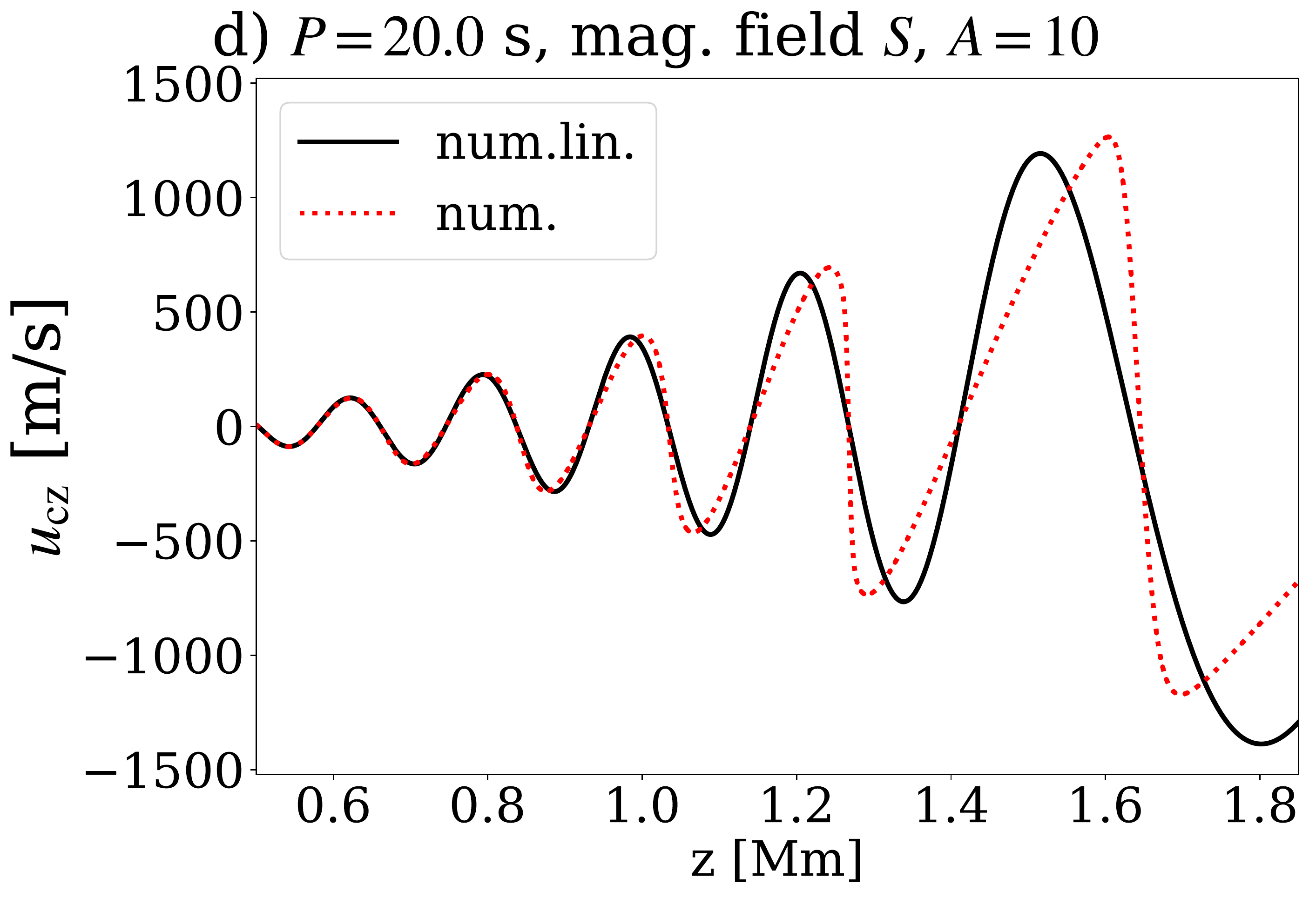} 
\caption{Comparison between numerical solution in linear regime (black lines) and nonlinear regime (red dotted line) with A=10. Individual panels show snapshots of the velocity of charges as a function of height at fixed time moments in the stationary regime of the simulations. Panels from left to right, from top to bottom are for wave periods of 1, 5, 7.5, and 20 s, and S magnetic field profile.}
\label{fig:l-nl-S10}
\end{figure*}

The nonlinear effects on the wave damping are shown in Figure \ref{fig:l-nl-S10} . This figure compares the numerical solutions done with the same initial amplitude (A=10), but in the linear and nonlinear regimes. The nonlinear effects significantly increase the wave damping for all considered wave frequencies. At lower heights, when the nonlinear steepening of the wave profile is not yet pronounced, the linear and nonlinear solutions match, both showing an increase of the wave amplitude with height due to the gravitational stratification. However, at higher heights, the amplitudes of the linear and nonlinear solutions become very different. The waves are significantly more damped in the nonlinear regime, and the effect strongly depends on the wave period. The effective shortening of the scale at the wave fronts facilitates the collisional damping. The nonlinear fronts form at lower heights for shorter period waves and, therefore, nonlinear effects for such waves are significantly more pronounced. One can compare the effective decrease of the amplitude between the nonlinear and linear cases in Figure \ref{fig:l-nl-S10}, showing how the progressively larger periods are less affected by the nonlinear effects. We conclude that nonlinear steepening of the wave fronts dramatically increases the collisional damping of waves in the chromosphere.

%%%%%%%%%%%%%%%%%%%%%%%%%%%%%%%%%%%%%%%%%%%%%%%%%%%%%%%%%
%%%%%%%%%%%%%%%%%%%%%%%%%%%%%%%%%%%%%%%%%%%%%%%%%%%%%%%%%
%%%%%%%%%%%%%%%%%%%%%%%%%%%%%%%%%%%%%%%%%%%%%%%%%%%%%%%%%%
%\begin{figure*}[t]
%\centering
%\includegraphics[width=8.5cm]{kI-M-1fl.pdf}
%\includegraphics[width=8.5cm]{kI-M-5-twofl-NOSTRATI.pdf} 
%\caption{Imaginary part of k as function of the  magnetic field (horizontal axis) and height (vertical axis) obtained after solving the dispersion relations.
%Left:  the one-fluid dispersion relation, Eq. (\ref{eq:d1})($k_I$ does not depend on the period in this case), 
%Right: the dispersion relation for the two fluid equations, with no stratification, Eq. (\ref{eq:d2}), for the wave with $P$=5 s. The lower and the upper
%limits on the horizontal axis are the $S$ and $B$ profiles used in the simulations.}
%\label{fig:str_col}
%\end{figure*}
%%%%%%%%%%%%%%%%%%%%%%%%%%%%%%%%%%%%%%%%%%%%%%%%%%%%%%%%%%
%
%%%%%%%%%%%%%%%%%%%%%%%%%%%%%%%%%%%%%%%%%%%%%%%%%%%%%%%%%%
%%%%%%%%%%%%%%%%%%%%%%%%%%%%%%%%%%%%%%%%%%%%%%%%%%%%%%%%%%

%%%%%%%%%%%%%%%%%%%%%%%%%%%%%%%%%%%%%%%%%%%%%%%%%%%%%%%%%
%\begin{figure*}[t]
%\centering
%\includegraphics[width=8.5cm]{{l-nl-S-1-100}.pdf}
%\includegraphics[width=8.5cm]{{l-nl-S-5-100}.pdf}
%\includegraphics[width=8.5cm]{{l-nl-S-7.5-100}.pdf}
%\includegraphics[width=8.5cm]{{l-nl-S-20-100}.pdf} 
%\caption{Comparison between the numerical solution in the linear regime (black lines) and the non-linear regime with A=100 (red dotted line). Individual panels show snapshots of the velocity of charges as a function of height at fixed time moments in the stationary regime of the simulations. Panels from left to right, from top to bottom are for wave periods of 1, 5, 7.5 and 20 s, and S magnetic field profile.}
%\label{fig:l-nl-S100}
%\end{figure*}
%%%%%%%%%%%%%%%%%%%%%%%%%%%%%%%%%%%%%%%%%%%%%%%%%%%%%%%%%

%%%%%%%%%%%%%%%%%%%%%%%%%%%%%%%%%%%%%%%%%%%%%%%%%%%%%%%%%
\section{Discussion and conclusions}
%%%%%%%%%%%%%%%%%%%%%%%%%%%%%%%%%%%%%%%%%%%%%%%%%%%%%%%%%

In this work we performed simulations of fast magneto-acoustic waves in the solar chromosphere using a two-fluid approach. We used stratified temperature distribution and number densities for charges and neutrals consistent with VALC solar model.\ Additionally, we considered two magnetic field profiles with an order of magnitude of different field strengths. Our main findings can be summarized as follows,
\begin{itemize}
\item 
Waves in neutrals and charges, initially coupled in the photosphere, become decoupled at some height in the chromosphere.\ This means that velocities of co-located fluid elements for neutrals and charges are different. In our simulations, the decoupling happens above 0.7 - 1 Mm height. The difference in the charges-neutral velocity can reach up to 30-50\% of the initial wave amplitude at the bottom of the domain (about 0.5 Mm in our model). 

\item 
The temperatures of neutrals and charges are effectively coupled by the thermal exchange and oscillate with the same amplitude and phase. 

\item 
The decoupling in the wave velocity is a function of the wave period, its amplitude, and the background magnetic field strength. In general, waves with smaller periods show greater decoupling at the same height. Decoupling happens at lower heights for stronger magnetic fields. 

\item 
Charge-neutral collisions cause significant wave damping. We observe significant damping of the shorter period waves, 1 and 5 seconds in our simulations. In such cases, the absolute value of the decoupling is smaller compared to the less damped case because the velocity amplitude strongly decreases after some height in the chromosphere and, in some cases, the perturbation associated with the waves completely disappears. Waves are damped at lower heights when the magnetic field is larger, but the damping length is shorter in the case of the smaller magnetic field.

\item 
The damping obtained in the simulations in the linear regime can be rather precisely described by the analytic theory solving linear two-fluid equations using the WKB approach. This level of agreement suggests that the damping observed in such simulations is determined by the linear effects due to decoupling. 

\item 
Nonlinear wave propagation effects and steepening of wave fronts largely enhance the effects of the linear collisional damping. Nonlinear effects for the damping are more pronounced for the shorter period waves because the shock formation happens at lower heights. 

\item 
Collisional damping produces an effective frictional heating and local background temperature increase. The rate of the temperature increase is constant in time and is proportional to the square of the velocity amplitude. 
\end{itemize}

%The kinetic energy of the waves is transferred into internal energy by the work done by the charge-neutral collisions, leading to an increase in the background temperature. This two-fluid effect is similar to the effect of ambipolar diffusion in the single-fluid description \citep{Khomenko+Collados2012}. The difference between the two approximations is that, in the single-fluid description, the ambipolar effect allows to convert the magnetic energy of a wave into heat via the dissipation of currents. This produces a local increase of temperature. In the two-fluid treatment, the kinetic energy of waves is converted directly into heat. The local temperature enhancement results in the plasma expansion, carrying with it the magnetic field lines. Therefore, the magnetic field locally decreases at the location of the temperature enhancement.

The charges-neutral velocity decoupling and wave damping are two related multi-fluid effects. The decoupling appears when the collisional time scales become similar or longer than the hydrodynamic time scales. The decoupling determines damping of the waves. The kinetic energy lost in the work done by the collisional terms is converted into internal energy through the frictional heating term, which is a positive quantity added in the same amount to the energy equation of both species. The damping of several types of waves by collisions is also suggested in other studies \citep[e.g.,][]{2011Zaq, 2013Soler, 2013Zaq, 2013Soler2}. 

In our case, the hydrodynamic time scales for the charges are determined by the ion cyclotron frequency and by the wave frequency. The wave frequencies considered here are much smaller than the ion cyclotron frequency at all heights. The collision time scales are determined by the ion-neutral and neutral-ion collision frequencies for the charges and neutrals, respectively. The relative importance of collisional terms, which couple the evolution of charges and neutrals, varies strongly with height due to the gravitational density stratification, as shown in Figure \ref{fig:freq}. For the small background magnetic field, $S$ profile, the ion-neutral collision frequency becomes smaller than the cyclotron frequency at $z \approx 0.8$ Mm in the atmosphere, and we expected the ions to decouple from the neutrals above this point, leading to wave damping. With the same argument, but for the larger magnetic field ($B$ profile), we expected the waves to be damped at even lower heights than for the $S$ profile because the ion cyclotron frequency increases and equals the collision frequency at the bottom of the atmosphere. Our numerical simulations show that this is indeed the case. However, there is an additional effect caused by the gravitational stratification of the atmosphere, which can produce strong variation of the perturbation wavelength with height, depending on the magnetic field strength. The perturbation wavelength is larger for larger fields. Consequently, since the perturbation gradient spatial scale is larger, the absolute value of the damping in the strong-field case is smaller (or, in other words, the damping length is larger). Both numerical simulations and the analytical WKB solution show similar values of the damping.

In the case of the stratified atmosphere, the effect of the damping competes with the effect of the wave amplitude increase due to the gravitational stratification. The exponential decrease of the background density with height results in the increase of the wave amplitude at lower heights. This effect is well reproduced in the analytical WKB solution. 

The nonlinear effects result in the steepening of the wave fronts and in the formation of shocks. 
The effective spatial scale at shock fronts  decreases and, depending on the wave period and amplitude, it can reduce to the collision mean free path between neutrals and charges, which  is
larger in the upper part of the atmosphere, and attains the value of 2 km at the height of 1.5 Mm. But even if the shock width  is several orders of magnitude larger 
than the mean free path, as in the case of \cite{2016Hillier}, its size, that is much smaller than the wavelength,  greatly enhances the damping of the waves. 
This effect is observed in the simulations when comparing the numerical solutions in the linear and nonlinear regimes in Figure \ref{fig:l-nl-S10}.

If the decrease in the amplitude of the shock due to collisional damping is large enough, the discontinuity at the shock front smoothes. This smoothing can be observed in Figure \ref{fig:l-nl-S10} (upper right and bottom left panels). The wave fronts have a marked saw-tooth profile at intermediate heights, but they are smooth at the upper part of the atmosphere. These smoothed shock profiles resemble a C-type shock. Shock waves in a weakly collisional interstellar medium were extensively studied by \cite{1971Mullan}, \cite{1980Draine}, and  \cite{1983Draine}.  It was found that when the Alfv\'en speed is larger than the shock speed, these shocks are preceded by a "magnetic precursor," which heats and compresses the medium ahead of the front where the neutral gas undergoes a discontinuous change of state. If the magnetic field is strong enough, the shock will be a C-type shock (when all the variables are continuous across the shock). Otherwise, the neutral gas variables are discontinuous (J-type shock). A substantial fraction of the energy would be dissipated in this magnetic precursor because of the ion-neutral streaming. The solar atmosphere differs from the interstellar medium due to significantly higher density and temperature; therefore, the decoupling is less pronounced.  Nevertheless, \cite{2016Hillier} find the existence of two both C-type and J-type shocks in their study of the reconnection-driven, slow-mode shocks in the solar corona. \cite{2016Hillier} observe  a continuous transition from C-type shocks for subsonic velocity upstream of the shock to J-type shocks, for all the variables, in the case of supersonic upstream velocity. 
 A further study by \cite{2019Snow}, who used   a magnetic field inclined with respect to the direction of propagation, reveal 
long-lived intermediate (Alfvén) shocks  within the slow-mode shock with 
a shock transition from above to below the Alfvén speed and a reversal of the magnetic field across the shock front.
Studying the structure of the shocks requires very good spatial resolution and will be the subject of our future investigations.

%Because the propagation speed of the charges is higher than the propagation speed of the neutrals, the neutrals will lag behind the  charges at the shock front where they decouple, but the decoupling is small, and we could barely see the transition a little bit smoother for the charges. 

\begin{acknowledgements}
This work was supported by the Spanish Ministry of Science through the project AYA2014-55078-P and the US National Science Foudation. Any opinion, findings, and conclusions or recommendations expressed in this material are those of the authors and do not necessarily reflect the views of the National Science Foundation. It contributes to the deliverable identified in FP7 European Research Council grant agreement ERC-2017-CoG771310-PI2FA for the project ``Partial Ionization: Two-fluid Approach''. The author(s) wish to acknowledge the contribution of Teide High-Performance Computing facilities to the results of this research. TeideHPC facilities are provided by the Instituto Tecnol\'ogico y de Energ\'ias Renovables (ITER, SA). URL: http://teidehpc.iter.es
\end{acknowledgements}

\bibliographystyle{aa}
%\bibliography{aa2Bib}

\end{document}